\documentclass{ws-b8-5x6-0}
 \usepackage{amssymb,amsmath}  
 \usepackage{graphicx}

\DeclareMathOperator{\Tr}{Tr}

\usepackage{graphicx,latexsym}

\usepackage{amssymb,amsmath}  
\usepackage{amsfonts}         
\usepackage{dcolumn}          
\usepackage{bm}               
\usepackage{graphicx}

\begin{document}
\chapter{Bridges from Lattice QCD to \mbox{Nuclear Physics}}
\author{A.M.~Green}
\addauthors{A.M.~Green}
\address{Helsinki Institute of Physics, P.O. Box 64, FIN--00014, Finland\\
E-mail:anthony.green@helsinki.fi}
\renewcommand{\theequation}{\arabic{chapter}.\arabic{equation}}

\begin{abstract}
A review is given of attempts to bridge the gap between everyday
particle and nuclear physics --- involving many quarks --- and the basic
underlying theory of QCD that can only be evaluated exactly for few
quark systems. Even the latter requires the original theory of QCD to be
discretised to give Lattice QCD --- but   this modification can still
yield {\it exact} results for
the original theory. These LQCD results can then be considered on a similar
footing to experimental data --- namely as cornerstones that must be
fitted by phenomenological models. In this way, the hope is that 
``QCD inspired" models can become more and more ``QCD based" models, by
fixing ---  in the few-quark case where LQCD can be carried
out --- the form of these models in such a way that they can be 
extended to multi-quark systems.
\end{abstract}

\section{Introduction}

Even though for over 30 years QCD has been  thought to be the theory of
strong interactions, it has had a rather limited impact on most other 
branches of physics  --- except for few-quark hadron  physics. 
Of course, the reason is well known ---  to write
down the Lagrangian that describes {\it exactly} the quark and gluon
interactions is easy, but actually performing calculations
{\it directly} with this Lagrangian has turned out to be extremely difficult.
The one exception to this last statement is at high energies, where --- due to
asymptotic freedom --- the interactions become sufficiently weak for
perturbation theory to be applicable and this has had much success
\cite{pertref}.
However, most of ``everyday" physics is far from this limit.
Furthermore, some quantities such as masses depend on the interaction
strength $g$
as $\exp(-1/g^2)$, which immediately rules out a perturbation
expansion in powers of $g$.

This inability to treat the QCD Lagrangian directly has led to several
different types of approximation being made. Essentially these fall into
\mbox{two} broad categories: the numerical and the effective  theory
approaches. Unfortunately, the latter 
 --- especially the Effective Potential Theories (EPTs), which
are the main subject of this chapter --- often have little
overlap with the numerical approaches.
 Those approaches that use the direct numerical way
concentrate mainly on the description of single mesons or baryons {\it i.e.}
$q\bar{q}$ or $qqq$ states, whereas most of the EPTs
tend to concentrate more on multiquark systems. However,
even though these EPTs are often advertized as being ``QCD motivated",
in most cases they are simply based on many-body ideas and techniques
that are well founded in nuclear physics --- but are not necessarily
justified for the description of multiquark states. The main purpose
of this chapter is to see to what extent these ``nuclear physics motivated"
methods are justified. But first a few general words should be said about
these two approaches.

\subsection{Numerical treatment of QCD}
\label{Numtreat}
The numerical treatment of the QCD Lagrangian --- Lattice QCD --- has been
the main subject of this volume and so, to avoid too much repetition,
only points relevant to this chapter will be mentioned. In Lattice QCD
the original exact Lagrangian is replaced by an approximate form that is
discretized on a 4-dimensional lattice with links of length $a$.
This discretization can be done in many ways, but in all cases the original QCD
Lagrangian must be recovered as $a \rightarrow 0$. The discretization
also reduces the subject to being mainly numerical.
However, as emphasized by   L\"{u}scher~\cite{lurev}:
"In general, numerical simulations have the reputation of being an
approximate method that mainly serves to obtain qualitative information
on the behaviour of complex systems. This is, however, not so in lattice
QCD, where the simulations produce results that are exact (on a given
lattice) up to statistical errors. The systematic uncertainties related
to the non-zero lattice spacing and the finite lattice volume then still
need to be investigated, but these effects are theoretically well
understood and can usually be brought under control."
Therefore, in order to recover results that are appropriate to
the original continuum Lagrangian, two main limits need to be studied:
\begin{itemize}
\item {\bf Limit 1. Are the results stable as 
${\bf a \rightarrow 0}$?}\\ 
Of course, this limit must be approached  with consideration of the
number of spatial sites in the  lattice $N^3$, since the volume $(V)$ of the
system being studied, {\it e.g.} 2-, 3- or 4-quarks, should be much smaller than
the physical lattice size $L^3=(aN)^3$. Furthermore, the Euclidean
time --- needed to extract observables --- should be much greater than $L$.
In practice, usually $T\approx 2L$ suffices. Therefore, the two inequalities that
need to be satisfied can be combined  as \mbox{$V^{1/3}\ll L \ll T $.}
This can instantly lead to problems for a meson--meson system
($q^2\bar{q}^2$) with mesons of size $\approx 1$ fm, since this would
need $V^{1/3}> 2$~fm, if the gradual separation of the mesons is of
interest. In this case, a lattice spacing of $\approx 0.1$ fm would
require $N>20$ and so $T>40$. Such large lattices are used by some
groups {\it e.g.} $32^3\times 60$ in the study of scattering lengths in
Ref.~\cite{ch6Aoki:2001hc} --- with even larger lattices $64^3\times 128$
now becoming feasible~\cite{lurev}. In fact, the progress in computer
technology allows the lattice extents to be doubled in all directions
roughly every 8 years~\cite{lurev}.
Unfortunately,  at present many of us have to be
satisfied with sizes more like $16^3\times 24$, which rules out the study
of completely separated mesons.
One way of partially overcoming this problem is to use  so called
Improved Actions. These incorporate modifications to the standard
lattice Lagrangian in order to remove the lowest order dependences
on $a$. This enables coarser lattices to be used, so that, in some cases,
$a\approx 0.5$ fm suffices compared with more usual values of 
\mbox{$a\approx 0.1$ fm} --- see Refs.~\cite{Lepage1}. 
In this way, the inequality $V^{1/3}\ll L$ is satisfied by
increasing $a$ and not $N$. Of course, there is often a price to pay. Since
improved actions are more complicated, they take more computer time to
implement compared with the standard actions. Sometimes this
is sufficient to remove the advantage of using a smaller lattice.
Having said that, there are some improved actions that seem to be
always advantageous.
Probably the most common of these  is the so-called
clover action for improving the quark part of the QCD Lagrangian
\cite{clover1,clover2} --- see Subsec.~\ref{EffFT} and also Appendix~B.5
of Chapter~4.

\vspace{0.3cm}

Often the $a\rightarrow 0$ limit is checked in three steps:
\begin{itemize}
\item i) First a {\bf benchmark calculation} is performed with, say, a lattice
$16^3\times 24$ and $a=0.12$ fm. This is the least time and storage
consuming of these three steps.
\item ii) Then the {\bf finite size} effect  $V^{1/3}\ll L $ is checked
with a larger lattice, say, $24^3\times 32$ but with the same $a=0.12$
fm. This is much more time and storage consuming by a factor of about 4,
but will show whether or not the process being treated ``fits" into the
lattice. It is concluded that the finite size effect is no problem,
if these latter results are directly the same, within error bars, as the
benchmark results.
\item iii) Finally, the calculation is repeated with the above larger
lattice but using, say,  $a'=0.08$ fm and the results compared with the above by
now including appropriate factors of $a'/a$. These final {\bf scaled} results
should now agree, within error bars, with the results from the above two
$a=0.12$ fm lattices.
\end{itemize}
A specific example of this {\bf scaling} procedure appears later in
Subsec.~\ref{4Qparam433}.
This raises the question: Given a fixed number of ``computer units" for a
calculation, then what is the most efficient way of spending these
units?
As pointed out by Kronfeld~\cite{Kronfeld}, it is much more efficient to run at
several lattice spacings than to put all the resources onto the finest
conceivable lattice.  Kronfeld gives the example of a computer budget of
100 units and
suggests using 65, 25 and 10 units on  a series of coarser
lattices with spacings $a_0$, $a_1=2^{1/4}a_0$ and $a_2=2^{1/2}a_0$.
The  time needed to create statistically independent
lattice gauge fields grows as $\tau_g\propto a^{-(4+z)}$, where
the 4 in the exponent arises because the number of variables to process
grows as $a^{-4}$ in a 3+1 dimensional world and $z\approx 1-2$
depending on the algorithm for updating the lattice.
In this case the three sets of lattices would have comparable error bars
that would only be slightly larger --- by a factor of about 1.25 ({\it i.e.}
$1/\sqrt{0.65}$) --- than the case of using all 100 units on
the finest lattice with $a_0$. However, using all three sets enables an
estimate to be made of the discretization effect. As Kronfeld says
``The slightly larger statistical error seems a small price to pay".\\

\item {\bf Limit 2. The mass of the light quarks should be realistic}. \\
This seems to be a
much more difficult limit to achieve. In practice, light quarks ($u,d$) with
a mass $\sim 100$ MeV are often used instead of the true values of
less than 10 MeV. This is reflected in the computed ratio of the $\pi-$
and $\rho-$masses
$R_{\pi \rho}=m_{\pi}/m_{\rho}$. The experimental masses give
$R_{\pi \rho}=0.2$, whereas $R_{\pi \rho}\approx 0.7$, if 
$m_{u,d}\approx m_s$ --- the strange
quark mass (a value used in many works). Unfortunately, extrapolating from results
using different bare quark masses to get the observed $R_{\pi \rho}=0.2$ is
not straightforward. At sufficiently low  values of the bare quark
masses  the effective field theory of Chiral Perturbation Theory becomes
applicable and shows that in these quark mass extrapolations logarithmic
terms arise in addition to simple power-law behaviour --- see
Subsec. \ref{EffFT}.
\end{itemize}

The above limits are discussed in more detail in Ref.~\cite{Kronfeld2}. There it
is pointed out that QCD is a multiscale problem. Not only is there
a  characteristic scale of QCD
($\Lambda_{\rm QCD}\sim 200 - 250 $ MeV~\cite{lurev}) but
also a wide range of quark masses with light quark masses $m_q\approx$
10 MeV up to \mbox{$m_Q\approx$ 5 GeV} for the $b$-quark, leading to the hierarchy
\begin{equation}
\label{hier1}
m_q \ll \Lambda_{\rm QCD} \ll m_Q.
\end{equation}
But two more scales are needed before QCD can be put on a lattice.
Firstly, for light quarks the lattice size ($L$) must be larger than
the size of a
light quark {\it i.e.} $m_q^{-1}\ll L$. Secondly, for heavy quarks the lattice
spacing ($a$) must be finer than the size of such quarks {\it i.e.} $a \ll
m_Q^{-1}$.
The hierachy in Eq.~\ref{hier1} then becomes
\begin{equation}
\label{hier2}
L^{-1}\ll m_q \ll \Lambda_{QCD} \ll m_Q \ll a^{-1}.
\end{equation}
However, for numerical reasons it is not possible to satisfy all these
conditions. So that, in practice, finite computer resources force the
hierachy
\begin{equation}
\label{hier3}
L^{-1} <  m_q \ll \Lambda_{QCD} \ll m_Q \sim a^{-1}
\end{equation}
instead of the idealized one.

In Ref.~\cite{AlanI} the dependence on $a$ and the appropriate value of
$m_{\pi}$ for the work required to obtain a ``new" configuration are
combined into the single approximate expression
\begin{equation}
\label{Gflops}
\frac{{\rm Gflops}}{{\rm config}}\approx 0.157 \left(\frac{L}{a}\right)^{3.41}
                \left(\frac{T}{a}\right)^{1.14}\left(\frac{1}{am_{\pi}}\right)^{2.77},
\end{equation}
where 1 Gflop is $10^9$ computer operations per second. 
This shows the $\sim a^{-8}$ dependence that is the main numerical
problem  for Lattice QCD to overcome and is the reason for being interested in using
coarser lattices with improved actions to be discussed in the next Subsection.
Estimates similar to Eq.~\ref{Gflops} are also made in Refs.~\cite{lurev,Farc}.

For those readers who would like a detailed development of Lattice QCD
the text books by Creutz \cite{Creutz}, Montvay and M\"{u}nster
\cite{Montvay}, and Rothe \cite{Rothe}  are recommended. Also there are
many review articles and summer school lecture series --- see
Ref.~\cite{Kronfeld} for a partial listing of these.
At the time of writing, some of the most recent reviews for a general 
audience are listed in Ref.~\cite{lurev}.\footnote{A  popular level
review in the February 2004 edition of 
Physics Today\protect\cite{DetGo2} has not been well received by
everyone \protect\cite{Neub2}.}

\subsection{Effective Field/Potential Theories}
Effective  theory formulations describing quark--gluon systems
fall into distinct categories. On the one extreme are the Effective Field
Theories \mbox{(EFTs)} that have a  rigorous basis, whereas at the other extreme
we have the Effective Potential Theories (EPTs), which are
essentially phenomenological being based on models  with potentials in
differential equations. It is important to discuss these two types of
theory separately,
since they play very different r\^{o}les in the present chapter and should
not be confused with each other.
\subsubsection{Effective Field Theories (EFTs)}
\label{EffFT}
Effective field theories play a crucial part in extracting
continuum (physical) results from the purely numerical lattice
techniques. A review of this topic has been given by Kronfeld~\cite{Kronfeld}.

In these theories an energy scale $(\Lambda)$ is introduced. This
essentially separates the
{\it short} distance effects  ({\it i.e.} less than $1/\Lambda$), which are lumped into
the coupling constants of the theory, from {\it long} distance effects, which
are described explicitly by the operators of the theory. Such theories are then
only applicable to  processes involving energies less than $\Lambda$.
In QCD the energy scale characteristic of non-perturbative effects is
$\sim 1$ GeV. There are several theories that fall into this category,
examples of which are:
\begin{enumerate}
\item {\bf Symanzik effective field theory }\\
The most obvious difference between Lattice QCD and the real life
situation of continuum QCD is the presence of the lattice with spacing
$a$. Only the $a\rightarrow 0$  limit has a physical meaning. A
systematic way of studying this lattice artifact was developed  by
Symanzik~\cite{Symanzik}. He showed how $O(a)$ effects could be removed in
a systematic way from lattice results by assuming $\Lambda a$ is small
and treating lattice artifacts as perturbations. He achieves this through
creating an effective field theory by adding terms with increasing
powers of $a$ (and containing parameters
$c_i$) to the basic lattice QCD Lagrangian of Wilson.
 The $c_i$ parameters are then adjusted (tuned)
to kill off the offending $O(a)$ effects. This has now been developed
into an industry for generating improved actions that only contain $O(a^2)$
lattice spacing corrections --- see Subsec.~4.1.7 and Appendix~B of
Chapter~4 for a more detailed description and for references.

A simple example of this is the quark--gluon coupling $\Gamma _{\mu}(p,p')$,
where $p, \ p'$ are the 4-momenta of the initial and final quark.
The above strategy is to first replace the continuum coupling
$\Gamma^C_{\mu}(p,p')$ by its lattice counterpart $\Gamma^L_{\mu}(p,p')$
and then to expand the latter in powers of $a$ {\it i.e.}

\begin{center}
$\Gamma^C_{\mu}(p,p')=c\gamma_{\mu}{\bf \longrightarrow}$

$ \Gamma^L_{\mu}(p,p')=
c\{ \gamma_{\mu}\cos[\frac{1}{2}(p+p')_{\mu}a]-i\sin[\frac{1}{2}(p+p')_{\mu}a]\}$

${\bf \longrightarrow}
c\{ \gamma_{\mu}-\frac{i}{2}a(p+p')_{\mu}+O(a^2)\}.$
\end{center}

To remove the $O(a)$ term,  Sheikholeslami and Wohlert \cite{clover1}
suggested adding
a lattice form of $\sigma_{\mu \nu}F^{\mu \nu}$ to the Wilson action
so that $\Gamma^L_{\mu}(p,p')$ became
\begin{multline}
\Gamma^L_{\mu}(p,p')=
c\{
\gamma_{\mu}\cos[\frac{1}{2}(p+p')_{\mu}a]-i\sin[\frac{1}{2}(p+p')_{\mu}a]+\\
\frac{1}{2}c_{SW}\sigma_{\mu \nu}\cos[\frac{1}{2}k_{\mu}a]
\sin[\frac{1}{2}k_{\nu}a]\}.
\label{Cloveract}
\end{multline}
Expanding  $\Gamma^L_{\mu}(p,p') {\bf \longrightarrow}
c\{ \gamma_{\mu}-\frac{i}{2}a[(p+p')_{\mu}+c_{SW}i\sigma_{\mu
\nu}k^{\nu}]+O(a^2)\},$
where $k=p'-p$ and $\sigma_{\mu \nu}=i[\gamma_{\mu},\gamma_{\nu}]/2$.
On the mass shell, if $c_{SW}$ is now ``tuned" to unity, then the two
terms of $O(a)$ cancel to leave corrections of only $O(a^2)$. When this
procedure is applied to  Wilson's fermion action the outcome is usually
referred to as the {\bf clover} action.

\item {\bf Chiral Perturbation theory  of
Gasser and Leutwyler~\cite{GasserL}}\\
Light quarks have a mass $m_l\approx 10$ MeV {\it i.e.} $m_l \ll \Lambda$.
This makes it numerically impractical to perform lattice QCD
calculations with such masses since the algorithms for computing
the quark propagators become slower and slower --- as seen from Eq.~\ref{Gflops}.
Therefore, the
procedure to reach this physical region is to first carry out the
lattice calculations with a sequence of masses ($m_q$) in a range of, say,
$0.2m_s<m_q<m_s$, where $m_s\sim 100$ MeV -- appropriate for the strange
quark. Given this sequence of results (masses or matrix elements), the
task is then to use a reliable method for extrapolating these results to
quark masses appropriate for the light quarks. By far the most
successful method for this extrapolation is based on  Chiral
Perturbation Theory ($\chi$PT), which
can be viewed as  an expansion in $m_q/\Lambda$. The numerical data,
with $m_q$ in the range $0.2m_s<m_q<m_s$,
can then be tested against the leading order (next-to-leading order
or next-to-next-to-leading order) prediction of  $\chi$PT. If this is
successful, then it gives confidence in extrapolating $m_q$ to the light quark
masses \cite{TT2002}.
\item {\bf Heavy-quark effective theory and
Non-relativistic QCD}\\
For a brief review of Heavy-quark effective theory (HQET) and
Non-relativistic QCD (NRQCD) see Ref.~\cite{Soto}.
In situations involving heavy \mbox{quarks} --- such as $B$-physics, where
some of the
quarks have a mass of $m_Q \approx 5$ GeV and are non-relativistic ---
it is appropriate to make 
expansions in terms of $\Lambda/m_Q$ or $\nu$, the relative velocity
between the quark and the  antiquark in the $B$-meson.
These two expansions
are usually referred to, respectively, as HQET
--- for  systems containing a single heavy-quark --- and
NRQCD for a heavy-quark heavy-antiquark system.
One way of deriving these
effective theories is to write down the heavy--quark theory as an
expansion  in terms of $\Lambda/m_Q$ or $\nu$ in the continuum and then 
replace the derivatives that arise
by their lattice counterparts to give Lattice HQET and Lattice NRQCD.
These ideas are still being developed. For example in Ref.~\cite{vario}
some of the irrelevant degrees of freedom in NRQCD are integrated out
to yield a theory called potential NRQCD (or pNRQCD), which is much
simpler to treat.

\end{enumerate}
EFTs are not only used with few-quark systems but also have a long
history in few- and many-nucleon systems --- see the works of
van Kolck {\it et al.}~\cite{Kolck,oller}. This approach was first advocated
by Weinberg \cite{Weinberg}, who illustrated how the nucleon--nucleon
and many--nucleon potentials could be qualitatively understood. For
example, these arguments show that, if the strength of the NN-potential is
$\sim 10$ MeV, then those of the NNN- and NNNN-potentials are $\sim 0.5$
and $\sim 0.02$ MeV respectively --- numbers that are in accord with
detailed few-nucleon phenomenology based on realistic potentials such as
that of Argonne \cite{Argonne}. However, it is not clear that this
approach --- in spite of its impressive {\bf qualitative} results --- could
ever compete {\bf quantitatively} with standard meson exchange models for
describing the NN-potential, where baryon resonances such as the
$\Delta (1236)$ and $N^*(1535)$ are included explicitly. As pointed out by
van Kolck himself, even the inclusion of the $\rho$- and $\omega$-mesons
give rise to interactions that  are ``at present an insurmountable
obstacle for a systematic approach".

In multi-particle systems EFTs are often converted into Mean Field
Theories (MFTs), in which the emphasis is on single particle properties
with all other particles being treated ``on the average".
Unfortunately, in some applications --- such as the equation of state of
high density nuclear matter, as encountered in relativistic heavy ion
collisions or in neutron stars --- there are serious questions
concerning their validity. Even advocates of the MFT approach 
(see, for example, Glendenning on pages 127 and 287 in Ref.~\cite{Glen}) express
reservations by writing
``In many ways it is not as good a theory as the Schr\"{o}dinger-based
theory of nuclear physics" and
``The status of an exotic solution of an effective theory is more
tenuous than from a fundamental theory".
Others are even more critical --- see for example Ref.~\cite{APR}:
``The Relativistic Mean Field approximation is very elegant and pedagogically
useful, but is not valid in the context of what is known about nuclear
\mbox{forces $\cdots$ .} \ It requires $\mu \ \langle r \rangle \ll 1$, where $\langle
r \rangle$ is the average
interparticle distance and $1/\mu $ a meson range. However, for pions
 \mbox{$\mu_{\pi} \ \langle r \rangle \sim 0.8 - 1.4$} but for vector mesons
$\mu_{v} \ \langle r \rangle \sim 4.7 -5.8$."
\subsubsection{Effective Potential Theories (EPTs)}
\label{sect.EPT}
The reason for briefly describing  the above Effective Field
Theories (EFTs) is to emphasize their difference from Effective Potential
Theories (EPTs), which are the main interest in most of this chapter.
The above EFTs are an integral part of the development of Lattice QCD
and play a crucial r\^{o}le in extracting precise continuum results for
few (2 or 3) quark systems {\it i.e.}
\begin{center}
Lattice QCD+{\bf EFTs} $\longrightarrow$ Continuum results for few quark systems.
\end{center}
On the other hand, the EPTs attempt to understand (interpret) these continuum
results for few quark systems in such a way that the theory can be
extended to the multiquark
systems of interest to nuclear physics {\it i.e.}
\begin{center}
Continuum results for few quark systems + {\bf EPTs} $\longrightarrow$\\
Descriptions of multiquark systems (Hadron Physics).
\end{center}
This step is here referred to as ``Bridges from Lattice QCD to
Nuclear Physics" --- the title of this chapter. In most cases this
step is mainly phenomenological.

The EFTs mainly concentrate on the properties of a {\it single}
particle, so that, for example, the energy of a multiparticle system is
expressed as the sum of the effective masses of the separate particles
with the effect of all other particles being treated in an average
manner. In this way symmetries of the fundamental Lagrangian can be
preserved. However, as said above, the extension of these ideas to more
complicated Lagrangians or many-body systems presents problems.
On the other hand, in the less ambitious and more phenomenological approach
of Effective Potential Theories (EPTs),
 the emphasis is first on the {\it two-body} system. In this case,
two-body potentials are the main ingredient.

For multi-nucleon systems
the NN-potential can be mainly phenomenological or based on EFTs
as with the Argonne  and Bonn potentials respectively
\cite{Argonne,Bonn}. Those based on EFTs can be generated with varying
degrees of ability to describe the two-body system. We have already
mentioned the works of Weinberg \cite{Weinberg} and van Kolck \cite{Kolck},
which  follow the procedure --- referred to by van Kolck
as Weinberg's ``theorem":
\begin{enumerate}
\item Identify the relevant degrees of freedom and symmetries involved.
\item Construct the most general Lagrangian consistent with item (1).
\item Do standard quantum field theory with this Lagrangian.
\end{enumerate}
The outcome is  qualitatively correct being within $\approx 10\%$
of the two-nucleon data --- but this is an accuracy that is often
insufficient for understanding nuclear phenomena. At the other
extreme we have the one-boson-exchange (OBE) potentials in which various
meson--baryon couplings are tuned to ensure a good fit to the NN
experimental data. These OBE potentials, even though they are based on
EFT-like Lagrangians, often incorporate couplings such as
$N\Delta(1236)\rho$ that can not be treated systematically by Weinberg's
``theorem".

In contrast, to implement EPTs for multiquark systems three ingredients are
necessary ---  a wave equation (differential or integrodifferential),
an interquark potential and effective quark masses:

\vspace{0.3cm}

\noindent{\bf A wave equation.}\\ Since EPTs are not derived from
more basic principles, the forms of the wave equations are not
predetermined and can vary considerably. Even for two-quark systems
there is a choice.
\begin{itemize}
\item {\it A non-relativistic Schr\"{o}dinger equation.}
This is suitable for heavy-quark mesons ($Q\bar{Q}$) with $m_Q\gg 1$ GeV
and was the form used in Refs.~\cite{Richardson,Cornell} for extracting
the $Q\bar{Q}$-potentials. The best cases for this are the Bottomonium
mesons such as the $\Upsilon(b\bar{b},9.5 {\rm GeV})$ since
$m_b\approx 5$ GeV.
\item {\it The Dirac equation.} Once the quarks are light,
{\it i.e.} \mbox{$m_q\ll 1$~GeV}, relativistic effects become important and we,
therefore, enter
the realm of large/small wave function components, pair-creation and
quantum field theory.
This means that the use of the Dirac equation is less ``clean"
phenomenologically than the Schr\"{o}dinger equation, since it deals
explicitly with large/small wave function components but not with the
related effect of pair-creation. In spite of this, for $Q\bar{q}$
systems such as the $B$-meson, where one quark is heavy, the Dirac equation
is essentially a one-body equation and so the full complications of the
two-body relativistic problem are avoided.
There are many references where this one-body Dirac equation has been
applied, {\it e.g.} \cite{BhBr,BakerBZ,PGG,DiP}.
\item {\it The Bethe--Salpeter equation.}
When the two quarks are both light the correct relativistic scattering
equation is  the \mbox{full} Bethe-Salpeter equation.
Unfortunately, direct use of this equation --- for physically interesting
cases --- presents severe problems (see Ref.~\cite{crater} for a recent
discussion).
Therefore, it is usually reduced from a four- to a three-dimensional
scattering equation by inserting appropriate \mbox{$\delta$-functions}
of the energy. This can be carried out in several ways and leads to a
number of  different  two-body \mbox{equations} that are covariant and satisfy
relativistic unitarity. Examples are the equations of
Blankenbecler--Sugar, Gross, Kadyshevsky, Thompson, Erkelenz--Holinde, \ldots
see Chapter~6 in Ref.~\cite{Brown+J}. Depending on the problem, these
alternatives have their separate advantages. For example, the Gross \mbox{form}
--- unlike those of Blankenbecler--Sugar and Thompson --- treats the two
quarks asymmetrically and has the
feature that it reduces to the Dirac equation when one of the quarks
becomes infinitely heavy. This suggests that this form is perhaps more
appropriate for describing the $B$-meson. On the other hand, when the
quarks are equal in mass, as in the $J/\psi (c\bar{c})$ system, the
Blankenbecler--Sugar or Thompson equations are probably preferable.

\item {\it Multi-quark wave equations.}
For multi-quark \mbox{systems} the choice of wave equation is very
limited with the Resonating Group approach being the most usual ---
see the review by Oka and Yazaki in Chapter~6 of Ref.~\cite{WWeise} 
and also more recently  \mbox{Refs.~\cite{SMBSY,Oka2}.}
This reduces
the interaction between quark clusters ($q^3+q^3$ for the
\mbox{NN-potential}
and $q\bar{q}+q\bar{q}$ for meson--meson interactions) to a {\it non-local}
Schr\"{o}dinger-like equation involving the relative distance between
the clusters. This is achieved by integrating out the explicit quark
degrees of freedom, which usually requires
the introduction of gaussian radial factors in order to carry out the
multiple integrals involved. At first sight such factors may seem to be
unrealistic with exponential or Yukawa forms being more physical.
However, it will be seen in Subsec.~\ref{Diracfit} that the one-body Dirac
equation with a linear confining potential can lead to gaussian
forms  asymptotically.  Unfortunately, the effective mass needed for
the quarks is $\approx 300$--400 MeV and so makes this non-relativistic approach
somewhat questionable. On the other hand,
 the r\^{o}le of relativity in many-body systems is still an open question ---
a recent summary being given by Coester \cite{coester}.
\end{itemize}

\vspace{0.3cm}

\noindent{\bf An interquark potential ($V_{Q\bar{Q}}$) --- the second
EPT ingredient.}\\
For two-quark systems the interquark potential is often taken to
have the form suggested by the static limit of infinitely heavy quarks,
namely
\begin{equation}
\label{VQQ}
V_{Q\bar{Q}}(r)=-\frac{e}{r}+b_sr+c,
\end{equation}
where the first term is that expected at short distances due to
one-gluon exchange and the second term that expected from quark
confinement --- \mbox{$c$ simply} being an additive constant.
However, it should be emphasized that the form in Eq.~\ref{VQQ}
is, strictly speaking, only appropriate for the interaction between static
quarks. At the present time, there seems to be no really convincing evidence for a
significant one-gluon exchange interaction in systems where only light
 constituent quarks (up to $\sim$400 MeV) are involved.
This becomes evident in the N and $\Lambda$ spectra.
There a one-gluon interaction is unable to describe the empirical ordering of
the positive and negative parity states. However, this ordering can be
accomplished by Goldstone-boson exchange mechanisms~\cite{glozdan},
even though that
model also has bad features --- some states in the $\Lambda$ and $\Delta$
spectra are poorly reproduced.
The reason why the one-gluon exchange becomes ineffective with light quarks
is because the use of  minimal relativity --- in, say, the
Blankenbecler-Sugar equation for describing the $Q\bar{q}$
interaction --- introduces relativistic square root factors. Indeed, if both
quarks are light, then the effective one-gluon interaction is essentially flat
and very weak for short distances. Since this damping of
one-gluon exchange  also enters in meson spectra, another mechanism is
needed for the
necessary short range attraction. A possible
candidate for this is an effective interaction generated by instantons, which is usually
expressed in terms of an attractive $\delta$-function in $r$ \cite{inst}.

More details on $V_{Q\bar{Q}}(r)$ can be found in Chapter~3, which
is devoted to this topic.
The above is a two-body potential.
However, for multi-quark systems there are strong indications
that multi-quark potentials and/or potentials involving excited gluon
states could also play a major r\^{o}le --- see
Secs.~\ref{potmodel} and \ref{fmodelext}.

For light quarks and for nucleons in dense matter, relativistic
effects enter and in some cases these can be
expressed as corrections to the non-relativistic potential involved. An
example of this is the so-called Relativistic Boost Correction shown to
be important in high density nuclear matter \cite{boost}.

\vspace{0.3cm}

\noindent{\bf Effective quark masses --- the third ingredient for an EPT.}\\
The masses of the quarks involved in
this  ``Wave Equation + Potential" approach are not those of bare quarks
{\it i.e.} not $\approx 10$ MeV for the
 \mbox{{\it u} and} {\it d} quarks.
They are essentially free
parameters, which are often taken to be 
\mbox{$\approx M_{\rm nucleon}/3 \approx$ 300 MeV.}
However, it has been suggested~\cite{shudan} that a more natural choice would
be $\approx M_{\Delta}/3 \approx$ 400 MeV.

\vspace{0.5cm}


So we see that the idea of EPTs covers an enormous number of theories,
models and approaches that are frequently used in nuclear physics.
The subject of the next section is to see how these EPT ideas can
possibly be utilized in the understanding of QCD. The magnitude of this
step should not be underestimated, since as the authors of
Ref.~\cite{Kolck} say:

``On the one hand, the concensus of the majority of the nuclear physics
community holds that in nuclei
\begin{itemize}
\item nucleons are non-relativistic
\item they interact via essentially two-body forces, with smaller
contributions from many-body forces
\item  the two-nucleon interaction generally possesses a high degree of
isospin symmetry
\item  external probes usually interact with mainly one nucleon at a
time.
\end{itemize}
By contrast, in QCD
\begin{itemize}
\item the $u$, $d$ and even $s$ quarks are relativistic
\item the interaction is manifestly multi-body, involving exchange of
multiple gluons
\item there is no obvious isospin symmetry
\item external probes can, and often do, interact with many quarks at once.
\end{itemize}
It should not be surprising, then, that some new ideas are required to
merge these two extraordinarily different bodies of theory."  Others
might simply say that this is a case of ``Mission impossible".
\section{What Is Meant by ``A Bridge"?}
\label{simplebr}
The main theme of this article is the study of bridges between lattice
QCD and nuclear physics. Unfortunately, what constitutes a possible bridge
is rather subjective, since the basic idea is to compare some quantity that
can be measured by lattice QCD with a ``corresponding quantity" that
arises in more conventional physics as the outcome of some EPT or is
directly connected with experiment.
Of course, the question instantly arises as to whether these two
quantities are indeed comparable {\it i.e.} to what extent are we confident that
we indeed have ``corresponding quantities". In this chapter the main
quantities to be related will be  energies or radial correlations.
This is probably best illustrated by the following simple example.

\subsection{A  simple example of a bridge}
Since it is the main goal  in this chapter, it may prove useful
to the reader  to first see a simple example of what is meant by 
 ``Bridges from LQCD to Nuclear Physics".

The results of any lattice QCD calculation are quantities expressed as
dimensionless numbers.
Therefore, to be able to make a connection with ``real life", one --- or
more --- of these numbers must be compared with its continuum
counterpart that can actually be measured experimentally. This then sets
the physical scale for lattice QCD.
\subsubsection{Setting the scale from the string tension}
\label{ssst}
For many years a quantity frequently used for this comparison was the string
tension ($b_s$),
which --- as its name implies -- is simply the energy/unit length
of the flux-tube ({\it i.e.} string) connecting two quarks and appears in
Eq.~\ref{VQQ}.
Experimentally, estimates can be made of this string tension
from the spectra of mesons and baryons with increasing
orbital angular momentum $(L)$ --- a series of
energies $(E)$ that depend crucially on the string increasing in length.
 This can be carried
out with varying degrees of sophistication. By simply plotting  $L$ versus
$E^2$ this  so-called Regge trajectory is found to be linear  for
both mesons and baryons --- the slope $(\alpha)$ in each case being about
0.9 GeV$^{-2}$ --- see, for example, Figs. 7.33 and 7.34 in Ref.~\cite{Perkins}.
As shown in Ref.~\cite{Olsson} using a simple classical model this slope is
 directly related to the string tension by the expression
$\alpha \approx 1/(8b_s)$. This results in a value of $\sqrt{b_s}\approx
380$ MeV --- a number that is somewhat smaller than the accepted value
of $\approx 440$ MeV, which is more in line with estimates from
string  models that give $\alpha \approx 1/(2\pi b_s)$~\cite{Rebbi}.

A less direct, but more precise, way to extract the string tension is to
first find an effective quark--antiquark potential ($V_{Q\bar{Q}}$) that
describes --- {\it by way of  a
non-relativistic Schr\"{o}dinger equation} --- the above meson energy spectra.
Naturally, for this non-relativistic approach to be realistic the mesons must
be constructed from quarks that are much heavier than the proton.
This, therefore, restricts the analysis to the Bottomonium $b\bar{b}$ mesons,
where the $b$ quark has a mass of about 5 GeV and
possibly the $b\bar{c}$ and $c\bar{c}$ mesons, where the $c$ quark has a
mass of about 1.5 GeV. In fact,  these spectra can be described by a
{\it single} effective potential of the form given in Eq.~\ref{VQQ}
so that a value for the string tension of $\sqrt{b_s}\approx 440$ MeV
results --- equivalent to
$b_s\approx 1$ GeV/fm or $\approx 5 \ {\rm fm}^{-2}$ in other units.
The potentials most frequently quoted are those of Richardson~\cite{Richardson}
and Cornell~\cite{Cornell}.
This value of $b_s$ is now the
experimental number to which the lattice estimate of $V_{Q\bar{Q}}(r)$ must be
matched.
Usually the latter is extracted by measuring a rectangular Wilson loop 
$W(l,t)$ of area $a^2lt$ for
two infinitely heavy quarks a distance $r=al$ apart on a lattice
and propagating a Euclidean time $at$ --- $a$
being the lattice spacing to be determined. Wilson loops will be
discussed in more detail in Sec.~\ref{Wloops}. A key observation, first
made by Wilson \cite{Wilson} in 1974, was that
\begin{equation}
\label{Willoop}
W(l,t)\rightarrow \exp[-tV_{Q\bar{Q}}(l)] \ \ \ {\rm as} \ \ t\rightarrow
\infty.
\end{equation}
Therefore, for sufficiently large $l$,  $V_{Q\bar{Q}}(l)\rightarrow b'_sl$,
where $b'_s$ is the dimensionless counterpart to the
experimental string tension $b_s$ defined in Eq.~\ref{VQQ}. The two are then matched by way of
the ``bridging equation"  $b'_s=a^2 b_s$ to give $a$.
A typical number for $b'_s$ is
$\approx 0.05$ giving $a\approx \sqrt{0.05/5} \approx 0.1$ fm.

\subsubsection{Sommer's prescription for setting the scale}
\label{sommersec}
In order to set the scale, the above has compared the lattice result with
experiment for the most simple of quantities --- the string tension.
However, the experimental data
is mainly probing distances of $r\approx 0.2$ fm to $r\approx 1$ fm and
{\it not} $r\rightarrow \infty$, since the rms--radii of the
$b\bar{b}$ mesons  cover the
range from about 0.2 to 0.7 fm and the $c\bar{c}$ mesons  the range from about
0.4 to 1 fm. Therefore, the experimental data encoded in the potential
$V_{Q\bar{Q}}(l)$ is not optimal for studying the string tension.
Furthermore, lattice calculations of the Wilson loop $W(l,t)$ require
$t\gg l$, so that those $W(l,t)$  dominated by the string tension need to
be evaluated for large values $t$. Unfortunately, as $t$ increases the
Signal/Noise ratio on $W(l,t)$ also increases --- eventually making
measurements for large $r=al$ meaningless. Therefore, on both the
experimental and lattice sides there are problems for making a reliable
estimate of $a$ from the string tension.

In an attempt to overcome this problem, Sommer~\cite{sommerch6} proposed comparing
the {\it potential} $V_{Q\bar{Q}}(l)$ as extracted from the lattice
with that from experiment.
However, to use the words of Sommer, ``We must remember that the
relationship between the static QCD potential and the effective
potential used in phenomenology is {\it not} well understood."
In spite of this, he suggests that the comparison be made at some value of
$r=al$ in the optimal experimental range
of $r\approx 0.5$~fm. Also at these values of $r$,  $V_{Q\bar{Q}}(l)$ can be
more reliably extracted on a lattice. In practice, it is the force, defined
essentially as
\begin{equation}
\label{Force}
F(l)=\frac{V_{Q\bar{Q}}(l)-V_{Q\bar{Q}}(l-a)}{a},
\end{equation}
that is compared through the expression
\begin{equation}
\label{Som}
r^2F(r)|_{r=R(c)}=c.
\end{equation}
 Sommer chose the dimensionless parameter $c=1.65$,  since --- for the
experimental potentials --- this corresponds  to a distance
$R(1.65)\equiv R_0\approx 0.5$~fm. Using the lattice forms of $F(r)$ that
fit the lattice potential,
Eq.~\ref{Som} can be solved for $r$ in lattice units $a$. Comparing this $r$
with $R_0$ then \mbox{gives $a$}
 --- see Ref.~\cite{sommerch6} for more technical details. However, it
should be added that the value of $c$ is somewhat uncertain with some
authors~\cite{PP1} preferring $c=2.44$, which corresponds to
$R_0\approx 0.66$ fm.
This second choice of $c$ gives values of $a$ that are a few percent
larger than before and also in better agreement with the string tension
estimate.

The reason for this rather lengthy description for extracting the scale $a$
is to show that ``A Bridge from Lattice QCD to Nuclear Physics" has
existed for many years.
It should be added that another way of extracting $a$, when dealing with
light quarks, is to use directly the mass of the $\rho$-meson as a
corner stone
and simply compare the experimental mass $m_{\rho}$ with the outcome of
the lattice QCD calculation --- the dimensionless combination
$am_{\rho}$~\cite{Allton2}. However, $m_{\rho}$ is known to be a sensitive indicator of
scaling violations {\it i.e.} how the lattice results depend on $a$ as
$a\rightarrow 0$. It is, therefore, sometimes reserved for this purpose with the
above method of Sommer being used to extract actual values of
$a$ \cite{Edwards}.  The reason for choosing  $m_{\rho}$ and not
$m_{\pi}$ is because  the $\pi$--meson, being so light, is more difficult to
treat on a lattice.

\subsection{Are there bridges other than $V_{Q\bar{Q}}$?}
\label{bridges}
The above simple example showed how the $Q\bar{Q}$ potential $V_{Q\bar{Q}}$
could be related to its lattice QCD counterpart and serve as a means
for extracting the lattice spacing $a$. The question then arises
concerning the possibility of there being other quantities that could be
compared. However, it must be noted that  $V_{Q\bar{Q}}$  and the
related force $F(r)$ are somewhat special and that
it is still true what Sommer wrote in 1993: ``As to today's knowledge,
the force $F(r)$ between two static quarks is the quantity which can
be calculated most precisely". There are several reasons for this and
they should be kept in mind in the following discussion. Firstly, both the
lattice and experimental determination of  $V_{Q\bar{Q}}$ can be done with
good statistical precision and, secondly, the two are what we think they
are. In contrast, the string tension can only be extracted at values of
$r$ that are not necessarily sufficiently asymptotic and where there could
be corrections from model dependent sub-leading terms.
So for the purposes of setting a scale
the use of $V_{Q\bar{Q}}$ is still the best.

Possibilities for bridges, in addition to the use of $V_{Q\bar{Q}}$,
 are listed in Table~\ref{table1}. This is essentially a ``Table of Contents"
for the rest of this chapter.
\begin{table}
\caption{Possible bridges between lattice QCD and nuclear physics }
\vskip 0.5cm
\label{table1}
\begin{tabular}{|c|c|c|c|}
 \hline
System&Quantity matched& Model& Refs. \\ \hline
($Q\bar{Q}$)&String Tension&Regge Trajectory&\protect\cite{Olsson}\\
 &$V_{Q\bar{Q}}$&Schr\"{o}dinger Equation&\protect\cite{sommerch6}\\ \hline
$[(Q\bar{Q})(Q\bar{Q})]$&Energies&Matrix diagonalistion&\protect\cite{GMS,JL}\\
&Flux tube structure&Discretized String&\protect\cite{IP}\\
&&Dual Potential Model&\protect\cite{BBZ}\\ \hline
($Q\bar{q}$)&Energies&Dirac Equation&\protect\cite{MP98,GIJK}\\
&Density distributions&Dirac Equation&\protect\cite{GIJK}--\protect\cite{G+K+P+M2}\\ \hline
$[(Q\bar{q})(Q\bar{q})]$&Energies&Variational&\protect\cite{GKP2} \\ \hline
($QQq$)&Energies&& \protect\cite{JonnaQQq} \\
&Density distributions&&\protect\cite{JonnaQQq} \\ \hline
\end{tabular}
\end{table}
The Lattice QCD $\leftrightarrow$ Nuclear Physics relationship changes as
we go through this list.
The first two rows for the $Q\bar{Q}$ static
quark system have been discussed above. Here the r\^{o}les of the string
tension and $V_{Q\bar{Q}}$ are to set a scale for QCD --- a necessary step
in order for lattice QCD to be compared with experiment {\it i.e.} the flow of
information is  Nuclear Physics $\rightarrow$ Lattice QCD. However, once
the results of lattice QCD can be expressed reliably with physical dimensions
then the information flow is completely Lattice QCD $\rightarrow$
Nuclear Physics. We can now consider the results of lattice QCD on the same
footing as experimental data --- assuming that the $a\rightarrow 0$
limit is under control and that the quarks are sufficiently light as
discussed in Subsec.~\ref{Numtreat}.

It is  lattice data that
models must attempt to fit. Many of these models resort to the use of
interquark potentials --- often the above $V_{Q\bar{Q}}$ --- in various forms
of wave equation. Now the lattice QCD data will possibly be able
to justify --- {\it or rule out} --- such models. At present these
models  are
often simply mimicking techniques that have proven successful in
Nuclear Physics. Hence my earlier statement that they are
``Nuclear Physics--inspired" and not ``QCD-inspired" as is often claimed.

Above I said that the results of lattice QCD can be considered on a similar
footing as experimental data. However, when setting up models, in some ways
lattice QCD data
can sometimes be superior to experimental data, since it can be generated in
``unphysical worlds".\footnote{In Sec.~3.1 of Chapter~3 these are
referred to as ``virtual worlds".} 
Such worlds can have the following unphysical
features that should, in some cases,  also be inserted into the
corresponding models to test the generality of these models:
\begin{enumerate}
\item%
The real world of three coloured quarks ({\it i.e.} SU(3)) can be replace by one
with two coloured quarks ({\it i.e.} SU(2)). Such a world is easier to deal with in
lattice QCD --- for example, there is essentially no distinction between
quarks and antiquarks. Also the system corresponding to a baryon now consists
of only two quarks.
However, this is not simply an academic exercise, since, in practice, it is
found that the ratio of many observables are similar
in both SU(2) and SU(3) --- but with a computer effort that is about an
order of magnitude smaller. An example of this is the ratio
$R=m_{\rm GB}/\sqrt{b_s}$, where $m_{\rm GB}$ is the glueball mass
 (see Chapter~2) and $b_s$
the string tension. It is found  for the glueball with the lowest
mass ($0^{++}$) that $R\approx$ 3.5 for both SU(2) and SU(3)
--- see Sec.~2.2.1 in Chapter~2  and Ref.~\cite{CMSU2}. In Ref.~\cite{SUNC} this is
extended to the general case of SU($N_C$) for several glueball states and
for the $0^{++}$ case results in
$R(0^{++})=3.341(76)+1.75/ N_C^2$.
\item%
 The real world with  3 space coordinates and 1 time coordinate (3+1)  can be
replaced by one with  1 or 2 space coordinates and \mbox{1 time} coordinate
(1+1 and 2+1).
On the lattice the latter are
easier to study so that results with such high accuracy can be achieved
that there is little ambiguity in any final conclusions.
Also the (2+1) world has interesting features in its own right
and enables comparisons  to be made between
SU(2), SU(3), SU(4),\ldots , SU($N_ C$) \cite{Teper}.
\item In the real world space is isotropic, but this need not be so on a
lattice, since the four axes can be treated differently by having unequal
lattice spacing --- in principle we could have $a_x \not= a_y \not= a_z \not= a_t$.
However, the most common choice is $a_x= a_y= a_z \not= a_t$. This is
appropriate for finite temperature systems \cite{anisolatt}, where the
temperature is defined to be
inversely proportional to the lattice size in the $t$-direction. For
high temperatures this would mean a lattice that contained fewer steps
in the $t$-direction and so lead to difficulties in extracting accurate
correlation functions. However, if $a_t$ is made \mbox{smaller} than the three
spatial $a$'s, then a given temperature is defined by more steps in the
$t$-direction and better correlation estimates --- see Ref.~\cite{Montvay}.
For studying high-momentum form factors such as
$B\rightarrow K^*\gamma,  \ B\rightarrow \pi l \nu$ or
$B\rightarrow \rho l\nu$ it has been suggested that the 2+2 anisotropic lattice
$a_x= a_y\not= a_z= a_t$ is more suitable~\cite{2+2}.

Explicit anisotropic forms of the  clover action --- see
Eq.~\ref{Cloveract} --- and the pure gauge action in Eq.~\ref{Willag}
can be found from Ref.~\cite{anclover}. More recent studies can be found
in Refs.~\cite{anisotropic2}.

\item   In the real world the vacuum (sea) contains $q\bar{q}$-pairs, where
the $q$ are sea--quarks that can have any flavour $u, \ d, \ s, \ c, \ b$
or $t$. These
pairs are being continuously created and annihilated. Lattice QCD
calculations that take this into account are said to involve
{\bf dynamical quarks}. In practice, only $u, \ d$ and, possibly, $s$
sea--quarks are included and these two cases are usually referred to as 
having $N_f=2$ or 3.
However, frequently this effect is neglected to give
the so-called {\bf quenched} approximation, which is numerically an order
of magnitude less demanding on computer resources. In view of this last
point, there has been  much work attempting to show how realistic
quenched results can be compared with their dynamical quark counterparts.
The conclusion seems to be that, although no formal connection has been
established between full QCD and the quenched approximation, the
similarity of the results ( $\approx 10\%$ differences) has led to the belief
that the effects of quenching are generally small, so that quenched QCD 
provides a reasonable approximation to the full theory~\cite{Aoki}. This has
been taken one step further in Ref.~\cite{Young}, where it is suggested
how quenched results can be corrected in a systematic way to retrieve
the corresponding full QCD prediction.

However, there are cases where the  quenched versus
unquenched  comparison can lead to qualitative differences.
In Ref.~\cite{Beane+S} the asymptotic potential between two quark
clusters is calculated algebraically in a quenched approximation effective field
theory and found to have
a {\it pure exponential} decay and not the usual Yukawa form. But it is
not clear that  this qualitative difference for large intercluster
distances leads to any overall quantitative differences. It must be
remembered that, at these distances, calculations involving only
quarks are expected to be incomplete with the introduction of, in
particular, explicit pion fields being necessary. The same authors also
study a {\bf partially--quenched} effective field theory, where the masses
of the sea--quarks are much larger than the valence quarks connected to
the external sources \cite{Beane+S2}. This they suggest would help
in the understanding of the NN--potential, when extrapolations are made of
NN--lattice calculations to realistic quark masses.
\item%
The real world of fixed quark masses can be replaced by one
where the quark masses take  on other values. In many cases, this is of
necessity, since for light hadrons the use of realistic light quark
masses of $\approx 10$ MeV is computationally too
heavy --- see subsection~\ref{Numtreat}.
However,  the variation of quark masses has interesting features in its own
right. In
particular, by carrying out lattice calculations with a series of light
quark masses, we can extract the differential combination
$J=m_{K^*}\frac{dM_V}{d(M^2_P)}$, where $M_V$ and $M_P$ are the 
corresponding vector and pseudoscalar meson masses and $m_{K^*}$ is the
mass of the $K^*$. This quantity $J$,
which can be shown to be independent of $a$ and the so-called hopping
parameter that is related to the bare quark masses, serves as a check
on the consistency of lattice QCD --- see Refs.~\cite{Lacock} and
\cite{Foster}.
Such  analyses can be performed by varying the valence-- and sea--quark
masses separately. In Subsection.~\ref{sectsumr} a similar
argument is used for estimating the matter sum rule.

In Ref.~\cite{Beane+SII} the authors have emphasized the importance of
the quark-mass dependence of the nucleon-nucleon interaction by saying: ``While
the $m_q$-dependence of the nuclear force is unrelated to present
day observables, it is a fundamental aspect of nuclear physics, and in
some sense serves as a benchmark for the development of a perturbative
theory of nuclear forces. Having this behaviour under control will be
essential to any bridge between lattice QCD simulations and nuclear
physics in the near future."

In Ref.~\cite{epel} the authors are more interested in the reverse
situation, namely, how to extrapolate nuclear forces calculated in
the chiral limit to larger pion masses pertinent for the extraction of
NN-observables from lattice calculations.
\item
In the real world,  hadron--hadron scattering is thought
to be described directly in terms of quark-gluon physics at small interhadron
distances, but at larger distances a description in terms of
meson-exchange is expected to be more appropriate. Both of these limits
must be included in a single model, if a direct comparison with
 {\it experiment} is
to be made. However, lattice QCD can concentrate on just the small
distance physics and generate ``data" that is exact not only for that limit
but, in principle, for larger interhadron distances --- until the
numerical signal becomes unmeasurable. In this way, models can be
constructed and compared directly with the lattice data ---
ignoring the effect of explicit meson exchange that enters in the real world.
However, it should be
added that, in some cases, lattice calculations based purely on quarks
and gluons
seem to be able to generate effects that resemble meson exchange.
An example of this is Ref.~\cite{BBpmg} discussed in Subsec.~\ref{sectBBlat}.
\item 
In the real world,  space is essentially infinite, whereas quantities
calculated on a lattice are restricted to volumes $L^3$ with 
\mbox{$L\approx 1-2$ fm}  often being comparable to the size of the object 
under study.
This leads to results that could depend on $L$. Usually this is considered
a negative feature and so lattices must be chosen sufficiently large to
avoid this problem. However, it was shown by L\"{u}scher in 
Refs.~\cite{lufinite} that this volume dependence can be utilized
to extract the interaction between hadrons. Recent examples of this
idea consider  the two-nucleon interaction~\cite{BBPS04,Bedaq} 
and $\pi \pi$ scattering~\cite{pipiYam}.
This approach is discussed in more detail in Chapter~4 Subsec.~4.1.5.
\item In the real world, the systems encountered contain a number of 
valence-quarks and antiquarks. However, on a lattice, systems consisting 
of only gluons can be studied --- with quarks only entering as sea-quarks
in the case of dynamical quarks mentioned in item 4 above. This
pure-glue world enables a cleaner study to be made of glueballs --- see
Sec.~2.2 of Chapter~2. 
\end{enumerate}
These possibilities of a different number of colours, spatial dimensions
and quark masses greatly expand the scope of lattice QCD and give
model builders much more data on which to test their models.
\section{The Energies of Four Static Quarks ($QQ\bar{Q}\bar{Q}$)}
\label{SectQ4}
\subsection{Quark descriptions of hadron--hadron interactions}
\label{factories}
Much of particle and nuclear physics studies the interaction between
\mbox{hadrons.} With increasing complexity in terms of the number of quarks
thought to be involved, this ranges from meson--meson scattering up to
heavy ion collisions.  Clearly, any attempt to describe such processes
at the quark level must begin with an understanding of meson--meson scattering.
Unfortunately, even with this system there are several major
complications preventing a direct comparison between theory and
experiment. Firstly, the only mesons for which there is suitable
experimental data are the pseudoscalars --- the $\pi$, $K$ and $B$,
since beams of these can now be generated at the various
$\pi$-, $K$- and $B$-factories [$(\pi)$ PSI (Villigen) and TRIUMF (Vancouver);
$(K)$ DAPHNE (Frascati) and KEK (Tsukuba); $(B)$
BaBaR at SLAC, Belle at KEK, Hera-B at DESY and
CLEO III at Cornell].
Of course, having beams of mesons does not lead directly to
obtaining data on meson--meson scattering. This can only be done
indirectly, as a final state interaction,
with the net result that essentially only $\pi \pi$ scattering data
(and considerably less $\pi K$ data) are at present available --- and even
those are
very  limited. This means that most theoretical attempts to understand
meson--meson scattering concentrate on the $\pi \pi$ system. However,
quark descriptions of this particular system are  then complicated
by the fact that the pion, being a Goldstone boson, does not have a
quark structure as simple as a single $q\bar{q}$ configuration. Models
of the pion (see, for example,  Weise in Chapter~2 of Ref.~\cite{WWeise})
suggest that it has
large multiquark components {\it i.e.}
$\phi_{\pi}=\sum_n a_n(q\bar{q})^n$. Furthermore, the total interaction
between the two pions can not be only due to interquark interactions
between the constituent quarks, since for large interpion distances it
is expected that meson exchange --- another multiquark mechanism ---
also plays a r\^{o}le
 \mbox{{\it i.e.} $\pi \pi$-scattering} involves much more than
a discussion of the $(q\bar{q})(q\bar{q})$ system.
Having said that, it should be added that these complications have not
detered the construction of models for  $\pi \pi$-scattering that are
 essentially nothing more than  a $(q\bar{q})_{\pi}$ interacting with a
$(q\bar{q})_{\pi}$ through an interquark potential of the form in
Eq.~\ref{VQQ}.
The references are too numerous to list here and, furthermore, they
often involve physicists who are my friends. In my opinion, these models
are, as yet, not justified. They are simply hoping that the success in
treating multi-hadron systems in terms of two-body potentials will
repeat itself.

\subsection{ The r\^{o}le of lattice QCD}
To make a bridge between quark and hadron descriptions of, say,
meson--meson scattering needs reliable experimental data. But, as said above,
this is not available --- and this is where lattice QCD enters.
The latter is based on QCD, which is thought to be the exact theory of
quark--gluon interactions, and its implementation on a lattice leads
(in principle) to exact results --- upto the lattice spacing, lattice
finite size and quark mass reservations mentioned in
Subsec.~\ref{Numtreat}. Therefore, if we want to study, for example,
$(q\bar{q})(q\bar{q})$  systems we simply calculate these on a lattice
and we get {\it exact} results that can now be considered as ``data". Model
builders then try to understand these data in terms of $(q\bar{q})(q\bar{q})$
states. Such a procedure guarantees one of the necessary requirements of
bridge building --- the need to compare like-with-like.
In this way, the lattice data generated for the $(q\bar{q})(q\bar{q})$
system can possibly be modelled with purely $(q\bar{q})(q\bar{q})$
configurations. In other words the conventional approach for model
building
\begin{center}
Experimental data $\stackrel{1}{\longrightarrow}$ Hadron description
$\stackrel{2}{\longrightarrow}$
Quark description\\
\end{center}
is replaced by the alternative
\begin{center}
Lattice data $\stackrel{3}{\longrightarrow}$ Quark description
$\stackrel{4}{\longrightarrow}$
Hadron description \\
\end{center}
{\it i.e.} by concentrating on step 3 we avoid: 
a) At step 1 the shortage of experimental data; b) At step 2 the
need to guess the hadron quark structure and  how a model based on this
structure matches on to models more appropriate at larger interhadron separation
where meson exchange dominates. However, there are several problems
when attempting to implement this second alternative:
\begin{itemize}
\item Step 4 is similar to step 2 each with their uncertainty in the
physical hadron structure. However, now it is less serious,  since step 3
enables a cleaner description to be made at the quark level.
This is in contrast
to the conventional approach, where  the quark description is ``shielded" from
the experimental data by needing to go via the hadron description,
which could well also involve explicit meson exchange.
\item Step 3 is only feasible technically for a very
restricted number of quark systems --- four quarks being essentially the
limit at present.  However, there are a few six quark lattice
calculations --- mainly studies of the NN--interaction for static
quarks (see Subsec.~4.5.2 in Chapter~4) and, in addition, attempts to determine whether or
not the H dibaryon
is bound. In Ref.~\cite{Hdib} the indications are that such a ($uuddss$)
bound state is ruled out. Also recently there have been multiquark
lattice calculations~\cite{Sasaki} with
$uudd\bar{s}$ quenched configurations in an attempt to describe the
$\Theta^+(1540)$ seen in several experiments and thought to be the first
observed pentaquark system~\cite{Nakano}.

\item The best lattice calculations are carried out with dynamical
\mbox{quarks} --- the so-called unquenched formulation. There the possibility
arises for the creation (and annihilation) of quark--antiquark pairs.
This is in contrast to the quenched approximation where such pairs do
not enter. This means that in the unquenched approximation the
configurations included do not have a fixed number of quarks and
antiquarks. In model building this effect is often ignored and only a fixed
type of configuration is used, {\it e.g.} $(q\bar{q})(q\bar{q})$ for meson--meson
interactions. Fortunately, it is found that in most cases of interest
here the refinement of dynamical fermions does not lead to significant
corrections to the quenched results. But this is only
known in hindsight and should, if possible, always be checked.
\end{itemize}
Ideally, in the above example of meson--meson scattering, we would want
to perform a lattice calculation with four light quarks {\it i.e.} where
the quarks in $(q\bar{q})(q\bar{q})$ are $u,d$ --quarks with masses
of less than 10 MeV. Unfortunately, with present day computers,
this is not yet possible. Therefore, the problem must be simplified ---
a process that can be done in several stages by making more and more
of the quarks infinitely heavy {\it i.e.}    $q\rightarrow Q$. This makes the
lattice calculations easier and easier. Examples are as follows:
\begin{itemize}
\item   $\bf{[(Q\bar{q})(Q\bar{q})]}$ configurations.
The energy of this system can be expressed as $V(R)$, where $R$ is the
distance between the two static quarks ($Q$). Ground and excited state
energies in $V(R)$ can then be calculated in the Born-Oppenheimer approximation by
assuming that the $Q$'s have some definite finite mass.
A specific
example would be the $B-B$ system, where $m_Q$ is the mass of the
$b$--quark {\it i.e.} about 5 GeV. This will be discussed in more detail later
in Sec.~\ref{sectBB}.
\item $\bf{[(Q\bar{q})(\bar{Q}q)]}$ configurations.
These are very similar to the above but with the added feature that we
can now have the $q\bar{q}$ annihilating to give simply a $(Q\bar{Q})$
configuration. Such a process is a model for the string breaking
mechanism $Q\bar{Q}(R)\rightarrow (Q\bar{q})(\bar{Q}q)$, where for some
sufficiently large $R$ it becomes
energetically favourable for a $q\bar{q}$-pair to be created from the
vacuum. This mechanism must occur in nature, but at present it has only
been conclusively demonstrated in simplified versions of QCD.
\mbox{A specific example} of this type of configuration would be the
$B\bar{B}$-system --- to be discussed later in Sec.~\ref{sectBbB}.
\item  $\bf{[(Q\bar{Q})(Q\bar{Q})]}$ configurations.
These are the most simple  four-quark configurations to deal with on
the lattice. Unfortunately, they are also the most distant from
appropriate experimental data --- the nearest being
$\Upsilon(b\bar{b},9.5 {\rm GeV}) \Upsilon(b\bar{b},9.5 {\rm GeV})$
scattering. However, this class of experiments is still far in the future.

\end{itemize}
Each of these three possibilities will be discussed separately in the
following sections.

The problem we are faced with is that the Lattice  and Nuclear Physics
approaches  are extremes that concentrate on two opposite aspects.
In the Nuclear Physics approach only the quark degrees of freedom are
introduced with the gluon degrees of freedom entering only {\it implicitly}
in the
interquark potential. In contrast, with the $\bf{[(Q\bar{Q})(Q\bar{Q})]}$
configurations --- the ones studied here in most detail --- the lattice
calculations treat the quarks  in the static
quenched limit, where they play no dynamical  r\^{o}le --- all the
effort being to deal explicitly with the gluon field.

\section{ $(Q\bar{Q})$ and $[(Q\bar{Q})(Q\bar{Q})]$ Configurations}
The most important observable for $Q\bar{Q}$ states is the interquark
potential $V_{Q\bar{Q}}$, which is the topic of Chapter~3 
 --- see also Ref.~\cite{Balirev} for a more detailed account. Here
only those aspects of the $(Q\bar{Q})$--system will be mentioned that are
relevant to the later $[(Q\bar{Q})(Q\bar{Q})]$ discussion --- the main
interest in this section.

When lattice calculations on four-quark systems were first attempted,
it was not clear whether  an acceptable signal could be achieved.
Therefore, various simplifications were made in an attempt to
maximize the possibility of success~\cite{GMS,JL,GMP,GMP93,GMPS}.
The most important of these was
the use of quarks with only two colours {\it i.e.} SU(2). In addition,
only configurations where the four quarks were at the corners of a
square were studied, since it was hoped that this degenerate situation would
lead to the maximum interaction --- a feature that in fact turned out to
be so and also one that had been assumed in the flip-flop model of
Ref.~\cite{flip}.
\subsection{Lattice calculations with  $(Q\bar{Q})$ configurations}
\label{importance}
Schematically, in analogy to thermodynamics, the expectation value of some
operator $O$ can be obtained
 with an action $S$ through  the chain
\begin{equation}
\label{Path}
\langle O\rangle=\frac{\int DU O(U)e^{-S(U)}}{\int DU e^{-S(U)}}
\stackrel{{\bf 1}}{\rightarrow}
\frac{\sum_i DU_i O(U_i) e^{-S(U_i)}}{\sum_i e^{-S(U_i)}}
\stackrel{{\bf 2}}{\rightarrow}
\frac{\sum_j^{N_M} O({U}_j)}{N_M},
\end{equation}
where, at the first stage, the variables $U$ are continuous. However,
for QCD, even though we know the exact form of the action $S(U)$, the
resulting \mbox{integrals} are singular. As discussed earlier, it is
necessary to carry out step 1, in which the variables $U$ are discretized
into variables $U_i$ that sit on the links ($i$) of a lattice.
This now removes the singularities in a consistent
manner and reduces the problem to a numerical
approximation for a multiple integral. Unfortunately, there are very
many $U_i$'s. For example, in QCD the $U_i$ are in fact  colour
matrices $U^{ab}_i$ --- defined by 8 independent real parameters.
Furthermore, for a $10^4$ space--time lattice there are about
$4\times 10^4$ links. In all this amounts to about 320000 integrations
to be done. If each of these were approximated by 10 points, we would
then be approximating the multiple integral with about $10^{320000}$
terms --- not a feasible task. However, many of the link configurations
lead to large values of $S(U_i)$ and so would have a negligible effect. These
configurations can be avoided by ``importance sampling", which
essentially generates configurations with the probability given by the
Boltzmann factor $\exp[-S(U_i)]$.
This automatically encodes the feature that the ratio of the values of
nearby links are exponential in form.
Now only a relatively few configurations (often $\sim 10$'s)
need to be generated to get a good estimate of
$\langle O\rangle$. This is depicted as step 2 in Eq.~\ref{Path},
where the Boltzmann factor has been replaced by the $N_M$ configurations
that tend to maximize this factor --- see, for example, p.252 in
Ref.~\cite{Rothe}.

The problem is, therefore, reduced to generating these ``important"
configurations and to finding an appropriate operator. These are the
topics of  the next two subsections.
\subsubsection{Generating lattice configurations}
Earlier, practitioners of lattice QCD generated their own lattice configurations.
However, nowadays there are groups that specialize in this very time
consuming computer task, {\it e.g.} the  UKQCD group in Edinburgh.
The rest of us are then able to simply perform
expectation values of our operators in the knowledge that we are using
configurations that are well tested and essentially independent of each other.
This also has the added benefit that different groups use the same
configurations, {\it i.e.} with the same lattice parameters, but with
different operators. At times, this can lead to useful direct comparisons
of the lattice results with experiment that
do not have to be scaled in any way beforehand, {\it i.e.} if one group
has extracted the lattice spacing for their particular problem, then the rest
of the lattice community can use this same
lattice spacing to convert all of their dimensionless lattice results
directly into physical quantities.

In spite of this modern trend, it is probably useful to remind the
reader of some of the problems that arise when generating configurations:

\vspace{0.1cm}

\noindent{\bf Equilibration}\\ If a lattice simulation is ever started from
scratch, then the first lattice must be simply ``guessed".
This could be ``cold", where all the $U_i$ are taken to be the same,
or ``hot", where the $U_i$ are random numbers. To carry out the above
``importance sampling" these lattices must first be equilibrated or, in the
terminology of spins-on-a-lattice, thermalized. This is where clever
programming enters to generate quickly configurations that are as independent of
each other as is conveniently possible. For example, in our early work
\cite{GMP},
where we generated all of our own configurations, the lattice was
equilibrated by the heat bath method.

\vspace{0.1cm}

\noindent{\bf Updating }\\ Once the lattice is equilibrated, measurements
of the operator $O$ can
begin. However, the lattice must be continuously updated after each
measurement to ensure these
measurements are made on lattices that are, as far as possible and
convenient, independent of each other.
In Eq.~\ref{Path} the number of such lattices is referred to
as $N_M$.
As an example, in Ref.~\cite{GMP} --- \mbox{after} equilibration  ---
 the lattice was further updated by a
combination of three {\bf over--relaxation} sweeps, each of which can change
the configuration a lot but leave the energy unchanged. This is followed by
 one {\bf heat bath} sweep, which changes the configuration less but can
change the energy as is necessary for ergodicity. Then
 a measurement is made of the appropriate correlation functions. In
that particular calculation, the lattice was updated 6400 times with
1600 measurements being made. The latter were divided into 20 blocks of
80 measurements for convenience in carrying out an error analysis.
In later work \cite{G+K+P+M2} 40 updates were made between measurements to
reduce possible correlations between successive measurements. However, it
should be added that, in our experience, there has never been
any sign of a significant correlation due to insufficient updating
between measurements, when dealing with the types of
problem we have been studying. This is usually checked by calculating
the autocorrelation  between blocks of measurements.
In the above, I mention spins-on-a-lattice --- the classical example
being the Ising model --- in order to remind the reader that lattice QCD
and the thermodynamics of spins-on-a-lattice have very much in common.
Over the years, many
ideas were first developed in the spin case before attempting to
implement them in the more complicated case of lattice QCD.

\subsubsection{Appropriate operators on a lattice}
\label{Wloops}
\begin{figure}[ht]
\centering
\includegraphics*[height=0.3\textwidth]{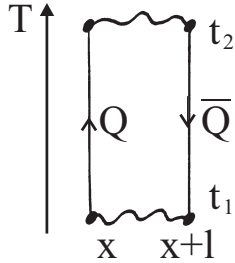}
\caption{The  Wilson loop for a $Q\bar{Q}$ state $\Psi[Q({\bf x})
\bar{Q}({\bf x}+{\bf l})]$
propagating from Euclidean time $t_1$ to $t_2$.}
\label{2QWiloop}
\end{figure}

Most of the operators evaluated on a lattice take the form of
correlations between different Euclidean times. The most simple example
is the Wilson loop involved in extracting $V_{Q\bar{Q}}$ in
Eqs.~\ref{VQQ} and ~\ref{Willoop}. Consider a $Q$ and a $\bar{Q}$ are at the
lattice sites ${\bf x}$ and ${\bf x}+{\bf l}$ --- with the notation
that ${\bf l}$  is the lower case letter corresponding to the  upper case $L$,
which is reserved for the spatial size of the lattice.
A $Q\bar{Q}(\bf{l})$-state, is then constructed from a sequence of connected
lattice links $U_i$, at a fixed Euclidean time $t_1$, as
\begin{equation}
\label{QQ1}
\Psi(Q\bar{Q}, \ {\bf l})=\phi^a_Q({\bf x})U^{ab}U^{bc}U^{cd}\ldots U^{za}
\bar{\phi}^a_{\bar{Q}}({\bf x}+{\bf l}).
\end{equation}
Here the colour indices $a,b,c, \ldots $ on successive links are coupled
to give overall a gauge invariant chain of $U$'s. Also the $Q$ and $\bar{Q}$
have the same colour to ensure the meson is a colour singlet.
To be more specific, if the  $Q$ and $\bar{Q}$ are  a distance of two
links apart {\it i.e.} they are at lattice sites
$(x,0,0,t_1)$ and $(x+2,0,0,t_1)$, then one possibility is the direct path
\begin{multline}
\label{QQ2}
\Psi(Q\bar{Q}, \ t_1)=\\
\phi^a_Q(x, \ t_1)
U^{ab}(x,x+1;t_1)U^{ba}(x+1,x+2;t_1)
\bar{\phi}^a_{\bar{Q}}(x+2,  \  t_1),
\end{multline}
where the two $U$-links are simply along the $x$-axis. Other choices of less
direct paths between the  $Q$ and $\bar{Q}$ are possible and, indeed,
necessary if excited  states are of interest.
To construct a Wilson loop a similar wave function is written down at time
$t_2=t_1+t$ as
\begin{multline}
\label{QQ3}
\Psi(Q\bar{Q}, \ t_2)=\\
\phi^{a'}_Q(x, \
t_2)U^{a'b'}(x,x+1;t_2)U^{b'a'}(x+1,x+2;t_2)
\bar{\phi}^{a'}_{\bar{Q}}(x+2,  \  t_2).
\end{multline}
This gives the two horizontal (wavy) sides of the Wilson loop in
Fig.~\ref{2QWiloop}.
To complete the loop we need to insert the propagators of the $Q$ and
$\bar{Q}$ from $t_1$ to $t_2$. For static quarks these are, for $t_2-t_1=2$,
\begin{equation*}
\phi^{a}_Q(x;t_1)U^{ab'}(x;t_1,t_1+1)U^{b'a'}(x;t_1+1,t_1+2)
\bar{\phi}^{a'}_Q(x;t_2)
\end{equation*}
and
\begin{equation*}
\phi^{a}_{\bar{Q}}(x+2;t_1)U^{ab'}(x+2;t_1,t_1+1)U^{b'a'}(x+2;t_1+1,t_1+2)
\bar{\phi}^{a'}_{\bar{Q}}(x+2;t_2),
\end{equation*}
where here the $U$'s are all in
the $T$ direction and, without loss of generality,  can be arranged by gauge
fixing to be simply
{\it unity}. The Wilson loop of Eq.\ref{Willoop} then reduces to the overlap
\begin{multline}
\label{Overl}
W(x,x+2;t_1,t_2)=
\delta^{aa'}U^{ab}(x,x+1,t_1)U^{ba}(x+1,x+2,t_1)\times \\
\left[U^{a'b'}(x,x+1,t_2)U^{b'a'}(x+1,x+2,t_2)\right]^{\dagger}.
\end{multline}
This is simply a number, which
in principle $\rightarrow \exp[-tV_{Q\bar{Q}}(l=2)]$ as $t\rightarrow \infty.$
The emergence of an exponential factor from this overlap should not be
surprising, since the essence of the importance sampling in
Subsec.~\ref{importance} was to encode the Boltzmann factor in
Eq.~\ref{Path} into the values of the links. Of course, a single overlap
$W(x,x+2;t_1,t_2)$ would not be very informative. Only when --- for
all three orientations $x,y,z$ --- this is
averaged over the whole $L^3T$ lattice and the $N_M$ different lattices
does one expect a reasonable numerical signal for $W(2,t)$ to emerge.
In general, this has the form
\[W(l,t)=\frac{1}{N_ML^3T}\sum_{N_M}\sum_{l_0=x_0,y_0,z_0}^L
\sum_{t_1}^TW(l_0,l_0+l;t_1,t_1+t)\]
\begin{equation}
\label{Wloopav}
\rightarrow \exp[-tV_{Q\bar{Q}}(l)] \ \ {\rm as} \ \  t\rightarrow \infty,
\end{equation}
where the average over the different lattices  includes:
\begin{enumerate}
\item All values of $x$ in the range $0\le x \le L$,
\item All three spatial directions $x,y,z$
\item All values of $t_1$ in the range $0\le t_1 \le T$
\end{enumerate}
and it is repeated for many as--independent--as--possible lattices $N_M$.

Having introduced the notion of a lattice link $U_i$ representing the gluon
field, we can combine four links to form a closed loop called an elementary
plaquette $U_{\Box }$. For example, a plaquette in the $xy$--plane would be
constructed from links $U(x\rightarrow x':y\rightarrow y')$ as
\begin{multline}
\label{elepla}
U_{\Box }^{xy}=
U^{ad}(x\rightarrow x+1:y)U^{dc}(x+1:y\rightarrow y+1)\times \\
U^{cb}(x+1\rightarrow x:y+1)U^{ba}(x:y+1\rightarrow y).
\end{multline}
Similarly there can be
plaquettes in the $xz,\ldots , zt$--planes. It was in terms of these plaquettes
that Wilson \cite{Wilson} first expressed the discretized  form of
the QCD action mentioned in Sec.~\ref{Numtreat}, which  for SU($N_C$)
has the form
\begin{equation}
\label{Willag}
S_W=\frac{2N_C}{g^2}\sum_{\Box }[1-\frac{1}{N_C} \ {\rm Tr} \ U_{\Box }].
\end{equation}
Usually the basic coupling ($g$) is expressed in terms of 
$\beta=2N_C/g^2$.

\subsubsection{Fuzzing}
\label{Fuzzingsec}
\begin{figure}[h]
\centering
\includegraphics*[width=0.65\textwidth]{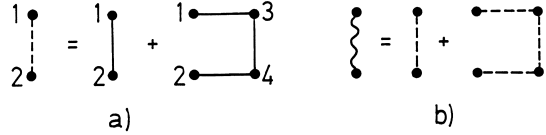}
\caption{Fuzzing. In a) the solid line 12-link is replaced by the
dashed line 12-link.\protect\\In b) this dashed line 12-link is itself now
replaced by the wavy line 12-link. }
\label{Figfuzz}
\end{figure}

The above Wilson loops were constructed from the basic lattice links
$U^0_{\mu}(n)$, where  --- using a slight change of notation for
convenience --- the earlier colour indices $a,b,\ldots$ are omitted and replaced by
an index $j=0$, $n$ is a lattice site and $\mu$ is either the $x,y$ or
$z$ direction.
These basic $(j=0)$ links can now be generalised
by  ``fuzzing, blocking or smearing"\cite{CM1,CM2} by means of which the
basic link is supplemented by a combination of neighbouring links. Here
I will concentrate on the the fuzzing option, but
blocking and smearing are similar. Fuzzing is depicted in
Fig.~\ref{Figfuzz}.

\noindent This illustrates the replacement
\begin{equation}
\label{fuzz}
U^0_{\mu}(n)\rightarrow U^1_{\mu}(n)=A^1_n\Biggl[
cU^0_{\mu}(n)+\sum_{\stackrel{\pm \nu \not = \mu}{\nu\not = 4}}
 U^0_{\nu}(n)U^0_{\mu}(n+\bar{\nu})U^{0{\dagger}}_{\nu}(n+\bar{\mu})
\Biggr]
\end{equation}
to give links $U^1$ constructed from the basic links $U^0$ by a single fuzzing $j=1$.
This is followed by
\begin{equation}
U^1_{\mu}(n)\rightarrow U^2_{\mu}(n)=A^2_n\Biggl[
cU^1_{\mu}(n)+\sum_{\stackrel{\pm \nu \not = \mu}{\nu\not = 4}}
 U^1_{\nu}(n)U^1_{\mu}(n+\bar{\nu})U^{1 {\dagger}}_{\nu}(n+\bar{\mu})
\Biggr]
\end{equation}
to give links $U^2$ constructed from the $U^1$ by a second fuzzing $j=2$ and so on.
In general
\begin{align}
U^{m-1}_{\mu}(n)\rightarrow U^m_{\mu}(n)=&\nonumber \\
A^m_n
\Biggl[cU^{m-1}_{\mu}(n)+
&\sum_{\stackrel{\pm \nu \not = \mu}{\nu\not = 4}}
U^{m-1}_{\nu}(n)U^{m-1}_{\mu}(n+\bar{\nu})
U^{m-1 \ {\dagger}}_{\nu}(n+\bar{\mu})
\Biggr].
\end{align}
Here the $A^j_n$ are normalisation factors chosen to project the
$U^j_{\mu}(n)$ into $SU(N_C)$ and $c$ is, in principle, a free parameter.
However, experience when measuring energies
\cite{CM1,CM2} has shown for SU(2)
that $c=4$ is a suitable value for the present class of problems.
The value of $c$ could be optimized, but it is found that results are never crucially
dependent on the precise value of $c$.
Another
degree of freedom is the amount of fuzzing ($m$). For correlations over large
distances the greater the fuzzing the better the efficiency of the
calculation in the sense that the state $\Psi(Q\bar{Q})$ in Eq.~\ref{QQ1},
generated by connecting together
a series of fuzzed links,  is expected to have a greater overlap
with the true ground state wave function. In some cases this has been carried
to great lengths {\it e.g.} in Ref.~\cite{Booth} $m=$110 fuzzing iterations were used, since
there the emphasis was on quark separations upto 24 lattice spacings.

In the calculation of the interquark potential in Eq.~\ref{Wloopav},
 the fuzzing procedure plays
the second r\^{o}le of generating different paths ($P_i$) between quarks,
 for use in the variational
approach of Ref.~\cite{CM1,CM2}. In this notation the basic path from
Eq.~\ref{QQ1} is
$P_0=U^{0,ab}U^{0,bc}U^{0,cd}.....U^{0,za}$ and the fuzzed paths
$P_i=U^{i,ab}U^{i,bc}U^{i,cd}.....U^{i,za}$ . The variational approach
then yields  the correlation matrix between paths with different fuzzing
levels $i,j$
\begin{equation}
\label{transfermat}
W^t_{ij}=<P_i|\bar{T}^t|P_j>.
\end{equation}
 Here $\bar{T}=\exp(-a\bar{H})$ is the transfer matrix for a single
time step $a$, with the basic QCD Hamiltonian $\bar{H}$, the $P_{i,j}$ are
paths constructed as products of fuzzed basic links and, as before, $t$ is the
number of steps in the imaginary time direction. As shown in
Ref.~\cite{CM1,CM2}, a trial wave function $\psi=\sum_ia_i|P_i>$
leads to the eigenvalue equation
\begin{equation}
\label{WN}
W^t_{ij}a^t_j=\lambda^{(t)}W^{t-1}_{ij}a^t_j.
\end{equation}
For a single path this reduces to
\begin{equation}
 \lambda^{(t)}=\frac{W^t_{11}}{W^{t-1}_{11}}=\exp(-aV_{Q\bar{Q}}),
\end{equation}
where $V_{Q\bar{Q}}$ is the potential of the quark system being studied.
Unfortunately,
in this single path case {\it i.e.} with only $P_0$ as in the previous 
subsection,
$t$ needs to be large and this
can lead to
unacceptable error bars on
the value of $V_{Q\bar{Q}}$ extracted. However, if --- in addition --- a
few fuzzed paths are included, it is found
that $t$ need only be small $(t\approx 5)$ to get a good convergence to $V_{Q\bar{Q}}$
with small error bars.

A further very important advantage of fuzzing is that not only
can the lowest eigenvalue be extracted but also higher ones, since a
matrix diagonalisation is involved. These higher states correspond to
excitations of the gluon field. However, with the direct paths from the
$Q$ to $\bar{Q}$ in Eq.~\ref{QQ2}, these excitations are purely S-wave.
To generate non-S-wave gluonic excitations combinations of indirect
paths are needed --- see Chapter~2.

\begin{figure}[h]
\centering
\includegraphics*[width=1.0\textwidth]{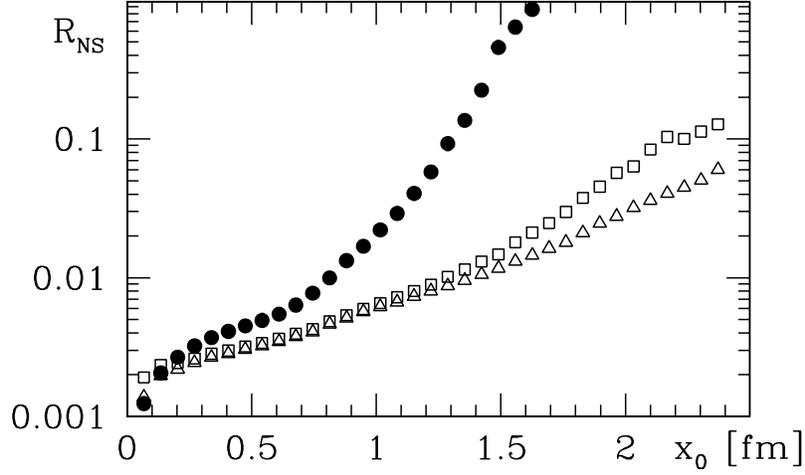}
\caption{Noise to signal ratio ($R_{{\rm NS}}$) of a $B-$meson correlation
function as a function of the time extent $x_0$.
The solid dots are when using  simply  $U^0_{\mu=0}(n)$. The open
squares are for $W^0_{\mu=0}(n)$ and the open triangles for $W_{{\rm HYP}}$
\protect\cite{DellaM}.}
\label{Rnsfig}
\end{figure}

The above fuzzing is an attempt to improve the overlap of the lattice
wavefunction with the true ground state wavefunction and so only involves
spatial links (N.B. $\nu \not= 4$ in the summations in Eq.~\ref{fuzz}
{\it i.e.} the staples in Fig.~\ref{Figfuzz} are purely spatial).
However, a
similar procedure can be applied to a time-like link~\cite{DellaM}.
In the notation of Eq.~\ref{fuzz}, the basic time-like link $U^0_{\mu=0}(n)$
is  now replaced by $W^0_{\mu=0}(n)$ --- the average of the six staples
in the planes $(\pm x,T), \  (\pm y,T)$ and $(\pm z,T)$. The benefit gained
from this is a reduction in the Noise/Signal ratio $R_{{\rm NS}}$ for configurations
involving a static quark. Without this time-like fuzzing, it has been 
observed for $B$-meson correlations, with time extent $x_0$, that
$R_{{\rm NS}}\propto\exp(x_0\Delta E)$, where \mbox{$\Delta E=E_0-m_{\pi}$} with $E_0$
being the ground state energy of the $B$-meson~\cite{Hash}. This leads to
very noisy signals as $x_0$ increases. However, using $W^0_{\mu=0}(n)$
results in almost an  {\it order of magnitude} reduction in $R_{{\rm NS}}$ for
\mbox{$x_0\approx  1.5$ fm.}
This is shown in Fig.~\ref{Rnsfig}.

An even greater reduction in $R_{NS}$ can be achieved, if the single
staple fuzzing in Fig.~\ref{Figfuzz} is replaced by the hypercubic
blocking of Ref.~\cite{hyperb}. There each level of fuzzing is described
by three equations (and so needing three free parameters $c_i$) similar to
Eq.~\ref{fuzz}. But these three equations mix links that are connected to the
original link only {\it i.e.} this ``fuzzing" only involves links in the
hypercube defined by the original link. This procedure yields a fuzzed
link $W_{{\rm HYP}}$ that is much more local than three fuzzing levels from
Eq.~\ref{fuzz} and so preserves short distance spatial structure.
Furthermore, when this is applied to a time-like link, then it reduces
$R_{{\rm NS}}$ even further than the use of $W^0_{\mu=0}(n)$. The use of
hypercubic links is still in its infancy and we should expect to see
much more of this development in the future --- see Refs.~\cite{hyperb2}.
This is also shown in Fig.~\ref{Rnsfig}.

\subsection{Lattice calculations with  $[(Q\bar{Q})(Q\bar{Q})]$
configurations}
\label{QQQQLC}
\begin{figure}[b]
\centering
\includegraphics*[width=0.95\textwidth]{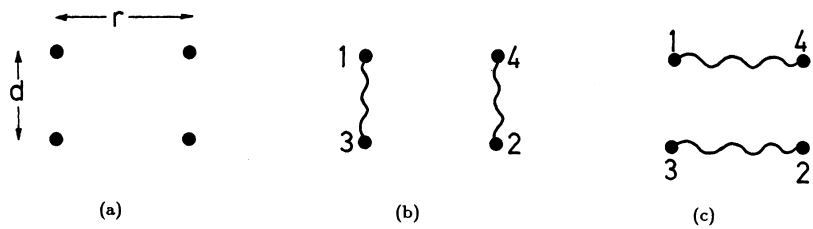}
\caption{a) Four quarks in a rectangle of sides $r$ and $d$:
\protect\\The two  partitions b) $A=[Q_1\bar{Q}_3][Q_2\bar{Q}_4]$ and
c) $B=[Q_1\bar{Q}_4][Q_2\bar{Q}_3]$}
\label{configs}
\end{figure}

\begin{figure}[b]
\centering
\includegraphics*[angle=-90,width=0.65\textwidth]{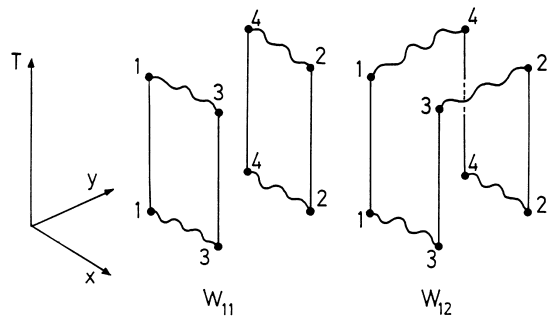}
\caption{The 4-Q Wilson loops $W_{11}$ and $W_{12}$.}
\label{4QWiloop}
\end{figure}

The previous section outlines the techniques for extracting the
two-quark potential. This is now generalised to
the $[(Q\bar{Q})(Q\bar{Q})]$ case, where the quarks are on the corners of
a rectangle with sides of length $d$ and $r$.
During the Monte  Carlo simulation, the correlations $W_{ij}(r)$
in Eq.~\ref{transfermat} --- appropriate for extracting the two-quark 
potential $V_{{Q\bar{Q}}}(r)$ --- and the correlations
$W_{ij}(d,r)$ for the four-quark potential
$V_4(d,r)$ are evaluated at the same time. As an example, in
Ref.~\cite{GMP} for $W_{ij}(r)$ three paths
were
generated by the three different fuzzing levels $m=$12, 16 and 20. On the
other
hand, for the $W_{i'j'}(d,r)$ only the level 20 was kept, with the
variational basis $(i'j')$ being the two configurations $A$ and $B$ in
Figs.~\ref{configs}.
These lead to the Wilson loops in Fig.~\ref{4QWiloop}.
 In the case of squares ({\it i.e.} $r=d$), Eq.~\ref{WN} gives the
potential energy of the ground state as
\begin{align}
\label{le0}
E_0=&aV_4(d,d)=aE_4(d,d)-2E(d)\nonumber \\
=&\log\left[\frac{W_{11}^{t-1}(d,d)+W_{12}^{t-1}(d,d)}
{W_{11}^{t}(d,d)+W_{12}^{t}(d,d)}\right] -2E(d)
\end{align}
and that of the first excited state
\begin{equation}
\label{le1}
E_1=\log\left[\frac{W_{11}^{t-1}(d,d)-W_{12}^{t-1}(d,d)}
{W_{11}^{t}(d,d)-W_{12}^{t}(d,d)}\right] -2E(d),
\end{equation}
where $E(d)$ is the result of diagonalising the $3\times 3$ variational basis
for $V_{Q\bar{Q}}$. In these equations, $t$ should tend to $\infty$. However,
in
practice it is found to be sufficient to have $t\approx 5$ for extracting
accurate values for $E_0$ and $E_1$.

\begin{figure}
\centering
\includegraphics*[width=0.75\textwidth]{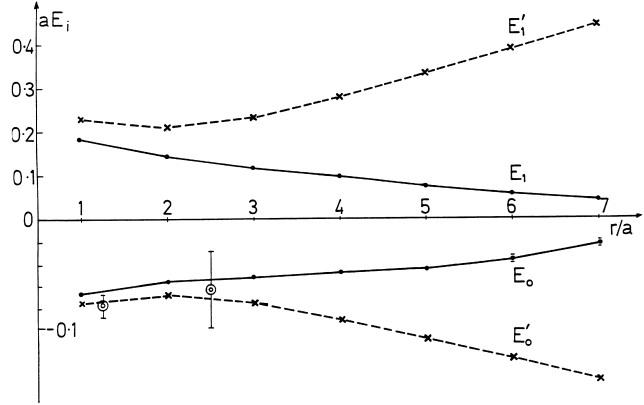}
\caption{The comparison of the  lattice data $E_{0,1}$ (continuous line)
with the $f=1$ model results $E'_{0,1}$ (dashed line) in Subsec.~\ref{Unmodified}
--- see Ref.~\protect\cite{GMP93}. Also the two lattice data points from
Ref.~\protect\cite{Ohta} are included.}
\label{f=1}
\end{figure}

The results are shown in Fig.~\ref{f=1}.
For comparison this also includes the
corresponding 1985 results of Ref.~\cite{Ohta} as the two
points with large error bars
at $r/a\approx 1.25$ and 2.5 --- remembering that the lattice spacing for
Ref.~\cite{Ohta} is $a'\approx 0.15$fm $\approx 1.25a$, where
$a\approx 0.12$ fm from Ref.~\cite{GMP93}.
It is seen that the present calculation gives
energies that have error bars which only become significant for
$r/a\geq 7$, whereas the 1985 work was unable to generate any meaningful
numbers beyond $r/a'=2$. In fact, for the present purpose, neither
of the 1985 points is of use, since the $r/a'=1$ result could
well suffer from lattice artefacts --- a problem afflicting all
calculations of
configurations involving a single lattice spacing --- and the $r/a'=2$
result has error bars that are too large for the analysis in
Subsec.~\ref{potmodel} to be carried
out. However, the following should be added in defence of
Ref.~\cite{Ohta}. Firstly, their calculation was for SU(3) and so it
was, at least,
an order of magnitude more demanding on CPU time --- especially in 1985.
 Secondly, it should be
remembered that the energies of interest ($E_{0,1}$) are very small
compared
with the total four-quark energy $E_4$ in Eq.~\ref{le0}. For example, at
$r=d=4a$ the value of $aE_0=-0.050(1)$ is obtained from the difference
between $aE_4$ and $2aE(d)$, which is 1.505--1.555 {\it i.e.} the error quoted
on $aE_0$ is less than $0.1\%$ of $aE_4$.

One of the most outstanding features of the results  is that for
$ r$ {\it equals} $d$ the value of $aE_0$ decreases smoothly in magnitude
from --0.07
for $r=d=a$ to --0.04 for $r=d=6a$. However, for the few cases where
$r\not=d$
the value of $E_0$ is {\it an order of magnitude smaller} than
the adjacent $r=d$ cases {\it e.g.} $E_0(2,3)\approx -0.006$, whereas
$E_0(2,2)$
and $E_0(3,3)$ are $\approx -0.055$. This result is reminiscent of the
conclusion found with the flux-tube model \cite{Paton} and the
ansatz made in the flip-flop model of Refs.~\cite{flip}.  In both of
these
models the interaction between the two separate two-quark partitions of
Fig.~\ref{configs} is very small (in fact zero in the flip-flop model)
except when
the unperturbed energies of the two partitions is the same {\it i.e.} at
$r=d$ in the case of rectangles.
\subsection{Lattice parameters and finite size/scaling check}
\label{4Qparam}
Most of the above (and later) calculations involving only static quarks were
carried out with the parameters in Table~\ref{tab.4Qparam}. There the 
coupling $\beta$ is defined after Eq.~\ref{Willag} and the lattice spacings
$a_n$ were determined by, for example, the Sommer method described in
Subsec.~\ref{sommersec}. Each of these three sets serve a specific purpose.

\subsubsection{Benchmark data}
Set 1 shows the original parameters with which most calculations are
performed ({\it e.g.} for the results in Fig.~\ref{f=1}) and against which the following results with Sets 2 and 3 are
compared. These parameters are chosen so that --- unlike Sets 2 and 3
--- the problems of computer memory and CPU time do not prevent generating
many ``independent" measurements to ensure good error estimates.
\subsubsection{Finite size effect}
\begin{table}[b]
\caption{Typical SU(2) lattice parameters used in $[(Q\bar{Q})(Q\bar{Q})]$
calculations. Much of the notation is explained in the text with
M/G being the number of Measurements per Geometry.}
\vskip 0.5cm
\label{tab.4Qparam}
\begin{center}
\begin{tabular}{|c|c|c|c|c|c|} \hline
Set&$L^3\times T$&$\beta$&$a_n$ (fm)&M/G& Refs.\\ \hline
1&$16^3\times 32$&2.4&$\approx 0.12$&720&
\protect\cite{GMP}--\protect\cite{GMPS}\\
2&$24^3\times 32$&2.4&$\approx 0.12$&160&\protect\cite{GMPS}\\
3&$24^3\times 32$&2.5&$\approx 0.082$&660&\protect\cite{GMPS}\\ \hline
\end{tabular}
\end{center}
\end{table}
With Set 2 the effect of the finite lattice size is checked. However, in
the present case it
was found that the results for the size of squares considered 
({\it i.e.} up to
$7\times 7$) were unchanged within error bars --- see Table 2 in
Ref.~\cite{GMPS}. This type of check is important for large squares
because of the spatial periodic boundary conditions at $0$ and $L$.
Two quarks can be connected by two paths --- a direct path  on the
lattice ({\it e.g.} that
in Eq.~\ref{QQ2}) or an indirect path that {\it encircles} the boundary. If the
two quarks are separated by $r>L/2$, then the indirect path is of length
$L-r < r$ and so is potentially more important than the direct path ---
the only one of the two that is explicitly treated. In the present case
with $L=16$, such problems should only begin to occur seriously with
$8\times 8$ squares. For this larger lattice, fewer configurations per square
were generated, since the storage and computer time increased by about a
factor of $\approx (24/16)^3\approx 3.$
\subsubsection{Smaller lattice spacing}
\label{4Qparam433}
In most cases the possible importance of finite size effects can be
seen beforehand by using simple arguments. We know, for example, that
the length $r$ of the direct
path should be smaller than the length $L-r$ of the indirect path.
However, the effect as $a\rightarrow 0$ is not obvious and needs
checking. In order to isolate this scaling effect from the finite
size effect, it is convenient to use lattices that have approximately
the same {\it physical} size. Therefore, since the spatial volume for
Set 1 is $(16 \times 0.12)^3\approx 7$ fm$^3$, the value of $a_3$ for a
$24^3$ lattice should be $\approx 7^{1/3}/24\approx 0.08$ fm --- a
lattice spacing that corresponds to $\beta \approx 2.5$. For the results
to show {\bf scaling}, the physical energies ({\it i.e.} in, say, MeV) at the same values
of $r$ (in fm) should be the same. In this case, it means
\begin{equation}
\label{scalinge}
E[{\rm Set} \ 1\ ({\rm or} \  2), \ r/a_1]\approx E[{\rm Set} \  3, \ r/a_3],
\end{equation}
 where the $a_n$E(Set n, $r/a_n$) are the {\it dimensionless} numbers
given by the lattice calculations and the $r/a_n$ are the number of
lattice links between the quarks. Of course, in general only $r/a_1$ or
$r/a_3$ is an integer, so that the comparison in Eq.~\ref{scalinge} must
be done by interpolation.
This approximate equality is sufficiently well satisfied
in the present problem (see Table~3 in Ref.~\cite{GMPS}).
In Ref.~\cite{PP1} a more complete test of the $a\rightarrow 0$ limit is
carried out with the addition of the non-rectangular geometries to be
discussed in Sec.~\ref{ComplQQ}. There the $\beta$ values (and lattice
sizes) used were $\beta$= 2.35, 2.4 ($16^3\times 32)$,
2.45 ($20^3\times 32)$, 2.5 ($24^3\times 32)$ and 2.55 ($26^3\times
32)$ and led to 4-quark energies that were essentially independent of
$a$ over this \mbox{range of $\beta$} {\it i.e.} scaling was achieved for 
$\beta\ge 2.35$.

It should be added that the above procedure is called {\it scaling} in
contrast to {\it asymptotic scaling}. The latter relates results from different
values of $a(\beta )$ by {\it perturbative} arguments and is not
expected to apply at the comparatively large values of $a$ in
Table~\ref{tab.4Qparam} --- see Ref.~\cite{asypsca}.
\section{Potential Model Description of the Lattice Data}
\label{potmodel}
The assumption often made by those who create models for multi-quark systems
is that  these systems can be treated
in terms of two-body potentials. This is the Nuclear Physics inspired
approach that is very successful for, say, multi-nucleon systems, where
 {\it three-} and {\it four-}body forces are small. Of course, the fact that the
latter were small was at first simply an assumption. However, later this
was found to be justified by Weinberg \cite{Weinberg} using various
low-energy theorems that force
the $\pi N\rightarrow \pi N$ interaction --- the main mechanism for
multi-nucleon forces ---
to essentially vanish in nuclei. At
present, there seems to be no such simplifying feature for multi-quark
interactions. In the following, I will first show the consequences of
using --- in the four-quark system --- simply two-quark potentials unmodified
 by the presence of the other two quarks. Then this model will be
improved by including also a direct four-quark interaction.
These are examples of what was referred to as  Effective Potential
Theories (EPTs) in  Subsec.~\ref{sect.EPT}. However, since the quarks
are now static, there is no kinetic energy and so the question of which
quark wave equation to use does
not arise. Similarly, the effective quark masses do not enter, since
the quantities of interest are the {\it binding energies} between the
two 2-quark clusters. Therefore, of the three ingredients usually needed
for an EPT of Subsec.~\ref{sect.EPT} only freedom with the interquark potential
remains. This is both a weakness,  by being a model that is unrealistic
and not comparable with any experimental data --- except for possible future
$\Upsilon(b\bar{b},9.5 {\rm GeV})\Upsilon(b\bar{b},9.5 {\rm GeV})$
scattering  --- and a strength,  in that only the potential ingredient
plays a r\^{o}le.

\subsection{Unmodified two-body approach}
\label{Unmodified}
Once the quark--quark potential $V_{Q\bar{Q}}$ is known and, if it is assumed
to be the
{\it only} interaction between the quarks, then the energy of a
multi-quark system can be readily calculated --- provided the wave function for
that system
is expressed in terms of a sufficient number of basis states. For the
present situation,
 the most obvious choices for such states are $A$ and $B$ in Figs.~\ref{configs}
b) and c).

In this extreme two-body approach, since the presence of the gluon
fields
 have been explicitly removed --- their only effect now being in the
colour
indices of the quarks --- the states $A$ and $B$ form a complete but
non-orthogonal basis. This implicitly assumes in Figs.~\ref{configs}
 the quark $(Q)$,
antiquark $(\bar{Q})$ assignment $A=[Q_1\bar{Q}_3][Q_2\bar{Q}_4]$. In SU(2)
the other quark assignment $[Q_1 Q_3][\bar{Q}_2\bar{Q}_4]$ is numerically
equivalent and so leads to nothing new.
In this approach, states are excluded in which the
gluon fields are excited. However, later this restriction is relaxed in
an extension to the model in Sec.~\ref{fmodelext} and Appendix A.3.
The energies ($E'_i$) of this static four-quark system can be extracted
from
the eigenvalues of the Hamiltonian
\begin{equation}
\label{Ham}
\left({\bf V}-\lambda_i {\bf N}\right)\Psi_i=0,
\end{equation}
where the normalisation matrix
\begin{equation}
\label{NV1}
{\bf N}=\left(\begin{array}{cc}
1&1/2\\
1/2&1\end{array}\right)
\end{equation}
and the potential energy matrix
\begin{equation}
\label{NV2}
{\bf V}=\left(\begin{array}{cc}
v_{13}+v_{24} & V_{AB}\\
V_{BA}&v_{14}+v_{23}\end{array}\right).
\end{equation}
Several points need explaining in these equations.
\begin{enumerate}
\item  The off-diagonal matrix element $N_{12}=\langle A|B \rangle =1/2$ shows the
non-orthogonality of the $A,B$ basis.
In general, for SU($N_C$) this becomes $N_{12}=1/N_C$ and arises simply by
recoupling the colour components of the quark terms {\it i.e.}
\begin{equation}
|1_{1\bar{4}}1_{2\bar{3}}\rangle=\frac{1}{N_C}|1_{1\bar{3}}1_{2\bar{4}}\rangle
+\frac{\sqrt{N_C^2-1}}{N_C}|A_{1\bar{3}}A_{2\bar{4}}\rangle,
\end{equation}
where $|1_{i\bar{j}}\rangle$ and  $|A_{i\bar{j}}\rangle$ are the
SU($N_C$) singlet and adjoint representations.

At this stage, the lack of
orthogonality
could have been avoided by simply using the basis $A\pm B$. However,
later it will be seen that the $A,B$ basis is in fact more convenient
and suggestive, when the gluon fields are reintroduced in a more explicit
manner.
\item  The interquark potential $V_{Q\bar{Q}}(ij)=v_{ij}$ in Eq.~\ref{VQQ}
has been extracted
 as the potential energy between a single static quark ($Q$) and a single
static antiquark ($\bar{Q}$). In order to evaluate general multiquark 
potential energy matrix elements, a further assumption is
needed concerning the colour structure  of $v_{ij}$. Here the usual
identification
\begin{equation}
\label{vcol}
V_{ij}=-\frac{1}{3} {\bf \tau}_i.{\bf \tau}_j v_{ij}
\end{equation}
will be made, where the ${\bf \tau}_i$ are the Pauli spin matrices appropriate
for SU(2). This choice ensures, for a colour singlet meson-like state
$[ij]^0$, that  $\langle [ij]^0|{\bf \tau}_i.{\bf \tau}_j|[ij]^0 \rangle =-3$ and
\begin{equation}
\langle [ij]^0|V_{ij}|[ij]^0\rangle =v_{ij}.
\end{equation}
Strictly speaking, the form in Eq.~\ref{vcol} is only true in the weak
coupling limit of one-gluon exchange, since this has replaced the local
gauge invariance --- ensured by the series of $U$-links of
Eq.~\ref{QQ1} connecting the two quarks --- by the global gauge
invariance reflected by the ${\bf \tau}_i.{\bf \tau}_j$ factor.
\item With the choice of $V_{ij}$ in Eq.~\ref{vcol}, the off-diagonal
potential matrix element becomes
\begin{align}
\label{AVB}
\langle A|V|B\rangle =&V_{AB}=V_{BA}\nonumber \\
=&\frac{1}{2}\left(v_{13} +v_{24} +v_{14}+v_{23} - v_{12}-v_{34} \right).
\end{align}
Since the following discussion only involves quark configurations in the
rectangular geometries of Figs.~\ref{configs}, it is convenient to use
the notation
\begin{equation}
\label{not}
v_{13}=v_{24}=v_{d}, \ \ v_{14}=v_{23}=v_{r} \ \ {\rm and} \ \
v_{12}=v_{34}=v_x,
\end{equation}
where the suffix $x$ refers to the diagonals of
the rectangles in Fig.~\ref{configs}.
\end{enumerate}
Even though the form of Eq.~\ref{AVB} is derived in the
one-gluon exchange
limit, it is now assumed that a more realistic model emerges if the
$v_{ij}$ are taken to be the complete potential of Eq.~\ref{VQQ} and
not just the one-gluon exchange component. This clearly has the correct
form when the distance between the two two-quark clusters of
Fig.~\ref{configs}
 are far
apart, since in this case the only interactions are those {\it within}
 the separate clusters --- due to the cancellation in Eq.~\ref{AVB} of 
$v_{12}, \ v_{34}$ with either $v_{13}, \ v_{24}$ or $v_{14}, \ v_{23}$.

The $2\times 2$ Hamiltonian of Eq.~\ref{Ham}
is easily diagonalised to give the eigenvalues
$\lambda _{0,1}$. Since it is the binding energy $E'_i$ of the
four-quark
system that is of interest, and also that was extracted in
Subsec.~\ref{QQQQLC} from the
Monte Carlo simulation, the internal energy of the meson-like state
with the lowest energy ({\it i.e.} $2v_d$ for $d\leq r$) is now
subtracted
from the $\lambda_i$ to give
\begin{equation}
\label{E'}
 E'_i=\lambda_i-2v_d
\end{equation}
in analogy with the lattice expressions in Eqs.~\ref{le0} and \ref{le1}.

Therefore, in this simplest version of the two-body approach, the
$E'_i$'s
should correspond to the $E_i$'s from the Monte Carlo simulation --- a
comparison which is made in  Fig.~\ref{f=1}. Since the
values
of $E_0$ for squares \mbox{({\it i.e.} $r=d$)} are much larger than those for neighbouring
rectangles,
only the results for squares are shown in Fig.~\ref{f=1}. In these cases
\begin{equation}
 E'_0=-\frac{2}{3}(v_x-v_d) \ \ {\rm and} \ \
 E'_1=2(v_x-v_d)=-3E'_0,
\end{equation}
In addition,
the corresponding normalised wave functions are of the form
$\psi(E'_0)=\frac{1}{\sqrt{3}}|A+B\rangle $ and $\psi(E'_1)=|A-B\rangle $.

\vspace{0.2cm}

Two comments should be made on these results:
\begin{itemize}
\item For the smallest squares and rectangles the agreement between
$E_i$ and $E'_i$ is best.
This is reasonable, since it is expected that at small interquark
distances
perturbation theory is adequate {\it i.e.} the lowest order gluonic effects
are
already incorporated correctly into the interquark potential.
\item As the squares and rectangles get larger the differences between the
$E_i$ and
the $E'_i$ grow until $E'_0$ is more than three times $E_0$, and $E'_1$
more than seven times $E_1$, for the largest squares
$d\approx 7a\approx 0.8$ fm. Since $E'_0$ is too attractive
and $E'_1$ too repulsive, this suggests that the off-diagonal matrix
element $V_{AB}$ in Eq.~\ref{AVB} is too large.
\end{itemize}
The main conclusion to be drawn from the above is that the two-body
potential of Eq.~\ref{vcol} does {\it not} give the potential energy of
the four-quark system --- the indication being that the off-diagonal
potential energy $V_{AB}$ is too large.
This is nothing more than the well known van der Waals effect of
Ref.~\cite{Gavela} --- see Oka and Yazaki in Chapter~6 of Ref.~\cite{WWeise}.
\subsection{The effect of multiquark interactions}
\label{mqi}
In the above model it is assumed that all of the gluonic effects
are incorporated into the two-body potential $v_{ij}$. However, this is
clearly an oversimplification that is at best only applicable in
situations
where perturbation theory holds,  namely at short distances, as
already
noted in the discussion of Fig.~\ref{f=1}. In more realistic models, the QCD
coupling
is sufficiently strong to constrain the gluon field into flux-tubes
connecting
the quarks in a given meson --- as visualised by the wavy lines between
the quarks in states $A$ and $B$ in Figs.~\ref{configs}. Therefore, the overlap of
states
$A$ and $B$, {\it i.e.} $N_{12}=\langle A|B\rangle$, is not simply the colour recoupling
factor
of 1/2, but should also involve the lack of overlap of the gluon fields
in
states $A$ and $B$. This can be incorporated by introducing an entity
$f$,
which simply multiplies the original $N_{12}$, and which is an unknown
function of the position coordinates of the four quarks. With this
interpretation of $f$ as a gluon field overlap factor, it is also
necessary
to multiply the off-diagonal potential matrix element $V_{AB}$ of
Eq.~\ref{AVB} by the {\it same} factor $f$. This factor must be the
same
in $N_{12}=f/2$ and $V_{12}=fV_{AB}$, since otherwise the binding energies
$E'_{0,1}$ would be dependent on the self-energy term $c$ in the form of $v_{ij}$
given in Eq.~\ref{VQQ} --- which would be unphysical.
The $E'_{0,1}$ are now extracted from Eq.~\ref{E'} after diagonalising
\begin{equation}
\label{Hamf}
\left[{\bf V}(f)-\lambda_i(f) {\bf N}(f)\right]\Psi_i=0
\end{equation}
with
\begin{equation}
\label{NVf}
{\bf N}(f)=\left(\begin{array}{cc}
1&f/2\\
f/2&1\end{array}\right)\ \ {\rm and}\ \ {\bf
V}(f)=\left(\begin{array}{cc}
v_{13}+v_{24} & fV_{AB}\\
fV_{BA}&v_{14}+v_{23}\end{array}\right).
\end{equation}

The two equations (\ref{Hamf}) and (\ref{NVf}) are the basis of the
following
analysis. They give a procedure for extending the model of
Eqs.~\ref{Ham} -- \ref{NV2}, which was justified in the weak coupling limit,
into the
domain beyond one-gluon exchange. The off-diagonal potential matrix
element performs this extension in two ways. Firstly, even though
$V_{AB}$
in Eq.~\ref{AVB}
still has the same algebraic structure in terms of the $v_{ij}$ as
dictated
by the one-gluon exchange limit, the $v_{ij}$'s themselves are the full
two-quark potential of Eq.~\ref{VQQ}. Secondly, in the off-diagonal
correlation $W_{12}$ of Fig.~\ref{4QWiloop}, the one-gluon exchange
model suggests the presence of the overall multiplicative factor $f$ due
to gluon exchange within the {\it initial} and {\it final} states $A$
and $B$
at euclidean times $T=0$ and $\infty$. In this interpretation, the terms
in
$V_{AB}$ arise during the period of propagation between $T=0$ and
$\infty$.

\begin{figure}[ht]
\centering
\includegraphics*[width=0.65\textwidth]{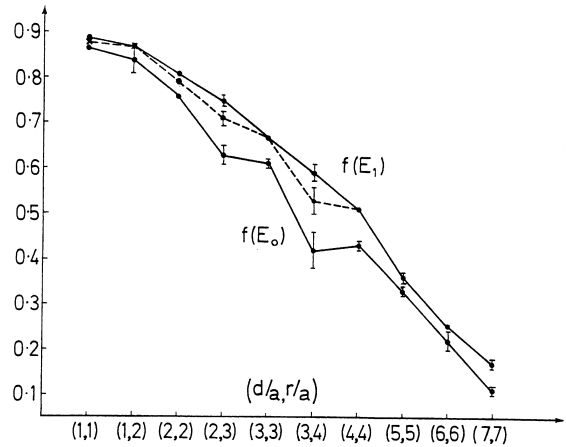}
\caption{The values of $f(E_0)$, $f(E_1)$ and $\bar{f}$ (from
Subsec.~\ref{mqi1}).}
\label{ffits}
\end{figure}

The strategy is now to adjust $f$ to get an exact fit to $E_0$ or $E_1$.
In
the case of squares, since
\begin{equation}
\label{e01}
E'_0=\frac{-f}{1+f/2}(v_x-v_d) \ \ {\rm and} \ \
E'_1=\frac{f}{1-f/2}(v_x-v_d),
\end{equation}
the appropriate expressions are
\begin{equation}
\label{f's}
f(E_0)=\frac{E_0}{v_d-v_x-E_0/2} \ \ {\rm and}    \ \
f(E_1)=\frac{E_1}{-v_d+v_x+E_1/2}
\end{equation}
for fitting $E_0$ and $E_1$ respectively --- the results being shown in
Fig.~\ref{ffits}. For general
rectangles the corresponding equations are somewhat more complicated.

Two points should be noted from this figure:
\begin{itemize}
\item All values of $f$
are less than unity as is expected from the interpretation of $f$ as
a gluon field  overlap factor. In addition, this idea is
supported by the fact that the values of $f$ decrease as the quarks
get further apart.
\item The values of $f(E_0)$ and $f(E_1)$ are rather similar, which 
suggests that a compromise value of $f$ would give a reasonable
description of both $E_0$ and $E_1$ --- as will be seen below.
\end{itemize}

\subsection{A compromise for the overlap factor $f$}
\label{mqi1}
The previous subsection suggests that for each geometry a {\it single}
value of $f$ could give a reasonable description of both $E_0$ and
$E_1$.
Here one such possibility is given by finding that value of $\bar{f}$
which minimizes the expression
\begin{equation}
\label{dmin}
D(\bar{f})=\left(\frac{E_0-E'_0(\bar{f})}{\Delta E_0}\right)^2+
       \left(\frac{E_1-E'_1(\bar{f})}{\Delta E_1}\right)^2,
\end{equation}
where the $\Delta E_i$ are the errors quoted for the $E_i$ from the
lattice calculation.
The result is  shown by the dashed line in Fig.~\ref{ffits}.

It is seen that indeed a single value of $f=\bar{f}$ suffices to explain
reasonably well
both energies. This is a non-trivial observation, since it indicates
that
the parametrization suggested in Eqs.~\ref{Hamf}, \ref{NVf} contains
the most important features of the more precise lattice calculation.

It should be added that the extraction of a compromise value of $f$ is
not simply a curiosity, since any model that needs different values of
$f$ for $E_0$ and $E_1$ would be more difficult to use in practice for
more complicated multi-quark systems.

At this point, even though for each geometry a single value of
$f=\bar{f}$
gives  values
of $E'_{0,1}$ that are in reasonable agreement with the $E_{0,1}$, it might be
asked
about the remaining small differences. Several possibilities are now
open:
\begin{itemize}
\item The lattice energies $E_i$ may not be sufficiently accurate due
to finite lattice size and scaling uncertainties. This was checked in
 Ref.\cite{GMPS} and found not to be a problem.
\item The  parametrization in Eqs.~\ref{Hamf}, \ref{NVf} may
be inadequate. One possibility would be to combine the notion of a gluon
overlap factor $f$ with a generalized form for the two-quark potential
in Eq.~\ref{VQQ}, since
this could introduce more free parameters.
\item  Any model based only on states $A$ and $B$ in
Fig.~\ref{configs} is incomplete and  other states are necessary in
addition. This point will be discussed further in
Sec.~\ref{fmodelext}.
\item The step from Eq.~\ref{Ham} to Eq.~\ref{Hamf} was motivated by one
gluon exchange. However, this could possibly be extended by
performing a {\it two-gluon} exchange calculation to see what new terms
arise and to
then be guided by this in making an improved parametrization --- the
topic of the next subsection.
\end{itemize}
\subsection{The effect of two-gluon exchange}
\label{2ge}
In Ref.~\cite{Morimatsu2} it was noted that the two-body $f=1$ models
discussed in  Subsec.~\ref{Unmodified} correspond to lowest order perturbation
theory in the  quark--gluon coupling {\it i.e.} to order
$\alpha=g^2/4\pi$ in the notation of Eq.~\ref{Willag}. This was extended
in Ref.~\cite{Lang}, where  a perturbative calculation to fourth order
in the quark--gluon coupling ({\it i.e.} to  $O(\alpha^2)$)
was made for the potential of the $QQ\bar{Q}\bar{Q}$
system. This was performed in the general case of colour SU($N$). 
Considering the
quarks to be at points $R_i$  define
\begin{equation*}
V_A=V(R_{13})+V(R_{24}), \
V_B=V(R_{14})+V(R_{23}), \
V_C=V(R_{12})+V(R_{34}),
\end{equation*}
where $V(R_{pq})$ is the two-body potential
between  the different quark and antiquark pairs  a
distance $R_{pq}$ apart.
For the two possible energy eigenstates, diagonalization yields
the following two potentials correct to $O(\alpha^2)$:
\[V_0=\]
\[\frac{
\left(N^2\!-\!2\right)
\left(V_A\!+\!V_B\right)
+2V_C
-N\sqrt{
N^2\left(V_A\!-\!V_B\right)^2
+4\left(V_A\!-\!V_C\right)\!\left(V_B\!
-\!V_C\right)
}
}{2\left(N^2-1\right)}\]
\[V_1=\]
\begin{equation}
\frac{
\left(N^2\!-\!2\right)
\left(V_A\!+\!V_B\right)
+2V_C
+N\sqrt{
N^2\left(V_A\!-\!V_B\right)^2
+4\left(V_A\!-\!V_C\right)\!\left(V_B\!
-\!V_C\right)
}
}{2\left(N^2-1\right)}.
\label{potentials}
\end{equation}
These potentials are {\it exactly equal} to those
given by the naive ($f=1$) two-body model in Eqs.~\ref{Ham} -- \ref{NV2}.
This fact that a straightforward two-body model is correct also to
next-to-leading order in the quark--gluon coupling
may be surprising in light of the non-abelian nature of QCD.
However, Ref.~\cite{Lang} does go on to  show that this two-body model fails
at $O(\alpha^3)$ as three- and four-body forces appear due to the onset of
three-gluon vertex effects.
In general, their nature seems to be complicated, but
for some geometries simplifications are possible; {\it e.g.}
for the four quarks on the corners of a regular
tetrahedron there will be no contribution from quark self--interactions
to four-body forces to $O(\alpha^3)$.

The overall conclusion from  Ref.~\cite{Lang} is as follows:
``Looking at the Monte Carlo lattice calculations for the
$QQ\bar{Q}\bar{Q}$-system
 in Refs.~\cite{GMP93,GMPS}, it is observed that for small interquark
distances
of a few lattice \mbox{spacings} (with  $a\approx 0.12$ fm)
the $f=1$ two-body model gives a reasonable approximation in the sense
that the four-quark potentials calculated from Eq.~\ref{potentials}
using the Monte Carlo two-body potentials are comparable to the
four-quark
potentials from the lattice simulation.
The agreement improves the smaller the distances get. By comparing the
perturbative ({\it i.e.}\ $1/R$) and
non-perturbative ({\it i.e.} linear) part in the usual parametrization of the
\mbox{$Q\bar{Q}$-potential} in Eq.~\ref{VQQ},
one would expect to start entering the perturbative regime
when  distances get down to  about two lattice spacings.
However, at that stage the approximation
provided by the two-body model is already very good. The
fact that the two-body model is correct to fourth order
in perturbation theory
certainly suggests that it should be a reasonable approximation
in the perturbative domain. This result supports the belief
that the results of the lattice simulations for small
enough distances indeed are correlated
to continuum perturbation theory,
and thus that continuum physics is extracted
from the Monte Carlo calculations."

\vskip 0.5 cm

So far in this section various models have been proposed in an attempt to
understand the results of the Monte Carlo simulations of lattice QCD.
The main outcome --- summarised
in Fig.~\ref{ffits} --- is the emergence of a function $\bar{f}$ that depends on the
coordinates of the four
quarks involved. This shows that the usual models based on purely 
two-quark
interactions need to be modified considerably --- essentially by the
factor $\bar{f}$, which becomes $\ll 1$ for large interquark distances.
This observation is in itself
of much interest, but at this stage it is not clear
how the effect can be incorporated into more realistic situations in
which
the quarks are not so restricted in their geometry. It is the purpose
of the next section to tackle this problem by first studying how $\bar{f}$
can be parametrized.

\subsection{ Parametrizations of the gluon-field overlap factor $f$ }
\label{paramoff}
In Subsecs.~\ref{mqi}, \ref{mqi1}, models were introduced in an attempt to
understand
the ground state binding energy $(E_0)$ and excited state energy $(E_1)$
emerging from a Monte Carlo simulation, in which four quarks were at the
corners of a rectangle. These models are summarised by
Eqs.~\ref{Hamf}, \ref{NVf}.
For each quark configuration, both of the energies $E_{0,1}$ are
described in terms of a function $f$ of the four quark positions
--- Eq.~\ref{e01}. As it stands,
this is not particularly useful, when wishing to extend these ideas to
systems containing more than four quarks, unless $f$ can be parametrized in some
sensible and convenient manner. In the literature, several
parametrizations
have been suggested. In Refs.~\cite{Morimatsu2,Morimatsu}, motivated by 
strong coupling arguments, the phenomenological
form is taken to be
\begin{equation}
\label{f1}
f_1=\exp[-\alpha b_s S],
\end{equation}
where $b_s=0.0736$ is the string tension in the  interquark
potential of Eq.~\ref{VQQ} and $S$ is the minimal area of the surface
bounded by the straight lines connecting the quarks and antiquarks.
The other form, the one proposed in \cite{Masud}, is
\begin{equation}
\label{f2}
f_2=\exp\big[-\frac{kb_s}{6}\sum\limits_{i<j}r^2_{ij}\big]
\end{equation}
{\it i.e.} the cut-down is governed by the average of the six links
present in a $Q^2\bar{Q}^2$ system.
In Eqs.~\ref{f1}, \ref{f2} the $\alpha$ and $k$ are at present free
parameters to be determined later.
Both of these parametrizations of $f$ accommodate the following two
extreme models.
\begin{itemize}
\item Weak coupling, which has $f=1$ when all $r_{ij}=0$.
This $f=1$ limit was assumed to apply for
all values of $r_{ij}$ in Subsec.~\ref{Unmodified}.
\item Strong coupling, which has $f=0$ when any $r_{ij}\rightarrow \infty$.
In this limit the flux-tubes between the quarks in the separate mesons
---  seen in Fig.~\ref{configs} --- are then very narrow and straight.
In this case the flux-tube overlap of configurations $A$ and $B$ tends
to zero in the limit that any $r_{ij}\rightarrow \infty$.
\end{itemize}
With squares $(r=d)$, for which the most accurate values of $f$ exist,
$k$ equals $3\alpha/4$. One measure of how meaningful these
parametrizations
really are, is given by extracting $\alpha $ and $k$ for each quark
configuration. The hope would then be that, for squares, $\alpha$ (and
therefore
$k$) would be independent of the separate configurations. Only the few
points
for non-square rectangles $(r\not = d)$ would be able to distinguish
between
$f_1$ and $f_2$.
\begin{figure}[ht]
\centering
\includegraphics*[width=0.65\textwidth]{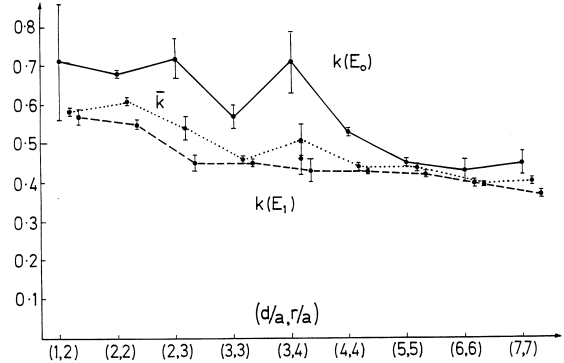}
\caption{The values of $k(E_0),k(E_1)$ and $\bar{k}$ corresponding to
the values of $f(E_0),f(E_1)$ and $\bar{f}$ in Fig.~\protect\ref{ffits}. }
\label{3kfits}
\end{figure}
The outcome is depicted in Fig.~\ref{3kfits} for the values of
$k(E_0),k(E_1)$ and $\bar{k}$ corresponding to
the values of $f(E_0),f(E_1)$ and $\bar{f}$ in Fig.~\ref{ffits}.
This shows that $k(E_0)$ for the squares appears to decrease slowly
from about
0.7 to about 0.5 as the sizes of the squares increase from ($2\times 2$)
to ($7\times 7$),
whereas for non-squares $k(E_0)$ appears to be stable at about 0.7$\pm$0.1.
On the other hand, $k(E_1)$ and $\bar{k}$ decrease somewhat less and
also the square and non-square values are consistent with each other.

The indications from this are that the {\it single} parameter
$\bar{k}\approx 0.5(1)$ is a suitable compromise value that results in
a reasonable fit to both $E_0$ and $E_1$ for a series of square and
near-square geometries.
\subsubsection{A reason for $f_1=\exp[-\alpha b_s S]$}
\label{Tiling}
\begin{figure}[t]
\centering
\includegraphics*[width=0.5\textwidth]{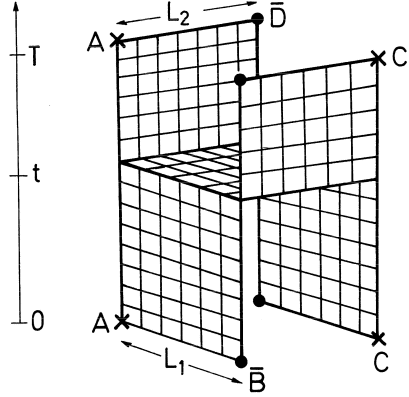}
\caption{Tiling the transition between the states
$[(Q_A\bar{Q}_B)(Q_C\bar{Q}_D)]$ and $[(Q_A\bar{Q}_D)(Q_C\bar{Q}_B)]$.}
\label{figtiling}
\end{figure}

In the limit of large $l$, Eq.~\ref{Willoop} becomes
\begin{equation}
\label{Willoopa}
W(l,t)\rightarrow \exp[-t b'_s l]=\exp[-b_s S] \ \ \ {\rm as} \ \
t\rightarrow \infty,
\end{equation}
where $S$ is the {\it minimal} space--time area enclosed by the loop in
Fig.~\ref{2QWiloop}. For this $Q\bar{Q}$ case, the meaning of $S$ is
clear --- it is simply $a^2lt$. But for the $[Q\bar{Q}][Q\bar{Q}]$ case in
Fig.~\ref{4QWiloop} the form of the appropriate minimal space--time area
is less clear. However, in the extreme strong coupling limit 
the \mbox{area $S$} is the one produced by the {\it minimum} number of 
elementary plaquettes
needed to tile the enclosed area in question. Any fluctuations
about this space-time surface would need more plaquettes and so be higher order
in the  strong coupling model. In this limit, the diagonal loops
$W_{ii}$ in Fig.~\ref{4QWiloop} are again simply two Wilson loops each
tiled separately by the minimum number of plaquettes.
However, the off-diagonal loops $W_{ij}$ are more complicated.
A model for this was suggested in Ref.~\cite{MatSiv} and developed in
Refs.~\cite{Morimatsu,Morimatsu3,GLW}. In the notation
of Fig.~\ref{figtiling} the Euclidean Green's function for a
$Q_A,\bar{Q}_B,Q_C,\bar{Q}_D$ system can be thought of as a \mbox{$2\times 2$}
matrix for a two channel problem with the $(A\bar{B})(C\bar{D})$ and
$(A\bar{D})(C\bar{B})$ configurations. The transition potential between
these two configurations may then be extracted from the expression for
the Wilson loop of this system,
\begin{equation}
\label{Strongce}
{\bf W}(T)=\left(\begin{array}{cc}
\exp[-2b_sL_1T] & \epsilon\\
\epsilon& \exp[-2b_sL_2T] \end{array}\right),
\end{equation}
where $L_1$ and $L_2$ are the minimum flux tube lengths in the basis
states and $b_s$ is the string tension of Subsec.~\ref{ssst}.
The diagonal terms are each simply the product of two propagators ---
$\exp[-b_sL_1T]$ for $W_{11}$ and $\exp[-b_sL_2T]$ for $W_{22}$.
In a language more familiar in nuclear physics, these are simply the
Green's functions
\begin{equation}
\label{Greensf}
W_{11}=G_{A\bar{B}}(T)G_{C\bar{D}}(T) \ \ {\rm and} \ \
W_{22}=G_{A\bar{D}}(T)G_{C\bar{B}}(T).
\end{equation}
The mixing term
\begin{equation}
\label{Strongmix}
\epsilon=\sum_t\exp[-2bL_2(T-t)]\exp[-bL_1L_2]\exp[-2bL_1t]
\end{equation}
corresponds to tiling of the off-diagonal loops $W_{ij}$.
This now resembles the standard expression for the lowest order
transition of a Green's function
\begin{equation}
\label{GFmix}
G_{fi}(T)=\int dt G_f(T-t)V_{fi}G_i(t),
\end{equation}
where here $i,f$ denote the initial and final channels
$[(Q_A\bar{Q}_B)(Q_C\bar{Q}_D)]$ and $[(Q_A\bar{Q}_D)(Q_C\bar{Q}_B)]$.
Remembering that $\sum_t \rightarrow a^{-1}\int dt$ we can identify
the transition potential as simply
\begin{equation}
\label{trapot}
V_{fi}=\exp[-b_sL_1L_2]/a\rightarrow \exp[-b_s S]/a
\end{equation}
{\it i.e.} the transition potential can be expressed in terms of a
{\it spatial} area --- as anticipated by the models in the previous subsections.
In more complicated geometries this should be the
minimum area in coordinate space associated with the given boundary
conditions.
\section{More Complicated $[(Q\bar{Q})(Q\bar{Q})]$ Geometries}
\label{ComplQQ}
\begin{figure}[ht]
\centering
\includegraphics*[width=0.65\textwidth]{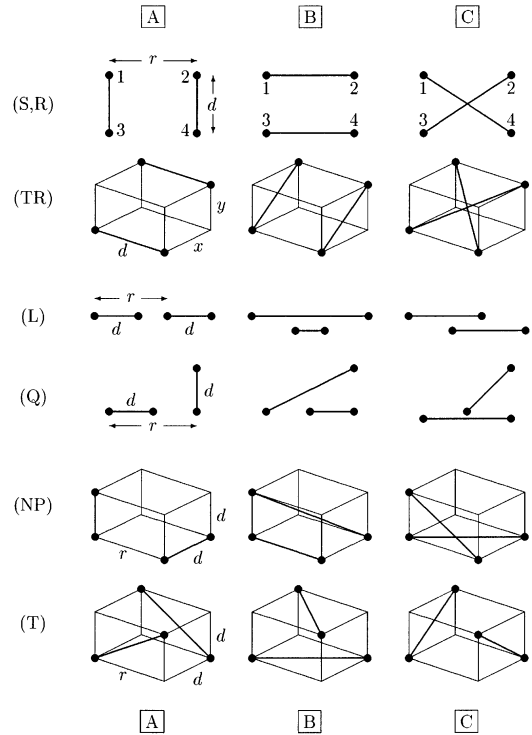}
\caption{Six four-quark geometries a) Squares (S) and  Rectangles (R),
b) Tilted rectangles (TR), c) Linear (L), d) Quadrilateral (Q),
e) Non-Planar (NP) and  f) Tetrahedra (T). Their energies are calculated
using Lattice QCD in Refs.~\protect\cite{PP1,GMS,JL,GMP,GMP93,GMPS} and
analysed in Ref.~\protect\cite{GPplb426}.}
\label{Sixg}
\end{figure}

So far the only $[(Q\bar{Q})(Q\bar{Q})]$ geometries considered above were
squares and rectangles. This suggested that the strongest interaction
between two separate two-quark clusters occurs when the clusters are
degenerate in energy --- namely for square geometries compared with
rectangles. To test this more, in Ref.~\cite{GMS} the six different
geometries in Fig.~\ref{Sixg} were studied.
Since this confirmed that
configurations degenerate in energy always gave the largest binding
--- not just in the square versus rectangle case --- in Ref.~\cite{PP1}
the study concentrated only on those configurations that were near degenerate.

In Ref.~\cite{JL} the case of
 tetrahedral configurations was considered in some detail, 
since in SU(2) this has
the {\it three} degenerate partitions seen in \mbox{Fig.~\ref{Sixg}(T)} when
$d=r$. Since tetrahedral and linear configurations have certain
interesting features, they will  be discussed separately below.

However, real life is an average over all 
spatial configurations, so why should we be so interested in such special
geometries for four static quarks? Firstly, it must be remembered that 
model (bridge) builders
should consider all spatial possibilities --- a failure with one
configuration indicating that the proposed model is faulty. Also, in
some ways the tetrahedral and linear configurations have  simplicities
and symmetries not present in other cases, which could make
``Lattice QCD $\leftrightarrow$ Model" comparisons easier.
\subsection{Tetrahedral configurations on a lattice}
\label{Tetcon}
Since tetrahedral configurations are so symmetrical, at first sight
there is no reason to consider only two of the three possible cluster
partions \mbox{$A+B,  \  \ B+C$ or $A+C$.}
In this case, using the notation that the suffices $1,2,3$ are any
combination of $A,B,C$ that forms a basis, then
the appropriate Wilson loop matrix is
\cite{JL}
\begin{equation}
\label{NVT}
{\bf W}^t=\left(\begin{array}{ccc}
W_{11}^t&W_{12}^t&W_{13}^t\\
W_{21}^t&W_{22}^t&W_{23}^t\\
W_{31}^t&W_{32}^t&W_{33}^t
\end{array}\right).
\end{equation}
For regular tetrahedra, the general symmetries $W_{11}^t=W_{22}^t=W_{33}^t$,
$W_{21}^t=W_{12}^t$, $W_{31}^t=W_{13}^t$, $W_{32}^t=W_{23}^t$ are
expected and, in addition, there are the equalities
$W_{13}^t=W_{12}^t$ and $W_{23}^t=-W_{13}^t$.
Therefore, in all, there are only {\it two}
independent Wilson loops $W_{11}^t$ and $W_{12}^t$.
Here the  minus sign
appearing in the last equation is a reminder that the quarks are in
fact fermions even though quarks and antiquarks transform in the
same way under SU(2). This detail is discussed more in the Appendix to
Ref.~\cite{GMS}.

\begin{figure}[ht]
\centering
\includegraphics*[angle=-90,width=1.0\textwidth]{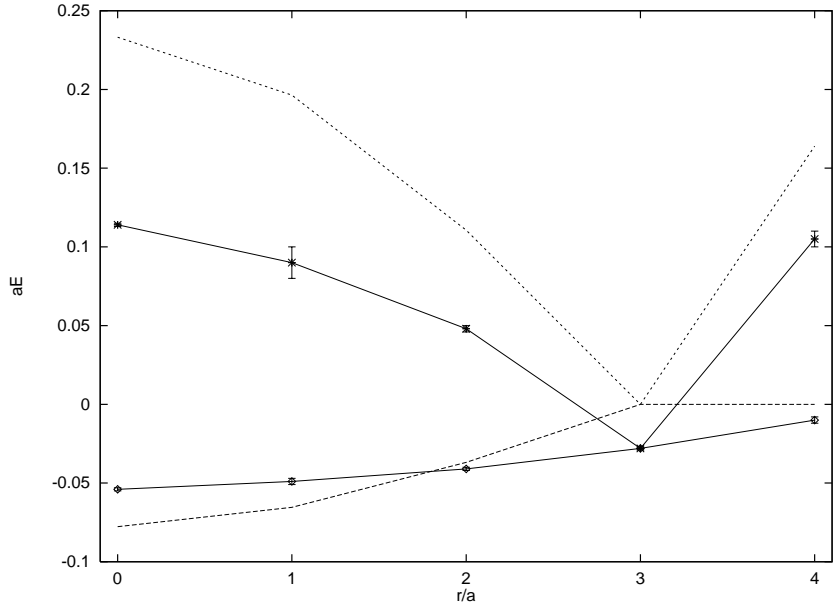}
\caption{The binding energies -- in units of the lattice spacing --
of the four-quark states for the tetrahedral geometry of
Fig.~\protect\ref{Sixg} (T) for $d/a=3$ and
$r/a=0,1,2,3,4$. \protect\\ Solid lines show lattice results:
$\diamond$ -- the ground-state binding energy $E_0$.
\protect\\ $\times$ -- the first excited state energy $E_1$.
\protect\\ Dashed and dotted lines show model results with $f=1$
from Eqs.\ \protect\ref{Ham} -- \protect\ref{NV2}, for $E_0$ and $E_1$
respectively --- see Ref~\protect\cite{JL}.
\label{f:energs}}
\end{figure}

Unfortunately, the inclusion of all three partitions  does not lead to even
more binding. However, it shows the curious feature that the ground
and first excited states become degenerate in this highly symmetrical limit,
since the eigenvalue equation (discussed earlier as Eq.~\ref{WN})
\begin{equation}
\label{Eig}
W_{ij}^t a_j^t=\lambda^{(t)}W_{ij}^{t-1} a_j^t,
\end{equation}
 is easily solved to give for the lowest energy
(occurring twice)
\begin{equation}
\label{331}
\lambda_{1,2}= \frac{W^t_{11}+W^t_{12}}{W^{t-1}_{11}+W^{t-1}_{12}}
\end{equation}
and for the excited state
\begin{equation}
\label{332}
\lambda_3= \frac{W^t_{11}-2W^t_{12}}{W^{t-1}_{11}-2W^{t-1}_{12}}.
\end{equation}
In comparison, using only two partitions gives
\begin{equation}
\label{NVS}
{\bf W}^t=\left(\begin{array}{ll}
W_{11}^t&W_{12}^t\\
W_{21}^t&W_{22}^t\end{array}\right),
\end{equation}
where not only is the general symmetry $W_{12}^t=W_{21}^t$ expected but
also for a regular tetrahedron $W_{11}^t=W_{22}^t$. 
In this case Eq.~\ref{Eig} is
again easily solved to give for the lowest energy
\begin{equation}
\label{221}
\lambda_1= \frac{W^t_{11}+W^t_{12}}{W^{t-1}_{11}+W^{t-1}_{12}}
\end{equation}
{\it i.e.} the {\bf same} as $\lambda_1$ in Eq.~\ref{331}. However,
 for the energy of the first excited state
\begin{equation}
\label{222}
\lambda_2= \frac{W^t_{11}-W^t_{12}}{W^{t-1}_{11}-W^{t-1}_{12}},
\end{equation}
which is quite different to the complete result in Eqs.~\ref{331}, \ref{332}.
So the degeneracy is easily explained as a feature of the more complete
$3\times 3$ lattice QCD simulation.
The effect is depicted in Fig.~\ref{f:energs} as $d\rightarrow r.$
This also shows that, whereas ground state
energies can be quite stable, those of excited states are more model
dependent.

The dominance of the cluster interaction by degenerate
configurations was  carried to the extreme in the so-called flip-flop model
of Ref.~\cite{flip}, which makes the ansatz that the 4-quark interaction
occurs {\it only} in the case of exact degeneracy. However, the model is
developed for a purely {\it linear} interquark potential {\it i.e.} with only
the $b_s$ term in Eq.~\ref{VQQ}, which means that the interaction only
occurs when --- in the notation of Fig.~\ref{configs} --- the spatial
distances $r_{13}+r_{24}$ and $r_{14}+ r_{23}$ are equal.

In several of these geometries it is not clear what is the best
``natural" partition into two-quark clusters. Therefore, in the lattice
simulation all three possibilities $A, \ B$ and $C$ in Fig.~\ref{Sixg}
should be
taken into account. However, in some cases it is found that one (or more)
of the combinations $A+B$ or
$A+C$ or $B+C$ is sufficient to give --- for both $E_0$ and $E_1$ ---
 similar results to the complete $A+B+C$ simulation.
This is particularly true if the configuration with the lowest energy
is one of the configurations used.
The study of tetrahedral and near-tetrahedral geometries on a lattice can be
summarized as follows:
\begin{itemize}
\item From earlier work in Refs.\ \cite{GMS,GMP,GMP93,GMPS}, for the
corresponding squares [{\it i.e.}\ $(d,0)$ in Fig.~\ref{Sixg} (T)], where two of
the basis states are degenerate, the binding energy of the lowest
state ranges from --0.07 to --0.05 as $d/a$ goes from 1 to 5.
 However, now --- even though at least two of
the basis states are always degenerate \mbox{($A$ and $B$)} --- the ground
state binding energy [$E_0(d,r)$] is always less than that of the corresponding
square [$E_0(d,0)$]. For a fixed $d$, $|E_0(d,r)|$ decreases as $r$
increases from $r=0$. Nothing interesting happens to $E_0$ at $r=d$, the point at which
 all the basis states are degenerate in energy.
\item For fixed $d$, as $r$ increases from 0 to $d$, the energy of
the first excited state $E_1$ decreases until $E_1(d,d)=E_0(d,d)$.
For $r > d$, $E_1(d,r)$ increases again. This degeneracy of
$E_{0,1}$ for the tetrahedron is a new feature compared with earlier
geometries. As will emerge in Sec.~\ref{fmodelext},
this is a severe
constraint on  models.
\item The choice of which $2\times 2$ basis to use depends on the
particular geometry, because
one of these two basis states must have the lowest unperturbed
energy. Since $A$ and $B$ are degenerate in energy, for a given $d$
this amounts to using $A+B$ for $r\leq d$. On the other hand, for $r > d$ it
is necessary to use $B+C$, since $C$ now has the lowest unperturbed
energy.
\item Except for the tetrahedra, the values of $E_1$ are essentially
the same in the $2\times 2$ and $3\times 3$ bases.
\item The values of $E_2(d,r)$ are always much higher than
$E_1(d,r)$. However, as will be discussed in the Sec.~\ref{fmodelext},
 this second
excited state is dominated by excitations of the gluon field and so
is outside the scope of the models introduced in that section.
\end{itemize}
The above is for colour SU(2). 
However, in the real world of SU(3) the state
$C=[Q_1Q_2][\bar{Q_3}\bar{Q}_4]$  cannot
appear asymptotically as  two clusters 
--- see Eq.~\ref{ABC3}. Even so this SU(2) lattice data 
 should be understandable in terms of models 
that are similar to those of SU(3)--- see Subsec.~\ref{bridges}.
\subsection{QCD in two dimensions (1+1)}
The above analyses unavoidably involve numerical results extracted
from Monte Carlo simulations on a 4-dimensional lattice. However,
 it is possible to study colinear colour sources in a simple
approximation for which exact theoretical results are known \cite{GMPS}.
This is QCD in two dimensions (QCD2) -- the one spatial direction
 allowing colinear (and only colinear) configurations. For quenched
QCD2,
the spectrum is known {\it exactly} even on a lattice. This
can be summarised as the requirement that each
link is in a representation of the colour group [{\it i.e.} SU(2) here].
Therefore, the links can be in the singlet ground state ($J$=0), or they
can be
excited
to a fundamental representation ($J$=1/2) or an adjoint representation
($J$=1) and so on. The energy per unit length for such links is given by
$b_J=\frac{4}{3}KJ(J+1)$.
For 4 colinear quarks, the lowest state [A in Fig.~\ref{Sixg} (L)]
 has energy $2Kd$, while the first excited state has, in the middle, an
adjoint
link of length $(r-d)$ resulting in an energy $2Kd+8K(r-d)/3$.  This
exact
result
applies at any $\beta$-value, {\it i.e.} strong coupling or weak coupling.

A comparison with the above $f$-mixing model of
Eqs.~\ref{Hamf}, \ref{NVf} shows that there is  agreement
with the exact QCD2 results  provided $f=1$ --- independent of which
combination of states is used for the analysis A+B, A+C or B+C.
 This is also true in the more general case when $d_{13}\not= d_{24}$.
If the interpretation of $f$ being a gluon-field overlap factor
is correct, then it is easy
to understand that here $f$ must be unity, since with only one spatial
direction
the colour flux must overlap fully.
Earlier, the main motivation for the $f$-mixing model had been from
weak coupling arguments, so that this agreement
between the mixing model and QCD2 does suggest that the model
is sensible even at large $\beta$. This adds support to
the claim that it may be a useful
phenomenological tool when the extension to $f\not = 1$ is made.

\section{Extensions of the $2\times 2$ $f$-Model}
\label{fmodelext}
In Sec.~\ref{potmodel} a model for describing 4-quark interactions
was developed as a $2\times 2$ matrix equation
for the two heavy-quark states $A=[Q_1\bar{Q}_3][Q_2\bar{Q}_4]$ and
$B=[Q_1\bar{Q}_4][Q_2\bar{Q}_3]$ in Fig.~\ref{configs} --- see
Eqs.~\ref{Hamf} and \ref{NVf}. There the only geometries studied were
those in which the quarks were at the corners of squares or
rectangles. However, when the four quarks are in more general geometries,
such as those in Fig.~\ref{Sixg}, the choice of which two partitions out
of the possible three to use is less clear 
{\it i.e.}  $C=[Q_1Q_2][\bar{Q_3}\bar{Q}_4]$ should be included. 
This is completely analogous to the
problem that arose in Subsec.~\ref{Tetcon} when deciding which
configurations to use in the corresponding lattice calculations.

In  Appendix A and in Ref.~\cite{GPprc57} the model of Sec.~\ref{potmodel}
is extended from being a $2\times 2$ matrix equation for states 
$A, \ B$ into a $3\times 3$ version, where all 3 basic
partitions $A, \ B, \ C$ are
included. Finally a $6\times 6$ extension is developed, in which interquark
excited states are introduced to give three additional basis states 
$A^*, \ B^*, \ C^*$.

Much of the  discussion will concentrate on the problems that 
arise in trying to model the energies of regular tetrahedral configurations.
This might be considered a minor point to worry about, since such a
configuration is very special. However, the philosophy is that, if {\em
any} configuration cannot be fitted, then the model fails, since then
there is no reason to  expect other configurations not checked explicitly to
be fitted.

A summary of the main points that emerge from  Appendix A are as
follows:

1) The $2\times 2 \longrightarrow 3\times 3$ extension, to some extent, 
clarifies the reason why the earlier $2\times 2$  f-model was, in many
cases, quite 
successful. Also it shows that an understanding of the tetrahedron
spectrum --- in particular the degenerate ground state --- requires a
generalisation of the two-quark approach.

2) The step  $3\times 3 \longrightarrow 6\times 6$ has the very positive
feature for tetrahedra of giving a ground state binding energy that initially 
{\it increases} with the size of the tetrahedron --- a result also seen
in the corresponding lattice data. This arises naturally, since the
energies of the additional states $A^*, \ B^*, \ C^*$ are each excited
by an energy of $\pi/R$ with respect to the $A, \ B, \ C$ states in the
$3\times 3$ model. Here $R$ is the interquark distance for that pair of
quarks containing the excited gluon field. Therefore, as the 
four-quark configuration
gets larger spatially, the energies of the $A^*, \ B^*, \ C^*$ states
approach from above the energies of the $A, \ B, \ C$ states. The
subsequent mixing between the two sets of states then manifests itself
as an additional overall attraction that also grows with the spatial size.

3) An even more interesting conclusion from the $6\times 6$ extension is
that it partially justifies the $f=1$ model of Subsec.~\ref{Unmodified}.
In  Appendix A it is seen that, by fitting simultaneously the
lattice  energies  for the geometries in Fig.~\ref{Sixg}, the $6\times 6$
model shows two features: 
\begin{itemize}
\item  Outside the range where perturbation theory
holds ({\it i.e.} beyond about 0.2 fm) the binding is dominated by the 
$A^*, \ B^*, \ C^*$ configurations. 
\item The overlap factor between the $A^*, \ B^*, \ C^*$ states
(corresponding to $f$ in Eq.~\ref{NVf} for the $A, \ B, \ C$ states)
 is essentially {\it unity}.
\end{itemize}
We, therefore, come to the following scenario for the four-quark
interaction:
At the shortest distances, up to about 0.2 fm, perturbation theory is
reasonable with the binding being given mainly by the $A,B,C$ states
interacting simply through the two-quark potentials with little
effect from four-quark potentials. However, for intermediate ranges,
from
about 0.2 to 0.5 fm, the four-quark potentials act in such a way as to
reduce the effect of the $A,B,C$ states so that the binding is dominated
by the $A^*,B^*,C^*$ states, which now  interact amongst themselves
again
simply through the two-quark potentials {\it with little
effect from four-quark potentials}. 
This suggests that models involving only two-quark potentials could be 
justified --- {\it provided excited gluon states (such as $A^*,B^*,C^*$) are
included on the same footing as the standard states $A,B,C$.}
The above result that excited states play an important r\^{o}le in the overall
binding is reminiscent of the nucleon-nucleon interaction, where  
nucleon excitations --- especially the $\Delta(1236)$ --- are also responsible
for a sizeable part of the attraction. 
 
\section{ Heavy-Light Mesons ($Q\bar{q}$)}
\label{Heavylight}
In the previous sections only static ({\it i.e.} infinitely heavy) quarks
with two colours were discussed. Even though this is a far cry from
the real world of finite mass quarks with three colours, it resulted in
several interesting conclusions. However, these were of a somewhat academic
nature useful for creating models, but could not be compared directly with experimental
data. In this section a compromise situation of two quarks is studied 
where one of the
quarks is still static but with the other being light. This is essentially
the ``hydrogen atom" of quark physics and is expected to be a reasonable
representation of the heavy-light $B$-mesons. This also means that in
the development of the Effective Potential Theories  of
Subsec.~\ref{sect.EPT}, it is possible that the {\it one-body} 
Schr\"{o}dinger or Dirac
equation is applicable. If this proves to be so, then it will be a
further reason for studying $B$-mesons in addition to the more basic
ones discussed below.

It should be added that the hydrogen atom analogy also partially holds for the
interaction ---  the coulomb potential $\propto e^2/r$
of the hydrogen atom versus the one-gluon exchange $\propto \alpha /r$
in the heavy-light meson. However, as mentioned in
Subsec.~\ref{sect.EPT}, for light quarks there are indications that this
$1/r$ attraction gets damped by form factor effects and that
much of the needed attraction could arise from a short ranged
instanton--generated interaction. Also, it must not be forgotten that
beyond $r\approx $0.2--0.3 fm the interaction  in Eq.~\ref{VQQ} becomes dominated
by the linear confining potential.

From Eq.~\ref{hier2}, for lattice calculations with
a heavy quark $Q$, we had the condition  that $a$ should satisfy
$a\ll m_Q^{-1}$. At
present, this rules out direct lattice calculations with $b$ quarks of
mass 5 GeV, since they would require $a\ll 0.04$ fm. This prompted the
weaker condition $a \sim m_Q^{-1}$ in Eq.~\ref{hier3}. One way to
partially avoid
this problem is to perform lattice calculations for lighter quarks
with $m_Q \sim 2$ GeV, which do not require such a fine mesh, and then to
{\it extrapolate } the results to the $b$ quark mass. However, if the
results from a {\it static}--light system are also included in the analysis
we end up having to {\it interpolate} to $m_Q \sim 5$ GeV. 
Since heavy quarks with such masses are essentially non-relativistic,
the appropriate interpolation parameter is $1/m_Q$, so that the
actual interpolation is between $1/m_Q=0$ to 0.5 GeV$^{-1}$.
This is usually a more reliable procedure \cite{cbern}.

\subsection{Bottom $(B)$-mesons}
Bottom mesons are the bound $q\bar{q}$ states
of a $\bar{b}$-antiquark of mass $\approx $5 GeV and a lighter quark.
These are listed in Table~\ref{Blist}. This century has opened with
there being renewed interest in $B$-physics --- see the Bottom meson
summary in the Particle Data listings of Ref.~\cite{PDG}. There are new
generations of $B$-meson experiments at BaBaR (SLAC), Belle, CLEO III and Hera-B.
These machines have started to accumulate $B$-mesons and the
long-awaited $B$-factory era has begun. The hope is that these
experiments will deliver the fundamental constants of the Standard Model
and also improve our understanding of $CP$ violation. However, having
seen that $B$-mesons are important objects, it must be confessed that
their study by lattice QCD is very incomplete. This will be the topic of
the present section, where we concentrate on their energies, and in
Sec.~\ref{sect.C3}, where we extract density distributions.

\begin{table}
\caption{Properties of the Bottom mesons \protect\cite{PDG}. The state
marked with $^*$ is from Ref.~\protect\cite{Nodestate} and those
marked with ? do not, at the time of writing, have $I(J^P)$ confirmed.}
\vskip 0.5cm
\label{Blist}
\begin{center}
\begin{tabular}{|c|c|l|c|l|}\hline
Meson&$q\bar{b}$&$I(J^P)$&$nL$&Mass(GeV)\\ \hline
$B^{+}$&$u\bar{b}$&$\frac{1}{2}(0^-)$&$1S$&5.2790(5)\\
$B^{0}$&$d\bar{b}$&$\frac{1}{2}(0^-)$&$1S$&5.2794(5)\\
$B^{'0}$&$d\bar{b}$&$\frac{1}{2}(0^-)$&$2S$&5.859(15)$^*$\\
$B^{*+}$&$u\bar{b}$&$\frac{1}{2}(1^-)$&$1S$&5.325(6)\\
$B^{**}_J$&$u\bar{b}$&$\frac{1}{2}(0,1,2^+)$&$1P$&5.698(8) ?\\
 \hline
$B^{0}_s$&$s\bar{b}$&$0(0^-)$&$1S$&5.3696(24)\\
$B^{*}_s$&$s\bar{b}$&$0(1^-)$&$1S$&5.4166(35) ?\\
$B^{**}_{Js}$&$s\bar{b}$&$0(0,1,2^+)$&$1P$&5.853(15) ? \\ \hline
$B^{+}_c$&$c\bar{b}$&$0(0^-)$&$1S$&6.4(4)\\ \hline \end{tabular}
\end{center}
\end{table}

\subsection{Lattice parameters}
\label{lpfti}
In the past few years there have been several detailed measurements
of \mbox{$B$-meson} excited state energies. 
Some of the parameters used in these studies are given in 
Table~\ref{lattpar}.
\begin{table}
\caption{Lattice parameters. The notation is explained in the text.
Refs.~\protect\cite{MP98,G+K+P+M} are in the quenched approximation and
Refs.~\protect\cite{G+K+P+M2,GKMMT03} are unquenched.}
\vskip 0.5cm
\label{lattpar}
\begin{center}
\begin{tabular}{|c|c|c|c|l|c|c|} \hline
Ref.&$L^3\times T$&$\beta$&$C_{SW}$&$ \ \kappa$&$a$ (fm)& $M_{PS}/M_V$\\ \hline
\protect\cite{MP98}&$12^3\times 24$&5.7&1.57&0.14077&$\approx 0.17$&0.65\\
\protect\cite{G+K+P+M}&$16^3\times 24$&5.7&1.57&0.14077&$\approx 0.17$&0.65\\
\protect\cite{G+K+P+M2}&$16^3\times 24$&5.2&1.76&0.1395&$\approx
0.14$&0.72\\
\protect\cite{GKMMT03}&$16^3\times 32$&5.2&2.02&0.1350&$\approx 0.11$
&0.70\\ \hline
\end{tabular}
\end{center}
\end{table}
 The quenched calculations of Refs.~\cite{MP98,G+K+P+M} are very similar to
each other
with the latter having a somewhat larger lattice. The parameter $C_{SW}$
has been tuned for the clover improved action mentioned earlier
\cite{clover1,clover2}. The hopping parameter $\kappa$
 essentially determines the mass $m_q$ of the light quark as  being
slightly
smaller than the accepted value of the strange quark mass {\it i.e.}
$m_{\bar{q}}=0.91(2) m_{{\rm s}}$.
The ratio of the pseudoscalar to vector masses ($M_{PS}/M_V$) --- {\it i.e.}
the ratio of the ``$\pi$"- and $\rho$-meson masses generated by those
particular configurations --- is another measure of $m_q$. Since this
ratio is much larger than the experimental value of 0.18, 
we again see that $m_q \gg m_{u,d}$.
Refs.~\cite{G+K+P+M2,GKMMT03} are expected to be a distinct improvement,
since these are unquenched  calculations with smaller lattice
spacings.
Now $m_{\bar{q}}=1.28,\ 1.12 m_{{\rm s}}$ respectively. However, in
practice, it is found that the energy splittings of the excited states
are only weakly dependent on $m_{\bar{q}}$ for the range of values used here.

\subsection{Maximal Variance Reduction (MVR) }
\label{MVR}
One of the reasons why the energies of $Q\bar{q}$-states can now be
calculated reliably is not only the improvement in computer
capabilities, but also by developments in formalism.
It has been demonstrated  that light-quark propagators can be
constructed in an efficient way using the so-called Maximal Variance
Reduction (MVR) method.  Since this has been explained in detail
elsewhere, for example Subsec. 4.5.3 of Chapter~4 and  in Ref.~\cite{MP98},
the emphasis here will be
mainly on the differences that arise when estimating on a lattice the
two- and three-point correlation functions $C_2, \ C_3$ needed for measuring
spatial
charge and matter densities ($C_3$) in addition to the energies ($C_2$).
In the MVR method the inverse of a positive
definite matrix $A$ is expressed in the form of a Monte Carlo
integration
 \begin{equation}
\label{Gauss1}
A^{-1}_{ji}=
\frac{1}{Z}\int D\phi \ \phi^*_i \phi_j \ \exp(-\frac{1}{2}\phi^*A\phi),
\end{equation}
 where the scalar fields $\phi$ are pseudofermions located on lattice
sites $i,j$. For a given gauge configuration on this lattice, $N$
independent samples of the $\phi$ fields can be constructed by Monte
Carlo techniques resulting in a stochastic estimate of $A^{-1}_{ji}$ as
an average of these $N$ samples  {\it i.e.} $A^{-1}_{ji}=\langle \phi^*_i
\phi_j \rangle$. The $N$ samples of the $\phi$ fields can be calculated
separately and stored for use in any problem involving light quarks with
the same gauge configurations. In practice, $N \approx 25$ is found to
be sufficient.

In LQCD the matrix of interest is the Wilson--Dirac matrix $Q=1-\kappa M$,
where $M$ is a discretized form of the Dirac operator $(\not\!\!\partial+m)$ and is
the mechanism for ``hopping " the quarks from one site to another.
However, $Q$ is not positive definite for those values of the hopping parameter
$\kappa$ that are of interest. Therefore, we must deal with
$A=Q^{\dagger}Q$, which is positive definite. Since $M$ contains only
nearest neighbour interactions, $A$ --- with at most next-to-nearest
neighbour interactions --- is still sufficiently local for effective
updating schemes to be implemented. In this case the light-quark
propagator   from  site $i$ to site $j$ is expressed as
 \begin{equation}
\label{Gauss2}
G_q=G_{ji}= Q^{-1}_{ji}=\langle (Q_{ik}\phi_k)^*\phi_j\rangle=
\langle \psi^*_i\phi_j\rangle.
\end{equation}
 This is the key element in the following formalism.
 The Wilson--Dirac matrix also leads to an alternative form for the above
light-quark propagator from site $i$ to near  site $j$
 \begin{equation}
\label{Gauss3}
G_q'= G'_{ji}=\gamma_5\langle(Q_{jk}\phi_k)\phi_i^*\rangle\gamma_5=
\gamma_5\langle \psi_j\phi_i^*\rangle\gamma_5.
\end{equation}
In practice both forms are used, since --- for the {\it same} correlation ---
they lead to independent measurements, which can then be averaged to improve
the overall statistical error.
 Later, it will be essential to use at some lattice sites operators that
are  {\it purely local}. This then restricts us to using at such sites
only  the $\phi$ fields that are located on single lattice sites.  In
contrast the $\psi_i$ fields, defined above as $Q_{ik}\phi_k$, are not
purely local, since they contain $\phi$ fields on next-to-nearest
neighbour sites.

In the above, the term ``Maximal Variance Reduction" comes from the
technique applied to  reduce the statistical noise in Eq.~\ref{Gauss2}.
The lattice is divided into  two boxes ($0 < t < T/2$ and $T/2 < t < T$)
whose boundary is kept fixed.  Variance of the pseudofermionic fields is
then reduced by numerically solving  the equation of motion inside each
box. This allows the variance of propagators from one box to the other
to be greatly reduced. However, in the case of a three-point correlation
in Subsec.~\ref{sect.C3} two  propagators are needed and this is best
treated by choosing one of
the points to be on the boundary of the boxes while the other two are
inside  their own boxes. Furthermore, the field at the boundary must be
local to avoid the two propagators interfering with each other. This
means that only the $\phi$ fields should be used on the boundary and there
they can couple to the charge, matter or any other one-body operator. For
the points in the
boxes,  the temporal distances from the  boundary should be approximately
equal to give the  propagators  a similar degree of  statistical
variance.

\subsection{Energies of heavy-light mesons ($Q\bar{q}$)}
\label{EQq}
\subsubsection{Two-point correlation functions $C_2$}
\begin{figure}[ht]
\centering
\includegraphics*[width=0.95\textwidth]{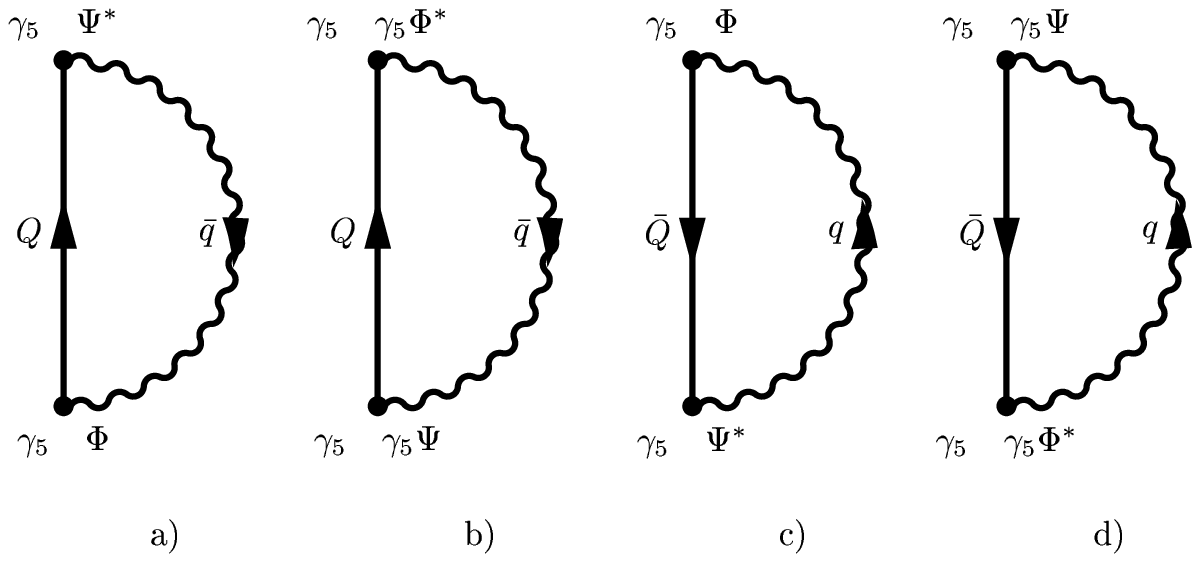}
\caption{The four contributions to the two-point correlation function
$C_2$.}
\label{c2diags}
\end{figure}

The  basic entities for measuring energies are the two-point
correlation functions $C_2$. These are  depicted
in Fig.~\ref{c2diags} and are seen to be
constructed from essentially two  quantities --- the heavy--quark
(static--quark)
propagator $G_Q$ and the light quark propagator $G_q$.
As discussed in detail in \mbox{Ref.~\cite{G+K+P+M},}
when the
heavy quark propagates from site $({\bf x}, \ t)$ to site $({\bf
x}, \ t+T)$,  $G_Q$ can be expressed as
 \begin{equation}
\label{GQ}
G_Q({\bf x}, \ t \ ; \ {\bf x}, \ t+T)=
\frac{1}{2}(1+\gamma_4) U^Q({\bf x},t,T),
 \end{equation}
 where $U^Q({\bf x},t,T)=\prod^{T-1}_{i=0} U_4({\bf x},t+i)$ is the
gauge link product in the time direction.
On the other hand, as  the light-quark
propagates   from  site $i$ to site $j$, it can be schematically
 expressed as one of the two alternatives \mbox{($G_q$ or $G_q'$)} in
Eqs.~\ref{Gauss2} and \ref{Gauss3}. Knowing $G_Q$ and $G_q$,
the general form of a two-point correlation can be constructed from a
heavy quark propagating from site
$({\bf x}, \ t)$ to site $({\bf x'}, \ t+T)$ and a light quark
propagating from site
$({\bf x'}, \ t+T)$ back to site $({\bf x}, \ t)$ as
\begin{align}
\label{C_2}
C_2(T)=&
\Tr \langle \Gamma^{\dagger}\, G_Q({\bf x},t;{\bf x'},t+T)
\, \Gamma \, G_q({\bf x'},t+T;{\bf x},t)\rangle\nonumber\\
=& 2 \langle
{\rm Re }\left[U^Q [\psi^*({\bf x},t+T)\phi({\bf x},t)+
\phi^*({\bf x},t+T)\psi({\bf x},t)]\right] \rangle.
\end{align}
Here $\Gamma$ is the spin structure of the heavy quark -- light quark
vertices at $t$ and $t+T$. In this case $\Gamma=\gamma_5$, since we are
only interested in pseudoscalar mesons such as the $B$-meson.  For
clarity, the Dirac indices have been omitted.
The four contributions to $C_2$ are depicted in Fig.~\ref{c2diags}.
Here the a) term uses the light quark propagator $G_q$ in Eq.~\ref{Gauss2}
and term b) the alternative $G_q'$ in Eq.~\ref{Gauss3} --- the two terms
in Eq.~\ref{C_2}. Terms c) and d) are the corresponding ones for a
heavy antiquark $(\bar{Q})$. It is necessary to include the
$\bar{Q}$-terms to ensure $C_2$ is real. It would be sufficient to use only
a)+c) or b)+d), since both combinations correspond to measuring $C_2$.
However, since these two combinations are independent measurements of the
{\it same} correlation, keeping both improves the statistics on the final
measurement.

The above  has been written down for a single type of gauge field.
The correlations can now be greatly improved by fuzzing as
discussed in Subsec.~\ref{Fuzzingsec}. This makes the two-point
correlation function into the fuzzing matrix $C_{2,ij}$.
Since $i,j$ usually take on 2 or 3 values, this means that \mbox{S-wave}
{\it excited} state energies and properties can now be studied
in addition to those of the ground state.
\subsubsection{Analysis of $C_2$ to extract energies}
\label{analysis}
There are several ways of analysing the correlation functions
$C_2$ in order to extract the quantities of interest {\it i.e.}
energies.  For a review of these methods see
Ref.~\cite{McN+M} --- with more details using the present notation  being
found  in Ref.~\cite{G+K+P+M}.

The actual analysis gives not only the
energies ($m_{\alpha})$ but also the  eigenvectors $({\bf v})$ for the
states of the $Q\bar{q}$-system.
These values of $m_{\alpha}$ and  ${\bf v}^{\alpha}$ are
later fixed when analysing the three-point
correlation  data $C_3$ to give the charge and matter densities
$x^{\alpha \beta}(r)$ in Sec.~\ref{sect.C3}. Each
element $C_{2,ij}(T)$ is  fitted with the form
\begin{equation}
\label{Cij}
C_{2,ij}(T)\approx \tilde{C}_{2,ij}(T)=\sum_{\alpha =1}^{M_2}v_i^{\alpha}
\exp(-m_{\alpha}T)v_j^{\alpha},
\end{equation}
where $M_2$ is the number of eigenvalues to be extracted (usually 3 or
4) and   $m_1$ is the ground state
energy of the heavy-light meson.
The values of $m_{\alpha}$ and $v_{i,j}^{\alpha}$ are then determined
by  minimizing the
difference between the $C_2$ data from the lattice and  the parametric form
$\tilde{C}_2$. The function actually minimized is the usual
\begin{equation}
\label{chim}
\chi^2=\sum_{i,j} \sum_{T_{2,{\rm min}}}^{T_{2,{\rm max}}}
\left[\frac{C_{2,ij}(T)- \tilde{C}_{2,ij}(T)}{\Delta C_{2,ij}(T)}\right]^2,
\end{equation}
where $\Delta C_{2,ij}(T)$ is the statistical error on $C_{2,ij}(T)$
and $T_{2,{\rm min}}, \ T_{2,{\rm max}}$ are the minimum and maximum values  of
$T_2$ used in the fit. The latter depend on the lattice size and the
future use to which the $m_{\alpha}$ and $v_{i,j}^{\alpha}$ are destined.
\begin{figure}[ht]
\centering
\includegraphics*[angle=-90,width=0.75\textwidth]{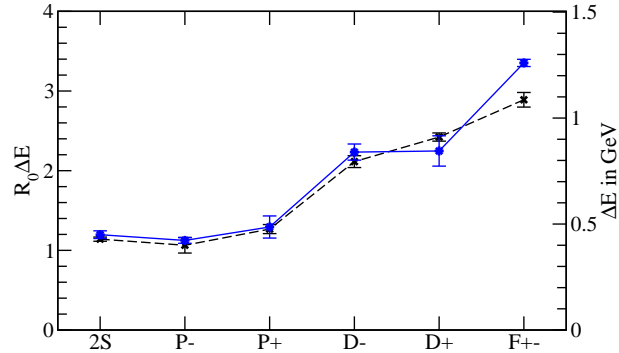}
\caption{ The energies of $Q\bar{q}$ states from
Ref.~\protect\cite{GKMMT03}. The solid line uses dynamical fermions and
the dashed line is a corresponding quenched calculation.
The energies of the L-wave excited states and
the S-wave radial excited state are relative to the ground state (1S)
 --- see text for notation.
These energies are given both in terms of GeV ---  right axis --- and in the more
usual Sommer units of $R_0\approx 0.5$ fm  defined in Subsec.~\ref{sommersec} 
--- left axis.}
\label{ch6-spdf}
\end{figure}
A typical outcome is depicted in Fig.~\ref{ch6-spdf} from
Ref.~\cite{GKMMT03}. This is for a dynamical fermion calculation 
and a corresponding quenched calculation  on a $16^3\times~24$
lattice. In both cases, for numerical reasons, the light quark has a mass
that is approximately that of a strange quark.
Since the heavy quark is
static, the energies can be labelled by $L_{\pm}$, where the coupling of
the light quark spin to the orbital angular momentum gives $j=L\pm 1/2$.
The total angular momentum ($J$) is then obtained by coupling
$j$ to the heavy quark spin giving $J=j\pm 1/2$. However, since the
heavy quark spin interaction can be neglected, the latter two states
are degenerate in energy {\it i.e.} the $P_-$ state will have $J^P=0^+,1^+$
and the $P_+$ state will have $J^P=1^+,2^+$ {\it etc.} 
The $D$-waves show the interesting feature that there appears to be
{\it little or no spin-orbit splitting} between the $D_-$ and $D_+$ states ---
contrary to some expectations~\cite{HJ1,HJ2} that there should be an
inversion of the level ordering (with $L_+$ lighter than $L_-$)  at
larger $L$ or for radial excitations. This has important implications
for phenomenological interpretations of the data. We return to this in 
Subsec.~\ref{Diracfit}.
For $F$-waves, 
only the energy  from a  spin independent mixed operator
$F_{\pm}$  is shown. The latter is expected to approximately correspond to
the usual spin-average of the \mbox{$F_-$ and $F_+$ states.} 
\section{Charge and Matter Distributions of Heavy-Light Mesons
($Q\bar{q}$)}
\label{sect.C3}
In many cases, when phenomenological models are constructed to describe
lattice data, the emphasis is on fitting the energies. However, there
are other observables that can be measured on a lattice. Of potential value
when constructing  models are lattice data for radial distributions of various
quantities such as the charge and matter densities, which can be measured
using three-point correlation functions $C_3$. Here we concentrate on
the radial distributions of the $\bar{q}$ in the $Q\bar{q}$ system, whereas
in Ref.~\cite{alex,alex2} the much more ambitious task of measuring radial
distributions in the $\pi, \ \rho, \ N$ and $\Delta$ is tackled.
The reasons why distributions in few-quark systems have received much
less attention are two-fold. Firstly, unlike energies, these distributions are
not directly observable, but only arise in integrated forms such as
sum rules, form factors, transition rates {\it etc.} Secondly, as will be seen
later, their measurement on a lattice is more difficult and less
accurate than the corresponding energies. In spite of this, it is of
interest to extract lattice estimates of various spatial distributions.
\subsection{Three-point correlation functions $C_3$}
\begin{figure}[t]
\centering
\includegraphics*[width=0.5\textwidth]{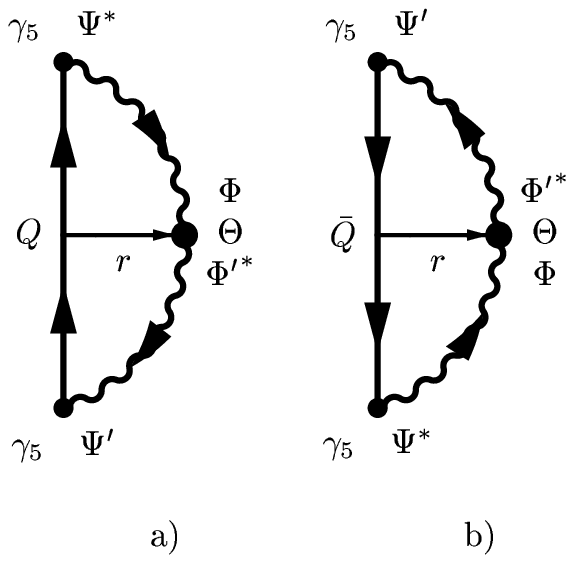}
\caption{The two contributions to the three-point correlation function
$C_3$.}
\label{c3diags}
\end{figure}
When the light-quark field is probed by an operator
$\Theta( {\bf r})$ at $t=0$ as the heavy quark propagates from $t=-t_2$
to $t=t_1$, the result is  the three-point correlation function
depicted in Fig.~\ref{c3diags}. This involves two light-quark
propagators --- one ($G'_q$) going   from $t=-t_2$ to $t=0$ and a second
($G_q$) going   from $t=0$ to $t=t_1$ and has the form
\begin{multline}
\label{AC_3Q}
 C_3(-t_2, \ t_1, \ {\bf r})= \\
\Tr \langle \Gamma^{\dagger} G_Q({\bf
x}, -t_2 ;{\bf x}, t_1) \Gamma G_q({\bf x}, t_1; {\bf x+r},
0) \ \Theta ({\bf r}) G'_q({\bf x+r}, 0; {\bf x}, -t_2) \rangle,
 \end{multline}
which can be expressed in terms of the pseudofermion fields
$\phi({\bf x},t)$ and $\psi({\bf x},t)$  --- similar to that for
$C_2$ in Eq.~\ref{C_2}.
Here the vertex $\Theta=\gamma_4$ for the charge (vector) distribution and 1
for the
matter (scalar) density. Again the $\bar{Q}$-term in Fig.~\ref{c3diags}
ensures that $C_3$ is real, but now there are only two terms compared
with the four for $C_2$ in
Fig.~\ref{c2diags}. This is because the fields connected to the probe
$\Theta$ must be local, since the purpose of the probe is to measure
the charge or matter distribution at a definite point ${\bf r}$.
Therefore, only those light quark propagators that involve the local
basic field $\phi$ at ${\bf r}$ can be used, since the  $\psi$ field
contains contributions from $\phi$ fields at next-to-nearest neighbour
sites and so is  {\it non-local}. This must also be kept in mind  when
fuzzing is introduced to give a matrix $C_{3,ij}(-t_2, \ t_1, \ {\bf r})$.
Here the fuzzing indices $i,j$ refer to the various fuzzing options of the
$\psi$'s at the $Q\bar{q}$ vertices. As with the energies extracted from
$C_2$, the fuzzing  permits the measurement of
excited state distributions.
\subsection{Analysis of $C_3$}
\label{analc3}
The analysis of the three-point correlation functions 
$C_3(\Theta,T=t_1+t_2,{\bf r})$
is performed using a generalisation of the one for $C_2$ in
Eq.~\ref{Cij}, namely, fitting $C_{3,ij}(\Theta ,T,{\bf r})$ with
the parametric form
 \begin{equation}
\label{C3fit}
\tilde{C}_{3,ij}(\Theta ,T,{\bf r})=
\sum_{\alpha =1}^{M_3}\sum_{\beta =1}^{M_3}v_i^{\alpha}
\exp[-m_{\alpha}t_1]x^{\alpha\beta}(r)\exp[-m_{\beta}(T-t_1)]v_j^{\beta}.
 \end{equation}
The $m_{\alpha}$ and ${\bf v}$-vectors are those obtained by
minimizing the $C_2$ in Eq.~\ref{Cij}
and, for each value of~$r$, the $x^{\alpha \beta}(r)$
are varied to ensure a good fit to $C_{3,ij}(\Theta,T,{\bf r})$ by the model
expression  $\tilde{C}_{3,ij}(\Theta ,T,{\bf r})$.

Two forms of  $x^{\alpha \beta}(r)$ have been used:
\begin{enumerate}
\item A non-separable (NS) form, where each $x^{\alpha \beta}(r)$ is treated as
a single entity.
\item  A separable (S) form
$x^{\alpha \beta}(r)=y_{\alpha}(r) y_{\beta}(r)$. It  is seen
from Eq.~\ref{AC_3Q} that this  appears to be a more natural parametrization,
since it contains the product of two light-quark
propagators $G_q$. Also the $y_{\alpha}(r)$, to some extent, resemble a
wave function for the state $\alpha$, since its square yields a distribution.
\end{enumerate}
\begin{figure}[t]
\centering
\includegraphics*[width=0.7\textwidth]{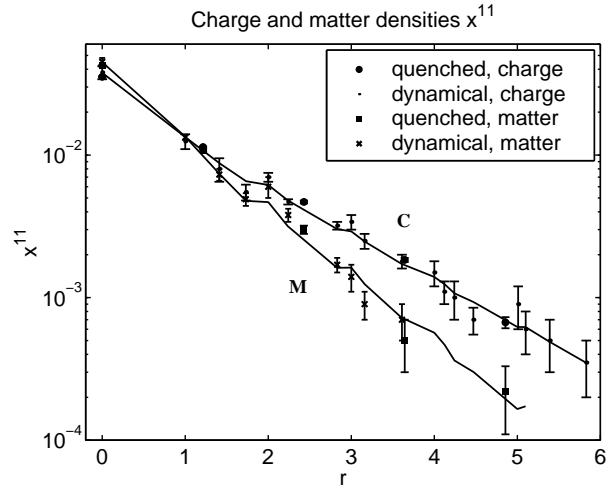}
\caption{The ground state charge (C) and matter (M) densities
[$x^{11}(r)$] as a
function of $r/a$ from Ref.~\protect\cite{G+K+P+M2}.
The lines shows a fit to these densities with a sum of
two lattice exponential functions --- see Subsec.~\protect\ref{formfits}.
The scaled quenched results of
Ref.~\protect\cite{G+K+P+M} are also shown by filled circles and
squares.}
\label{x11vs2LY}
\end{figure}
The outcome is shown in Fig.~\ref{x11vs2LY} for the ground state
and in Fig.~\ref{x12x11} for the excited states.
Several points are of interest in  Fig.~\ref{x11vs2LY}:
\begin{itemize}
\item At small values  of $r$ the two densities are comparable {\it i.e.}
$x^{11}_C \approx x^{11}_M$.
\item As $r$ increases from zero the matter density drops off faster
than the charge density. A similar difference has also been seen
in Ref.~\cite{alex2}, where the authors measure these densities for the
$\pi, \rho, N $ and $\Delta $ on a $16^3\times 32$ lattice with $\beta=6.0$ for
both quenched and unquenched configurations.
\item The densities calculated with the quenched approximation in \\
Ref.~\cite{G+K+P+M} are the same, within error bars, as those
for the full dynamical quark calculation of Ref.~\cite{G+K+P+M2}.
However, as will be discussed in Subsec.~\ref{sectsumr}, the matter
{\it sum rule} does seem to differ.
\item The densities do not have a smooth variation with $r$, but, as
will be shown in Subsec.~\ref{formfits}, many of the kinks can be
understood in terms of latticized forms of standard Yukawa, exponential
or gaussian functions.
\end{itemize}
 \begin{figure}[t]
\centering
\includegraphics*[width=0.99\textwidth]{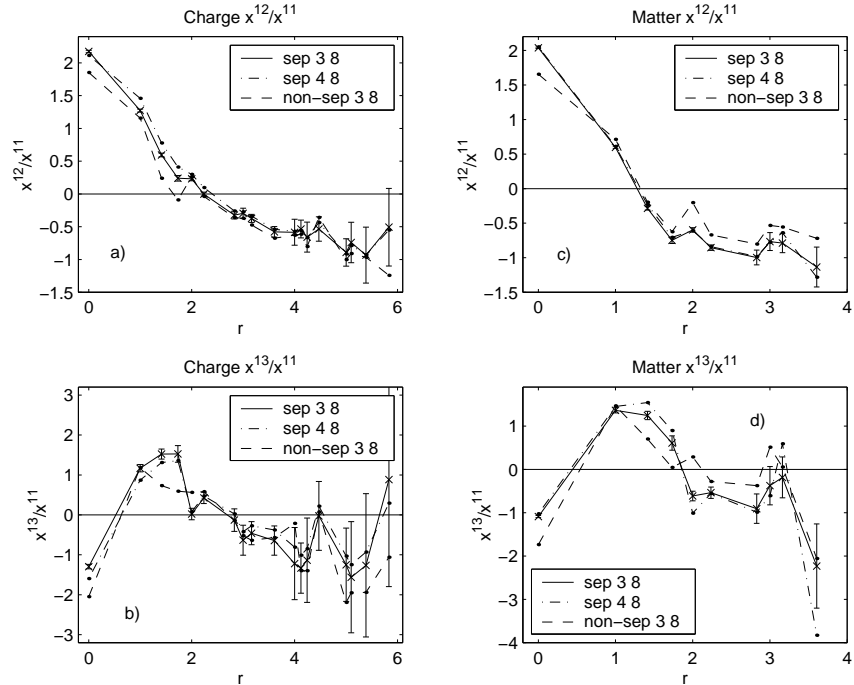}
\caption{a) and b): The ratios $y_2(r)=x^{12}/x^{11}$ and
$y_3(r)=x^{13}/x^{11}$
for the charge distribution. c) and d): These ratios for the matter
distribution --- see Ref.~\protect\cite{G+K+P+M2}.}
\label{x12x11}
\end{figure}
In Fig.~\ref{x12x11} the excited state results are shown for different
types of analysis. The label ``sep" refers to the separable assumption
for $x^{ij}$ and ``non-sep" to the non-separable assumption.
The two numbers associated with each set of data
refer to $T_{2,{\rm min}}$ and $T_{3,{\rm min}}$ in Eq.~\ref{chim} and
the corresponding equation  for Eq.~\ref{C3fit}.
In this figure the main point of interest is the
appearance of nodes.
 The presence of these nodes is very clear and also the number of nodes
is as expected. The first excited state has a single node at about 0.3 fm
in the charge case and about 0.2 fm for the matter.
The second excited state seems to have  two nodes with one being at
about 0.1 fm and a second at about 0.4 fm for the charge and 0.3 fm
for the matter. It will be a challenge for phenomenological models to
explain these data.

When discussing the use of the separable form
$x^{\alpha \beta}(r)=y_{\alpha}(r) y_{\beta}(r)$,
it was stated that  $y_{\alpha}(r)$ can possibly be
interpreted as a wave function for the state $\alpha$. However, there are
other radial distributions  associated with the $Q\bar{q}$ system that
can also be interpreted as wave functions. These are the Bethe--Salpeter
wave functions $w_{\alpha}(r)$ discussed in Ref.~\cite{MP98}.
They are extracted by assuming  the  hadronic operators
$C_{\alpha \alpha}(r_1, \ r_2, \ T)$ to be of the
form $w_{\alpha}(r_1)w_{\alpha}(r_2)\exp(-m_{\alpha}T)$, where the sink
and source operators are of spatial size $r_1$ and  $r_2$.
Qualitatively, the forms of  $y_1(r)$ and $y_2(r)$ plotted in
Fig.~\ref{x12x11} and the corresponding
ones for $w_1(r)$ and $w_2(r)$ are found to be similar.  However, even
though they  do bear some similarities, it
should be added that there are several reasons why these two types of wave
function should {\it not} agree in detail with each other. In particular, the
$[w_{\alpha}(r)]^2$ cannot be identified as a charge or matter
distribution. In addition, they are found by using an explicit fuzzed path
between $Q$ and $\bar{q}$ and so are dependent on the fuzzing
prescription, whereas the $y_{\alpha}(r)$ are defined in a 
path-independent way.

The above considers only S-wave distributions for both ground and
excited states. The extension to other partial waves is now in 
progress~\cite{JonnaQQq}. Preliminary results for the $P_-$ state indicate
that there the distributions are qualitatively of the form expected from a Dirac
equation description {\it i.e.} the
charge and matter distributions are {\it not} zero for $r=0$.

\subsection{Fits to the radial forms}
\label{formfits}
\subsubsection{Fitting data with Yukawa, exponential and gaussian forms}
\label{YEGfits}
In Figs.~\ref{x11vs2LY} and \ref{x12x11} the results are presented as a
series of numbers. However, even though these are the actual lattice data,
they are, in practice,  not very convenient  to use or interpret.
To overcome this it is, therefore, useful to
parametrize the data in some simple form.
Furthermore the results in Fig.~\ref{x11vs2LY} do not follow smooth curves but
exhibit several kinks. If the latter are first ignored, then average fits can
be reasonably well achieved with simple Yukawa (Y), exponential (E) or
gaussian (G) forms giving $\chi^2/n_{{\rm dof}}\approx 1.4$.
The reason for using exponential and Yukawa radial functions is that they
arise naturally as propagators in quantum field theory --- usually
in their momentum space form $(q^2+m^2)^{-1}$, where $m$ can be
interpreted as the mass of a meson being exchanged between the heavy quark
and the point at which the light quark is probed. On the other hand, when
going away
from  quantum field theory and attempting to understand the radial
dependences in terms of wave functions from, for example, the Dirac
equation,   then gaussian forms can arise naturally --- see
Subsec.~\ref{Diracfit}.
However, this fitting can be greatly improved by using lattice versions
(LY, LE, LG) of the above Yukawa, exponential or
gaussian forms, namely
\begin{equation}
\label{LY}
\left[\frac{\exp(- r/r^{{\rm LY}})}{ r}\right]_{{\rm LY}}=
\frac{\pi}{aL^3}\sum_{{\bf q}}
\frac{\cos({\bf r}.{\bf q})}{D+0.25[a/r^{{\rm LY}}]^2},
\end{equation}
\begin{equation}
\label{LE}
\left[\exp(-  r/r^{{\rm LE}})\right]_{{\rm LE}}=
\frac{\pi a}{2r^{{\rm LE}}L^3}\sum_{{\bf q}}
\frac{\cos({\bf r}.{\bf q})}{[D+0.25(a/r^{{\rm LE}})^2]^2},
\end{equation}
\begin{equation}
\label{LG}
\left[\exp[- ( r/r^{{\rm LG}})^2]\right]_{{\rm LG}}=
\left[\frac{r^{{\rm LG}}\sqrt{\pi}}{aL}\right]^3\sum_{{\bf q}}
\cos({\bf r}.{\bf q})\exp[-(r^{{\rm LG}}/a)^2D].
\end{equation}
Here $L$ is the lattice size along one axis
 and $D=\sum^3_{i=1}\sin^2(aq_i /2)$, where
$ aq_i=0, \ \frac{2\pi}{L}, \ \ldots, \  \frac{2\pi( L-1)}{L}.$

\begin{figure}
\includegraphics*[width=0.8\textwidth]{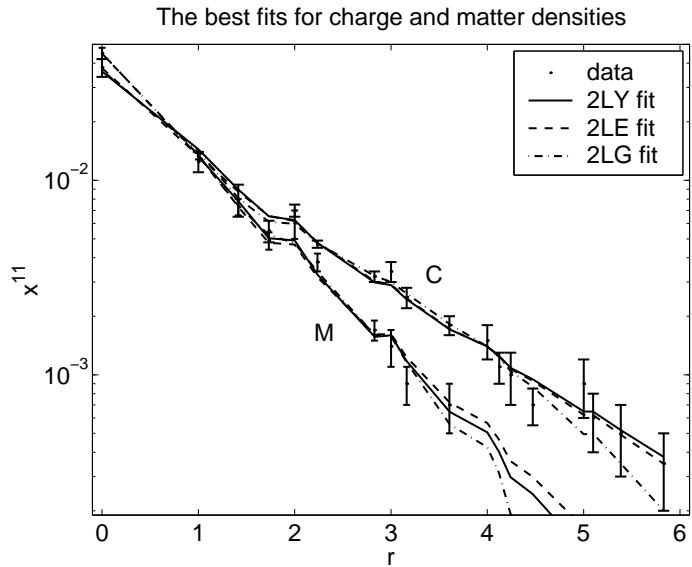}
\caption{Fits to the lattice data in Fig.~\protect\ref{x11vs2LY}
 with  lattice exponential (2LE), Yukawa (2LY) and gaussian (2LG)
forms from Ref.~\protect\cite{G+K+P+M2}.}
\label{LYEf}
\end{figure}

 These are able to give much of the kink structure in
the data --- as is seen in Fig.~\ref{LYEf}, where in each case two 
Yukawas,  exponentials or gaussians are
used to give the fits 2LY, 2LE and 2LG with $\chi^2/n_{{\rm dof}}\approx 1$.
All three of these forms are equally acceptable \cite{G+K+P+M2}.
In the 2LE and 2LY fits it is of interest to express the range
parameters $r^{{\rm LY}}$ and $r^{{\rm LE}}$ in terms of the mass ($1/r^i$)
of the particle producing this range. Both types of fit find that the
longest range corresponds to the exchange of a vector meson of mass
$m^{v}_0 \approx$1 GeV for
the charge density and a scalar meson of mass
$m^{s}_0\approx$1.5 GeV for the matter density.
These masses of the mesons have been extracted in a rather indirect manner.
However, in the literature there have been direct calculations
of the energies of these $q\bar{q}$ states using the
same lattice parameters and lattice size as those employed here. In
Ref.~\cite{Allton} they got $m^{v}_0=1.11(1)$ GeV
and in Ref.~\cite{qmass} $m^{s}_0=1.66(10)$ GeV
--- numbers  consistent with  the above indirect estimates.

It should be added that the expression LY in Eq.~\ref{LY} is often used
in the Coulomb limit $r^{{\rm LY}}\rightarrow \infty$ for discretizing
the $1/r$ term of $V_{Q\bar{Q}}(r)$ in Eq.~\ref{VQQ}. However, in that
limit the condition ${\bf q}\not=0$ is necessary.
\subsubsection{Fitting $Q\bar{q}$ data with the Dirac equation}
\label{Diracfit}
The above fits are purely phenomenological with there being no
connection between the forms fitting the charge and matter distributions.
However, there is an alternative approach in the spirit of an Effective
Potential Theory (EPT), in which the data are fitted by the solutions
of the Dirac equation.
In the present situation using the Dirac equation and not the Schr\"{o}dinger
equation does not seem unreasonable on three counts:
\begin{enumerate}
\item The light-quark propagators are generated by a discretized form of
the Dirac operator --- see Subsec.~\ref{MVR}.
\item The mass (or any effective mass) of the light quark is
$\ll 1$ GeV, so that relativistic effects are expected.
\item  Figs.~\ref{x11vs2LY} and \ref{LYEf} show that the charge and matter
distributions are different --- a feature not
easy to understand in a non-relativistic Schr\"{o}dinger  approach,
where the wave functions have only one component $\phi$, so that both the
charge and matter distributions would be proportional to $|\phi|^2$.
In comparison the Dirac equation has \mbox{two} components. Of course,
one could also resort to the argument familiar in nuclear physics, when
discussing the difference between the charge and matter RMS radii
of nuclei, namely,  that the matter distribution is the basic one with the
charge distribution obtained by folding in  the charge radius of an 
individual nucleon. We return to this in the next subsection.
\end{enumerate}
In the notation that $G$ and $F$ are the large and small components
of the solution to the Dirac equation  then the charge ($C$) and matter
($M$) distributions can be expressed as
\begin{equation*}
x_C^{\alpha\beta}(r)=G_{\alpha}(r)G_{\beta}(r)+F_{\alpha}(r)F_{\beta}(r)
 \ \ {\rm and} \ \ 
x_M^{\alpha\beta}(r)=G_{\alpha}(r)G_{\beta}(r)-F_{\alpha}(r)F_{\beta}(r)
\end{equation*}
respectively. Attempts are now underway \cite{GIJK} to study to what extent the
above distributions can indeed be interpreted by these two relationships.

Since we are now in the realm of EPTs, we need the three ingredients
discussed in Subsec.~\ref{sect.EPT} --- a wave equation (now the
Dirac equation), a potential $V_{Q\bar{q}}$ and an effective quark mass (a
free parameter). The main problem is the form of  $V_{Q\bar{q}}$, since
it is not at all clear whether the ``natural" generalisation of the form
appropriate for static quarks in Eq.~\ref{VQQ} to
\begin{equation}
\label{VQQrel}
V_{Q\bar{q}}(r)=-\frac{e}{r}+b_sr\gamma_4,
\end{equation}
is correct. Here the first term is treated as  a four-vector simply by the
analogy between one-gluon exchange and one-photon exchange. However, if
this term for large $r$ is in fact the so-called L\"{u}scher term
\cite{luterm},
a vibrating string correction of $-\pi/(12r)$ to the leading string term
$b_s r$, then it should be a {\it scalar}.
On the other hand, there are   theoretical reasons that
partially justify the use of the Dirac equation for heavy-light mesons
with the form for the potential in Eq.~\ref{VQQrel}. For example,
in Ref.~\cite{Mur} the authors derive
a Dirac equation for a heavy-light meson by starting from the
QCD Lagrangian and taking into account both perturbative and
nonperturbative effects. The Coulomb-like effect is
treated rigorously and the confining potential heuristically. The
outcome is that the confining potential is a scalar and the Coulomb
part is the fourth component of a 4-vector. 
However, the lack of any significant $D$-wave spin-orbit splitting, as is seen in 
Fig.~\ref{ch6-spdf}, does suggest that the confining potential can not be
purely scalar. This follows from the simple argument that, in a 
heavy(static)-light quark system,  a central
potential of the form $V(r)=a/r+br$ should give rise to a spin-orbit
potential
\begin{equation}
V_{SO}(r)=-\frac{1}{4m^2r}\frac{dV}{dr}=
\frac{1}{4m^2}(\frac{a}{r^3}-\frac{b}{r}).
\end{equation} 
Such a potential would lead
to inversion of the level ordering (with $L_+$ lighter than $L_-$)  at 
larger $L$ or for radial excitations and this is {\it not} seen
in Fig.~\ref{ch6-spdf}. The authors of Ref.~\cite{PGG}, in fact, give
arguments why the interquark confining potential should have the form 
\begin{equation}
\label{extreme}
V(r)=a/r+br{(1+\gamma_4)}/2
\end{equation}
 leading to simply
\begin{equation}
V_{SO}(r)=\frac{1}{4m^2}\frac{a}{r^3},
\end{equation}
which would rapidly vanish at larger $L$. In Ref.~\cite{Paris} the
electromagnetic structure functions of a heavy(static)-light quark
system interacting via the potential in Eq.~\ref{extreme} are calculated.  
This ambiguity between the
vector {\it vs} scalar structure of $V_{Q\bar{q}}$ is an ongoing argument ---
see Ref.~\cite{PGG} for a list of references on this controversy.
Also, as mentioned in Subsec.~\ref{sect.EPT}, form factor effects reduce
the r\^{o}le of any one-gluon exchange and that this loss of
vector-like attraction could be replaced by a very short ranged
instanton--generated {\it scalar} interaction.

An interesting property of the solutions to the Dirac equation  is that, for a
linearly rising potential, both the large $(G)$ and small $(F)$ components
decay asymptotically as {\it gaussians}. This follows from the observation
that the coupled Dirac equations for large $r$ can be written as
\begin{equation}
-m(r)G(r)=-F'(r) \ \ {\rm and} \ \ m(r)F(r)=G'(r),
\label{Diracasym}
 \end{equation}
where $m(r)=m+cr\rightarrow cr$  as $r \rightarrow \infty$
giving asymptotically  the simple harmonic oscillator equation
$G''+(cr)^2G=0$. This was the reason for considering
not only Yukawa and exponential  but also gaussian forms in the
last subsection.

\subsubsection{Fitting $Q\bar{q}$ data with the Schr\"{o}dinger equation}
\label{Schrfit}
In the previous subsection three reasons were given for attempting
to fit the lattice data with the Dirac equation. However, two of these
reasons are far from compelling --- only the fact that the effective
light-quark mass being \mbox{$\ll 1$} GeV seems unavoidable. The comparison
with  the discretized Dirac operator used in the lattice
simulation, as mentioned in  Subsec.~\ref{MVR},
is nothing more than an analogy without theoretical basis.
Secondly, the fact that the charge distribution has a longer range than
that for the matter is reminiscent of the difference between the charge
and matter radii of nuclei. There this is usually expressed as
\begin{equation}
\label{RMSeq}
\langle r^2\rangle _{\rm charge}=\langle r^2\rangle _{\rm matter}+
\langle r^2\rangle _{\rm proton},
\end{equation}
 relating the Mean Square Radius of the nuclear charge distribution
with that for the matter distribution and the proton charge radius.
This suggests that there is only {\it one} basic distribution --- that of the
matter --- with the charge distribution arising from a finite size
correction to the charge of the light-quark. This should not be surprising since
the effective quark mass $(m_{q, \ {\rm effective}})$
is quite different (larger) compared with
 that used in the lattice
calculation, which is  about that of the strange quark.
Here we are now saying that, in addition to a mass renormalisation,  the
light-quark
 develops a {\it charge form factor}.
A direct application of Eq.~\ref{RMSeq}
on the fits to the data in Fig.~\ref{LYEf} results in  $\bar{q}$
RMS charge radii of 0.51(4), 0.49(7) and 0.35(3) fm for the Yukawa,
exponential and gaussian fits respectively. These are sizes consistent
with  $m_{q, \ {\rm effective}} \sim 500$ MeV 
\mbox{({\it i.e.} $1/m_{q, \ {\rm effective}}\approx 0.4$ fm)} and 
are surprisingly
large being 2-4 lattice spacings.
Also $\bar{q}$ charge form factors of this size qualitatively
explain why in Fig.~\ref{x12x11} the node in the $x^{12}/x^{11}$ charge
distribution is
at a larger value of $r$ than that for the matter.

 Since  the problem has been reduced to
only one basic distribution, we are now able to use an Effective
Potential Theory (EPT) based on  a non-relativistic Schr\"{o}dinger equation.
As before, the other two ingredients for an EPT are
$m_{q, \ {\rm effective}}$ (a free parameter)  and
the  potential $V_{Q\bar{q}}$  presumably based on the form
appropriate for static quarks in Eq.~\ref{VQQ}.

\begin{figure}[ht]
\centering
\includegraphics*[height=0.5\textwidth]{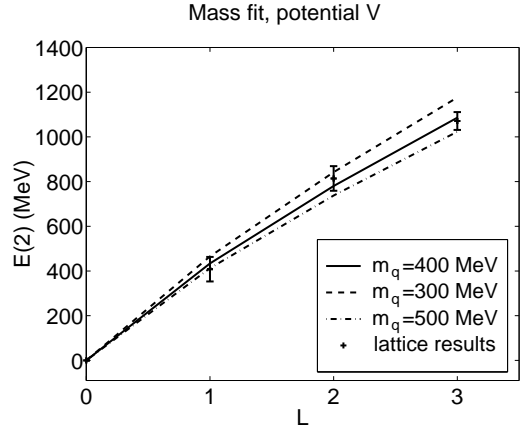}
\caption{Fits to the spin-averaged energies in
Fig.~\protect\ref{ch6-spdf}
for a series of values for $m_q$.}
\label{e2fit}
\end{figure}

In Ref.~\cite{GKP2}, when analysing the $Q^2 \bar{q}^2$ lattice data of
Ref.~\cite{BBpmg}, this strategy was used first  to extract a value of
$m_{q, \ {\rm effective}}$ by fitting the spin-averaged energies of the
$Q \bar{q}$ system --- see Fig.~\ref{e2fit}.
The outcome was $m_{q, \ {\rm effective}}\approx 400$ MeV --- a value
consistent with the above size estimate. We return to this in
Subsec.~\ref{sectBBmod}.
The above approach has also been carried out in  Ref.~\cite{alex}, where
nucleon charge correlations are measured and
then fitted with a non-relativistic Schr\"{o}dinger equation.
First the authors extract two entities $C(r_{\Delta})$ and $C(r_Y)$, which they
interpret as the nucleon wave functions in either the coordinate
$r_{\Delta}=(r_{12}+r_{23}+r_{13})$ --- the so-called $\Delta$-Ansatz
--- or the coordinate $r_{Y}={\rm Min}(r_{1\epsilon}+r_{2\epsilon}+r_{3\epsilon})$,
where ${\bf \epsilon}$ is the junction at which the three flux tubes meet.
They then go on to fit this data with Airy functions
that decay as $\exp(-cr^{3/2})$. These are the wave functions expected
from  a non-relativistic Schr\"{o}dinger equation when using linearly
rising potentials $V(r_{\Delta})\propto r_{\Delta}$ or
$V(r_Y)\propto r_{Y}$.  In this way, fitting the $C(r_{\Delta},r_Y)$ data
seemed to slightly favour the $\Delta$-Ansatz.
However, it should be added that this conclusion is far from being universally
accepted. For example, in Ref.~\cite{Kuzs} the authors say:``In particular the
\mbox{$\Delta$-shape} configuration debated in the literature is shown to be
impossible and the well-known Y-shaped baryon is the only possibility."
This latter result is supported in Refs.~\cite{Ichie,Zeng}.
\subsubsection{Fitting $Q\bar{q}$ data with semirelativistic  equations}
\label{Semrelfit}
The equations considered in the above subsections are two extremes
with the Dirac equation being fully relativistic and the  Schr\"{o}dinger
equation completely non-relativistic. However, if --- as in the ``real"
$B$-meson --- the heavy quark is not
static, then other possibilities arise when
the basic Bethe--Salpeter equation is reduced in a systematic way to a
Lippmann--Schwinger form. As mentioned in Subsec.~\ref{sect.EPT} this
can be carried out in a variety of ways, which
give rise to the Blankenbecler--Sugar, Gross, Kadyshevsky, Thompson,
Erkelenz--Holinde and other equations \cite{Brown+J}.
Unfortunately, these equations are not so easy to treat because of
the unavoidable presence of the typical relativistic factors
$\sqrt{m_i/E_i}$, where $E_i=\sqrt{m_i^2+ {\bf p_i}^2}$ is the energy of a
{\it single} particle of momentum ${\bf p_i}$. This automatically leads
to a non-local interaction in coordinate space, since a local
interaction requires a function of the {\it relative} momentum
${\bf q}={\bf p_a}-{\bf p_b}$. When dealing with one-gluon exchange this
does not present a problem, since there the basic interaction is
$\propto 1/q^2$ and so the equations can be formulated directly in
momentum space. However, as seen in Eq.~\ref{VQQ}, a crucial part of the
interquark interaction is the confining term $b_s r$ --- an interaction
that can not be conveniently  treated directly in momentum space.
In the literature there are
several attempts to fit directly the meagre experimental $B$-meson data
with these forms. For example, in Refs.~\cite{Zeng} and \cite{Timo}
the Gross and Blankenbecler--Sugar equations, respectively, are employed
to interpret the $B$-meson spectrum and a series of available transition
rates.

Some of the complications that arise when dealing with the above equations
can be  partially overcome  by
 using  instantaneous interactions. Shortly after formulating the
Bethe-Salpeter (B-S)
equation in 1951~\cite{Bethe-S}, Salpeter in 1952~\cite{Salpeter}
replaced the interaction $G(q^2={\bf q}^2-q_4^2)$ in the B-S equation by its
three dimensional counterpart $G({\bf q}^2)$. This resulted in an equation
for two particles $a$ and $b$, which could be written in the
centre--of--mass system as
 \begin{equation}
\label{Salequ}
[E-H_a({\bf p})-H_b({\bf p})]\phi({\bf p})=
P\int d^3q \gamma_4^a G({\bf q})\gamma_4^b\phi({\bf p}+{\bf q}),
 \end{equation}
where $H_i({\bf p})=m_i\beta^i+\bm{ p. \alpha^i}$ with $\beta$
and $\bm {\alpha^i}$ being the usual Dirac matrices. The operator
$P=\Lambda^a_+\Lambda^b_+ - \Lambda^a_-\Lambda^b_-$ is a combination of the
projection operators $\Lambda^a_{\pm}=[E_a(p)\pm H_a({\bf p}) ]/E_a(p)$,
where $E_a(p)=\sqrt{m_a^2+{\bf p}^2}$.
It should be added that Eq.~\ref{Salequ}  differs from an earlier one by
Breit in 1929~\cite{Gbreit} by the presence of the $P$ factor. In the
nonrelativistic limit both equations reduce to the same form, but
in general the Breit equation has a more limited applicability than
Eq.~\ref{Salequ} --- as discussed in Ref.~\cite{Salpeter}.

In Refs.~\cite{Hallref},  Eq.~\ref{Salequ} is  further simplified by
first considering only  positive-energy solutions {\it i.e.} omit the
$\Lambda^a_-\Lambda^b_-$ terms in $P$.
This results in the reduced Salpeter equation
\begin{equation}
\label{redsp}
[E-\sqrt{m_a^2+{\bf p}^2}-\sqrt{m_{\smash[t]{b}}^2+{\bf p}^2} \ 
]\phi({\bf p})=
\int d^3q \Lambda^a_+\gamma_4^a G({\bf q})\gamma_4^b \Lambda^b_+
\phi({\bf p}+{\bf q}).
\end{equation}
Then the formalism is further restricted to the positive
energy components to give the semirelativistic spinless--Salpeter equation
\begin{equation}
\label{Hall}
[\sqrt{m_a^2+{\bf p}^2}+\sqrt{m_{\smash[t]{b}}^2+{\bf p}^2}+V(x)]\psi=E\psi,
 \end{equation}
where $V(x)$ is the Fourier transform of the $G({\bf q})$.
Finally the authors of Refs.~\cite{Hallref} go one more step  by
studying the equal mass case
\begin{equation}
\label{Hall2}
[\beta \sqrt{m^2+{\bf p}^2}+V(x)]\psi=E\psi,
 \end{equation}
where, instead of fixing $\beta$ at 2, they show that $\beta >0$ can simulate
the effect of several particles all of
mass $m$. They interpret this equation as ``the generalization of the
nonrelativistic Schr\"{o}dinger Hamiltonian towards relativistic
kinematics" and study algebraically its properties for a variety of
forms for $V(x)$. The reason why these authors go through explicitly
the simplification of Eq.~\ref{Salequ}  is to show that Eq.~\ref{Hall2}
--- the obvious extension of the  Schr\"{o}dinger equation --- can
indeed be derived in a systematic manner and is not just an educated guess.
\subsection{Sum rules}
\label{sectsumr}
In addition to measuring $C_3({\bf r})$ for various values of
${\bf r}$, the correlation where ${\bf r}$ is {\it summed}
over the whole lattice is also obtained. This leads to
the charge sum rule as discussed in Ref.~\cite{G+K+P+M2}.
For the charge distribution, the outcome is that
$\sum_{\bf r}x^{11}({\bf r})=X^{11}$
is $\approx 1.3(1)$, which is  consistent with the earlier
quenched result~\cite{G+K+P+M}.
The fact that $X^{11}$ is not unity, as expected in the continuum
limit, can be qualitatively understood by introducing a
renormalisation factor of $\approx 1/1.3 \approx 0.8$ into the basic
$\gamma_4$
vertex used to measure the charge density. Such a factor of this
magnitude is reasonable as shown in Ref.~\cite{SRule}. It is also reassuring
that the $X^{\alpha \beta}$ with $\alpha \not=\beta$
are, in general, consistent with zero --- as expected in the continuum
limit.

The matter sum rule has a
somewhat wider spread of values with 0.9(1) being a reasonable
compromise --- a number that is about twice the estimate of 0.4(1)
for the quenched calculation of Ref.~\cite{G+K+P+M}. Perhaps this is an
indication that, unlike the corresponding matter radial distributions in
Figure~\ref{x11vs2LY},
 the quenched and unquenched results, due to the effect of
disconnected processes,  can differ even with
the present sea--quark masses of about that of the strange quark.
Certainly differences should appear with very light sea--quarks, since then
the contribution from the disconnected processes, that only enter for the
matter distributions, can become significant.
However, we do not have the data to cross check with
 Refs.~\cite{Foster}, which
advocate the existence of such differences for the matter sum rule.

The fact that the matter sum rule is considerably less than that for
charge can be qualitatively understood
 by  employing data from different hopping
parameters ($\kappa$) and  using the identity
 \begin{equation}
\label{ident}
X^{11}=\frac{d(am_1)}{d\kappa ^{-1}},
 \end{equation}
 where $am_1$ is the ground state energy and $\kappa$ the hopping
parameter --- see Subsec.~\ref{lpfti} and Ref.~\cite{Foster}.  When the $m_1$'s correspond to the
cases where the light quark is of  about one and two strange  quark
masses, Refs.~\cite{Foster} and ~\cite{MP98} give  
$X^{11}\approx$ 0.34(8) and
0.31(6) respectively --- consistent with the present value of 0.4(1).
These  values are also consistent with the following simple
estimate: If the
$Q\bar{q}$-meson mass $(am_1)$ is taken to be simply the sum of the
quark masses {\it i.e.} $am_Q+am_q$ and $\kappa ^{-1}=8+2am_q$, 
then Eq.~\ref{ident} gives
 \begin{equation}
\label{ident2}
X^{11}=\frac{d(am_Q+am_q)}{d(8+2am_q)}=0.5.
 \end{equation}
 Another reason for expecting $X^{11}_M< X^{11}_C$ also follows from 
a potential
approach using the Dirac equation as in Subsec.~\ref{Diracfit}.
This results in $X^{11}_C \sim G_1^2+F_1^2$ and $X^{11}_M \sim G_1^2-F_1^2$.
Here $G_1^2$ and $F_1^2$ are  integrals of the large and small
components of the solution to the Dirac equation.
\section{The $B-B$ System as a $[(Q\bar{q})(Q\bar{q})]$ Configuration.}
\label{sectBB}
In Secs.~\ref{Heavylight} and \ref{sect.C3} the energies and some
radial distributions of a {\it single} heavy-light meson were studied.
In this section the interaction between two such mesons is extracted
using lattice QCD and the outcome is fitted with an extension of the
$f$-model developed earlier for the $[(Q\bar{Q})(Q\bar{Q})]$ system in
 Secs.~\ref{potmodel} and \ref{fmodelext}.
However, it should be added that the study of $[(Q\bar{q})(Q\bar{q})]$
configurations is much less academic than their $[(Q\bar{Q})(Q\bar{Q})]$
counterparts.  Many years ago simple
multi-quark systems have been proposed to exist as bound
states~\cite{vol:76,jaf:76,gut:79}. Also four quarks forming colour
singlets or as bound states of two mesons are candidates for particles
lying
close to meson--antimeson thresholds, such as $a_0(980),\ f_0(980)$
($K\bar{K}$), $f_0(1500),\ f_2(1500)$ ($\omega\omega,\ \rho\rho$),
$f_J(1710)$ ($K^*\bar{K}^*$), $\psi(4040)$ ($D^*\bar{D}^*$),
$\Upsilon(10580)$ ($B^*\bar{B}^*$)~\cite{PDG}.
Systems involving $b$ quarks are particularly interesting
since they should be more easily bound provided the potential
is attractive, since the repulsive kinetic energy of the quarks
is smaller, while the attractive two-body potential remains the same.
In so-called deuson  models~\cite{tor:91} the long-range potential
between two mesons  comes from one-pion exchange, suggesting that
meson--meson systems are significantly less bound than meson--antimeson
systems.
 Other models used for realistic four-quark systems include string-flip
potential  models (see Ref.~\cite{boy:96} for a review), bag
models~\cite{jaf:76}, and a model-independent approach~\cite{lip:86}.
In fact four-quark states  with two heavy quarks have been
predicted to be stable~\cite{zou:86}.  Most models give stability
for systems where the heavy quarks have the  $b$ mass, but long range
forces might push the required heavy-to-light mass  ratio down
sufficiently
 so that $cc\bar{q}\bar{q}$ states would be bound as well.

\subsection{Lattice calculation of the $[(Q\bar{q})(Q\bar{q})]$ system}
\label{sectBBlat}
\begin{figure}[h]
\centering
\includegraphics*[height=0.22\textwidth]{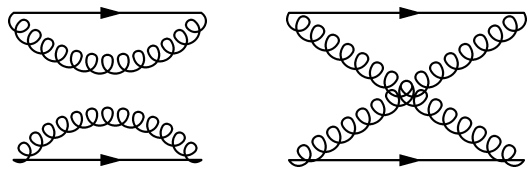}
\caption{The interaction between two $(Q\bar{q})$ states --- the
so-called uncrossed and crossed diagrams. The solid lines represent $Q$'s
and the wavy ones $\bar{q}$'s.
\protect\\NB Compared with earlier
figures such as Figs.~\ref{c2diags} and \ref{c3diags} the Euclidean time
direction is now horizontal.}
\label{FigBB}
\end{figure}

The interaction between two $(Q\bar{q})$ states is depicted in
Fig.~\ref{FigBB} as a sum of two terms --- the  uncrossed and crossed
diagrams. In the latter diagram the $\bar{q}$ hops from one $Q$ to the other.
Exploratory studies of
two-meson systems have been made for the cross  diagram only
 for SU(3) colour~\cite{BBold} and for both  diagrams  in
Refs.~\cite{BBcanada},~\cite{mih:97} for SU(2), SU(3) colour
respectively. This topic is also discussed in Sec.~2.4 of Chapter~2
and Subsec.~4.4.3 of Chapter~4.

In Refs.~\cite{BBpmg,BBpmg2} quenched lattices are used with SU(3)
colour and static  heavy quarks  with light quarks of  approximately the
strange quark mass. Also, when  the two $Q$'s are at the same point
({\it i.e.}  $R=0$) and by
using the SU(3) colour relationship
$3\otimes 3=\bar{3}\oplus6$, then the two $Q$'s behave like a $\bar{Q}$.
This equivalence implies that the $I_q=1,\ S_q=1$ state  will have the
same  light quark structure as the $\Sigma_b$ baryon and the $I_q=0,\
S_q=0$ state  as  that of the $\Lambda_b$ baryon. The other two allowed
 states at $R=0$ correspond  to a static sextet source.
When the  results for $R=0$ are  compared with the known spectrum
of {\it baryons} with one heavy quark ($\Lambda_b$
 and $\Sigma_b$), good agreement is found.  Note that this  link to baryons at
small separation $R$ {\it cannot} be explored using  SU(2) of colour.

Below, the mass of one quark in  each meson is taken to be very heavy --- the
prototype being the $B$-meson.  The static limit is then the leading term
in the heavy quark effective theory for a heavy quark  of zero velocity
and there will be   corrections of higher orders in $1/m_Q$, where $m_Q$
is the heavy quark mass.  In the limit of a static heavy quark, the
heavy quark  spin is uncoupled since the relevant magnetic moment
vanishes which implies  that  the pseudoscalar $B$-meson and the vector
$B^*$-meson  will be degenerate. This is a reasonable  approximation
since they are split by only 46 MeV experimentally, which is less than 1\% of
the mass of the mesons --- see Table~\ref{Blist}. Since it is often convenient
 to treat these  two mesonic states as if they were degenerate,
here they are  described  collectively as the ${\cal B}$-meson.  Because of the
insensitivity to the heavy quark spin,  it is then  appropriate to
classify  these degenerate ${\cal B}$-meson states by the light quark spin.
 The  system of two heavy-light mesons at spatial separation $R$ will
be referred to as the ${\cal BB}$ system. With
both heavy-light mesons static, this ${\cal BB}$ system is then
described  by the two  independent spin states of the two light quarks 
in the two mesons {\it i.e.} $S_q=0$ and 1.
Thus there are four possible states  and it is necessary to classify the
interaction in terms of these spin states.

This situation is very similar to that of the hydrogen molecule in the
Born--Oppenheimer approximation --- with, however,  the additional
possibility that the two ``electrons" can have different properties.
  Another similarity is with  the potential between   quarks which has a
central component and also scalar and tensor spin-dependent
contributions.

Each ${\cal B}$-meson will have a light quark {\it flavour} assignment. For
the ${\cal BB}$  system, it will be appropriate to classify these
states according to their symmetry under interchange of the light quark
flavours. For  identical flavours ({\it e.g.} ${s} s$ or ${u} u$), we have
symmetry  under interchange, whereas  for non-identical flavours ({\it e.g.}
${s}u$  or ${d} u$), we may have either symmetry or antisymmetry. For
two light quarks, it is convenient to classify the states according to
isospin as $I_q=1$ (with $uu$, $ud+du$ and $dd$) or $I_q=0$ (with
$ud-du$).
To ensure overall symmetry of the wave function  under interchange and,
assuming symmetry for spatial interchange, the  flavour, total light
quark spin ($S_q$) and total heavy quark spin ($S_b$) must be  combined
to achieve  this. Thus in the limit of an isotropic spatial wave function
\mbox{{\it i.e.} $L=0$,} there will be
 four different ground state levels of the ${\cal BB}$ system
 labelled by $(I_q,\ S_q)$ in the following discussion.
These are shown in Table~4.2 of Chapter~4.

To check for possible finite size effects,
the numerical analysis was carried out on  quenched lattices of
sizes $12^3 \times 24$ and $16^3 \times 24$, at
$\beta=5.7$, corresponding to $a\approx 0.18$ fm. The bare mass of the light quark
was near that of the strange mass and light quark propagators were
generated using the Maximal Variance Reduction method in
Subsec.~\ref{MVR}. This enabled measurements of   the
strength of the interaction to be made out to
separations of  $R \approx 8$, which  corresponds roughly to 1.4 fm.

As seen in Figs.~2.5, 2.6  of Chapter~2 and Figs.~4.21  of Chapter~4,
attraction between two ${\cal B}$ mesons
 is found   at small values of $R$ for $(I_q,S_q)$=(0,0) and
(1,1) and  at moderate $R$ ($\sim$0.5 fm) for (1,0) and (0,1).
For very  heavy quarks, this will imply binding of the ${\cal BB}$
molecules with these quantum numbers and $L=0$ --- see Sec.~2.4 of
Chapter~2 for more details.

\begin{figure}[ht]
\centering
\includegraphics*[width=0.7\textwidth]{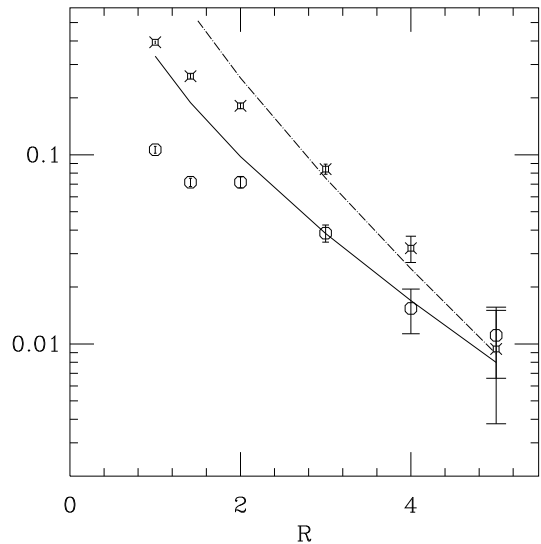}
\caption{The ratio of the crossed-diagram contributions to  the
spin-averaged uncrossed contribution to the ${\cal B B}$ correlation.
 The meson exchange expressions,\ $\exp(-MR)/R$,  are compared
with these results   --- using pion exchange with $M=580$ MeV
(continuous line) and  rho  exchange with $M=890$ MeV (dash-dotted line).
Note that the pion exchange expression is normalised as described in the
text, whereas the rho exchange contribution has an {\it ad hoc} normalisation.
Here $R$ is in units of $a\approx 0.18$ fm~\protect\cite{BBpmg}. }
\label{figpirho}
\end{figure}

It is also possible  to extract from the lattice data quantities
that can be identified as $\pi$- and $\rho$-exchange between the two
${\cal B}$-mesons. This is shown in Fig.~\ref{figpirho}.
For the $\pi$-exchange at large $R$ the  potential is expected to be
of the form
\begin{equation}
\label{OPEPlatt}
V(R)= \vec{\tau}_1.\vec{\tau}_2 \frac{ g^2 M^2}{ 4 \pi f_{\pi}^2} 
\frac{e^{-MR}}{R},
\end{equation}
where $g/f_{\pi}$ is the pion coupling to quarks~\cite{tor:91}. 
From the lattice
studies of $BB^*\pi$ coupling ~\cite{Divitiis} the
value  of $g=0.42(8)$ is predicted and
$f_{\pi}$  is the pion decay constant (132 MeV). Because the  comparison
with the  lattice results is for  light quarks,  the pion mass is
taken to be $Ma=0.53$ {\it i.e.} a ``pion" of mass $\approx 580$ MeV --- a value appropriate for this lattice.
Therefore, in Fig.~\ref{figpirho} the solid curve being compared with
that data containing one-pion-exchange is from Eq.~\ref{OPEPlatt} and is
seen to have the correct features --- giving support, at these large values
of $R$, for the deuson model of Ref.~\cite{tor:91}.
 However, the corresponding
$\rho$-exchange comparison with the
dash-dotted line in Fig.~\ref{figpirho} is less informative, since the
normalisation is {\it ad hoc}.

Even though  the agreement with one-pion-exchange is seen to be very good,
the authors of Ref.~\cite{BBpmg}  are quick to point out that
 their result is ``better than should be expected" and give arguments why
  ``This  implies that we should not take our estimate of the
magnitude of one  pion exchange as more than a rough guide at the
$R$-values we are able  to measure."
It is possible that here we are seeing
an effect reminiscent of the Chiral Bag Model described by Myhrer in
Chapter~4 of Ref.~\cite{WWeise}. In such models there are pion fields
outside some radius $r_b\approx 0.5$ fm and only quarks inside $r_b$.
Calculations of one-pion-exchange potentials (OPEP) between,
say, two nucleons then naturally give the usual form of OPEP for $r>r_b$.
However, for some smaller values of $r$ the interaction  seems to be simply a
continuation of this ``usual form of OPEP" {\it i.e.} there is a precocious
onset of OPEP in a region where there are no pions. Perhaps it is this
that is being seen in Ref.~\cite{BBpmg} and
Fig.~\ref{figpirho}.
It should be added that in some  studies of interacting
clusters, the OPEP is expected to emerge with an {\it exponential}
dependence  and not the Yukawa form in Eq.~\ref{OPEPlatt}. This has been
demonstrated in Ref.~\cite{Beane+S}, where the authors show that this
unconventional form of OPEP is due to the use of the {\it quenched}
approximation. To avoid this problem with the long range part of the 
interaction, in a later paper \cite{BeaneQQ} these authors study the 
$\Lambda_Q\Lambda_Q$ potential, which does not contain  a one-pion
or one-eta contribution.

\subsection{Extension of the $f$-model to the $[(Q\bar{q})(Q\bar{q})]$ system}
\label{sectBBmod}

In Sec.~\ref{potmodel} a model was developed for understanding the
lattice energies of four static quarks
$Q({\bf r_1})Q({\bf r_2})\bar{Q}({\bf r_3})\bar{Q}({\bf r_4})$
in terms of two-quark potentials and is summarised by
Eqs.~\ref{Hamf}, \ref{NVf}.
This f-model, although very simple, contains the same basic
assumptions made in the more elaborate many-body models that incorporate
kinetic energy {\it e.g.} the Resonating Group Method described by  Oka and
Yazaki in Chapter~6 of Ref.~\cite{WWeise}.
It is, therefore, reasonable that this simplified f-model can to some
extent check the validity of these  more elaborate counterparts.
In Ref.~\cite{GKP2} the model in Sec.~\ref{potmodel} was extended as
below  to study the interaction between two $Q\bar{q}$ mesons.
This resulted in a non-relativistic  Schr\"{o}dinger-like equation
\begin{equation}
\label{KVN2}
|{\bf K}'(R)+{\bf V}'(R)-E(4,R){\bf N}'(R)|\psi=0
\end{equation}
--- for more details see Appendix B.

\begin{figure}[h]
\centering
\includegraphics*[width=0.7\textwidth]{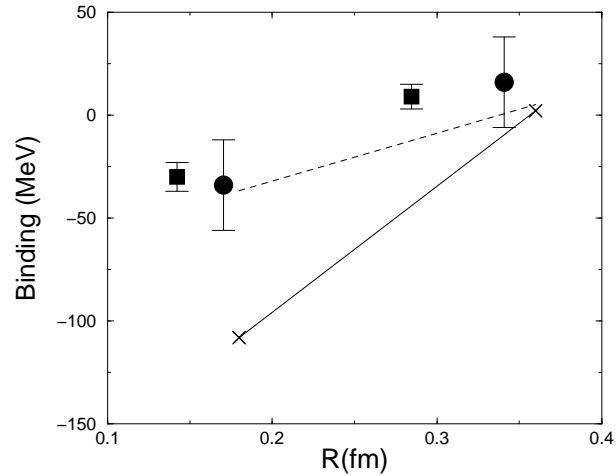}
\caption{Comparison between the spin independent part
of the $Q^2\bar{q}^2$ binding energies
calculated on a lattice  \protect\cite{BBpmg} (solid circles -- quenched
approximation with \mbox{$a=0.170$ fm)}~\protect\cite{UKQCDmp}
(solid squares -- with dynamical fermions and $a=0.142$ fm).
The crosses, with the solid  line to guide the eye,
show the model in the weak coupling limit ($k_f=0$).
The dotted line shows the result with $k_f$ =0.6 and $m_q$=400 MeV.
The dynamical fermion
data was not used in any fit. They are simply included to show that it is
qualitatively consistent with the quenched data but with considerably
smaller error bars.}
\label{Q2q2f1}
\end{figure}

So far this model has only been developed for spin independent
potentials, which means that it should only be compared with the
spin-averaged results of Ref.~\cite{BBpmg} shown in Fig.~4.21 in Chapter~4.
The outcome from Refs.~\cite{GKP2,GKP3} is shown in Fig.~\ref{Q2q2f1}, where
it is seen by the solid line that the use of only two-quark potentials
({\it i.e.} in the weak coupling limit $f=1$ or $k_f=0$) results in a
considerable overbinding at $R=0.18$ fm. The dashed line shows the
effect of using a form factor with $k_f=0.6$.
Admittedly  this is less convincing than the earlier four static quark case,
since the conclusion depends essentially on only the
two data points corresponding to the two $Q$'s being 1 and 2 lattice
spacings apart. However,  the fact that $k_f$ is consistent with
$\bar{k}\approx 0.5(1)$ --- see Fig.~\ref{3kfits} --- the
corresponding parameter  needed  in Subsec.~\ref{paramoff}
for $[(Q\bar{Q})(Q\bar{Q})]$ configurations in squares or near-squares
lends support to the general approach of this model.

In Refs.~\cite{Barnes1} $BB$-scattering has been treated in the weak
coupling limit of $k_f=0$ --- a limit that appears to be ruled out by
the comparison in Fig.~\ref{Q2q2f1}. However, it is possible that this
limit can, to some extent, be salvaged if the model is 
extended by including states with excited glue  --- as in Sec.~\ref{fmodelext}
and Appendix~A.3 for the $[(Q\bar{Q})(Q\bar{Q})]$ system. 

\section{The $B-\bar{B}$ System as a $[(Q\bar{q})(\bar{Q}q)]$ Configuration}
\label{sectBbB}
The $B$-factories discussed in Subsec.~\ref{factories} are not able to study
directly $BB$ reactions. However, the related $B\bar{B}$ system is
accessible as a final state in the decay of the
$\Upsilon(4S, 10580\ {\rm MeV})$, whose main
branching ($\geq 96\%$) is into $B\bar{B}$.

\begin{figure}[b]
\centering
\includegraphics*[height=0.45\textwidth]{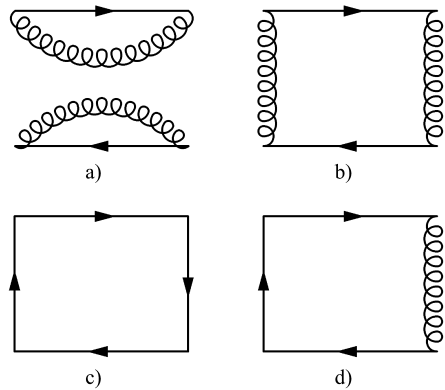}
\caption{The $[(Q\bar{q})(\bar{Q}q)]$ interaction and string breaking.
\protect\\ The a) uncrossed and b) crossed $[(Q\bar{q})(\bar{Q}q)]\rightarrow
[(Q\bar{q})(\bar{Q}q)]$ contributions to the interaction;
 c) The $(Q\bar{Q})\rightarrow (Q\bar{Q})$ Wilson loop and
d) the $[(Q\bar{q})(\bar{Q}q)]\rightarrow
(Q\bar{Q})$ off-diagonal correlation. }
\label{FigBBb}
\end{figure}

At first sight it may be
thought   that the $[(Q\bar{q})(\bar{Q}q)]$  and
$[(Q\bar{q})(Q\bar{q})]$ configurations  have  similar properties.
However, this is not so, since now the $q$ and $\bar{q}$ can annihilate
each other. This means that there is a coupling between
$(Q\bar{Q})$  and $[(Q\bar{q})(\bar{Q}q)]$ states {\it i.e.} the
string (flux) connecting the $Q$ and $\bar{Q}$ in the $(Q\bar{Q})$ state
can break into two mesons $(Q\bar{q})$ and $(\bar{Q}q)$.
This becomes very clear if the mechanism  for the interaction between
two $(Q\bar{q})$ states, shown in the Fig.~\ref{FigBB}, is compared with
the corresponding ones for the present $[(Q\bar{q})(\bar{Q}q)]$ case
on the first row of Fig.~\ref{FigBBb} --- only the uncrossed
contributions look similar.
The diagram b) represents the two step process involving the annihilation and
creation of a $q\bar{q}$ pair 
\begin{center}
$[(Q\bar{q})(\bar{Q}q)]{\bf \longrightarrow}(Q\bar{Q}){\bf
\longrightarrow} [(Q\bar{q})(\bar{Q}q)].$
\end{center}
This breaking of a long flux tube between two static quarks into a
quark--antiquark pair is one of the most fundamental phenomena in QCD.
Because  of  its highly non-perturbative nature it has defied
analytical calculation, while its large scale, {\it e.g.} when compared to
the sizes of composite particles in the theory, has caused difficulties in
standard nonperturbative methods. Thus string breaking has remained a
widely publicized feature of the strong interaction  that has never, apart
from rough models, been reproduced  from the theory. 

\begin{figure}[t]
\centering
\includegraphics*[height=0.55\textwidth]{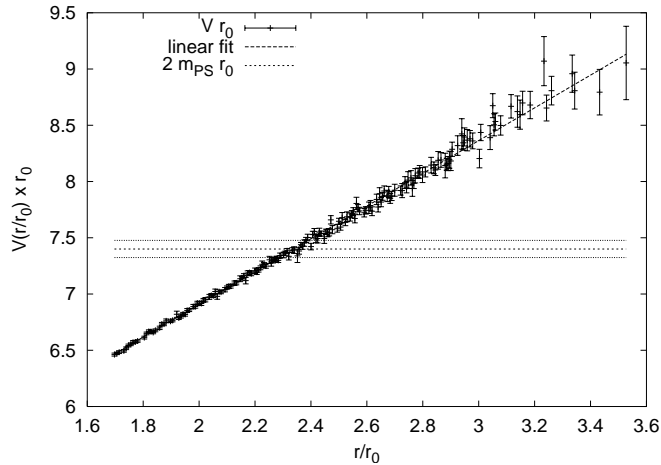}
\caption{The static potential $V(Q\bar{Q})$ obtained from simply the Wilson
loop of Fig.~\protect\ref{2QWiloop}.   N.B. There is no sign of the expected
flattening at $r/r_0\approx 2.4$, where it becomes energetically
favourable to create two mesons\protect\cite{Bolder00}.}
\label{Bolderfig}
\end{figure}

String breaking can occur  in hadronic decays of $Q\bar{Q}$ mesons
and is especially relevant when this meson is lying close to a
meson--antimeson ($Q\bar{q})(\bar{Q}q$) threshold. Its effect should be seen
most directly by measuring the $Q\bar{Q}$ potential, since the onset
of string breaking would change the form of the standard $Q\bar{Q}$
potential in Eq.~\ref{VQQ} so that the confining term --- $br$ for all $r$
--- would become $br$ only for $r<r_c$ and have the constant value
$br_c$ for $r>r_c$.
Unfortunately,  this direct approach of trying to see the flattening in
the static  $Q\bar{Q}$ potential at large separation has only had
limited  success.
An example\cite{Bolder00} of the usual outcome is seen in
Fig.~\ref{Bolderfig}.
There the potential $V(Q\bar{Q})$ continues to rise linearly way past the
value of $r/r_0 \approx 2.4$, where the breaking into two mesons should
occur --- denoted by the horizontal dotted lines. 
There is clearly no sign of the expected flattening.
The failure  of this Wilson loop method seems to be
mainly due to the poor overlap of the operator(s) with the
$[(Q\bar{q})(\bar{Q}q)]$   state~\cite{phi:98b,ste:99}.
 In three-dimensional SU(2) with
staggered fermions an improved action approach has been claimed to
be successful with just Wilson loops~\cite{tro:98}.
However,  in full QCD with  fermions, effective operators for
$[(Q\bar{q})(\bar{Q}q)] $
systems are hard to implement; part of the problem is the
exhausting computational effort  required to get sufficient statistics
for light quark propagators with conventional techniques for fermion
matrix inversion. In Ref.~\cite{ppcmsb} this problem was to a large extent
overcome by
applying the Maximal Variance Reduction method of Subsec.~\ref{MVR}
 using SU(3) on a $16^3\times 24$ lattice with the Wilson
gauge action plus the  Sheikholeslami--Wohlert quark action  and $\beta=5.2$
({\it i.e.} $a\approx 0.14$ fm) with two degenerate flavours of both
valence-- and sea--quarks.

Since  Fig.~\ref{FigBBb} b) is a
two-step process it is natural to consider the process as a coupled
channels  problem between the $[(Q\bar{q})(\bar{Q}q)]$ and $(Q\bar{Q})$
 configurations. This leads to the two new processes in Fig.~\ref{FigBBb}
\mbox{d) and c)} --- the off-diagonal
term $[(Q\bar{q})(\bar{Q}q)]{\bf \longleftrightarrow}(Q\bar{Q})$
and the corresponding diagonal term $(Q\bar{Q}){\bf \longleftrightarrow}(Q\bar{Q})$,
which is nothing more than the Wilson loop discussed in Subsec.~\ref{Wloops}.

\begin{figure}[h]
\centering
\includegraphics*[height=0.6\textwidth]{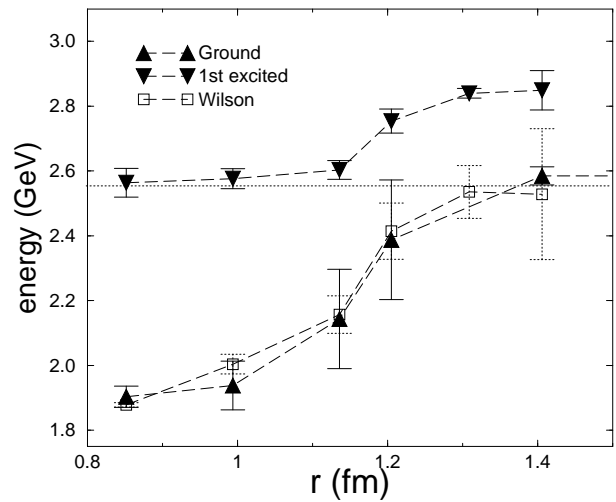}
\caption{Ground and excited state from a variational calculation including
$Q\bar{Q}$ and $Q\bar{q}\bar{Q}q$ operators~\protect\cite{ppcmsb}. 
The highest fuzzed basis
state for both is used here. The ground state of the Wilson loop and
$2m_{Q\bar{q}}$ are also shown. }
\label{FigBBbsb}
\end{figure}

In Fig.~\ref{FigBBbsb} the results from a calculation using just the most
fuzzed basis states for both $(Q\bar{Q})$ and $[(Q\bar{q})(\bar{Q}q)]$  (a
$2\times 2$ matrix) are shown. Unfortunately, the statistics are not
sufficient to give accurate plateaux for the energies. However, the
authors of Ref.~\cite{ppcmsb} are able to extract a quantity they call
the mixing matrix element $x$, which can be interpreted as an indirect 
measure of possible string breaking. They find that $x=48(6)$ MeV.
At $r_c$ one would expect the ground and
excited state energies to be separated by
$2x$. However, in Fig.~\ref{FigBBbsb}  a larger separation is observed,
 which is presumably again due
to insufficient  statistics.
In Ref.~\cite{ppcmsb0}, utilizing the SU(3) colour relationship
$3\otimes \bar{3}=1\oplus8$, the $Q\bar{Q}$ at $R=0$ behave as a
singlet(vacuum), so that the remanent $q\bar{q}$ component with
$(S_{q\bar{q}},I_{q\bar{q}})=(0,1)$ can be compared successfully to a pion with
non-zero momentum.

In the above, the failure of the $Q\bar{Q}$ correlations alone to
give a flattening  of $V(Q\bar{Q})$ was considered a negative
feature. However, in the Conclusion it will be seen that this ``failure"
could be useful for constructing models.

So far no one has attempted to understand this $B\bar{B}$ data with the
extended $f$-model of Subsec.~\ref{sectBBmod} for $BB$ states. Such
an extension would also have to incorporate $q\bar{q}$ creation and
annihilation using some model such as the so-called Quark Pair Creation or
$^3P_0$ model \cite{LeY}. When this  $^3P_0$ model was combined with
the harmonic oscillator flux tube model of Isgur and Paton~\cite{IP}, it
proved successful for describing flux tube breaking and formation ~\cite{IPK}.

\section{Conclusions and the Future}
\label{Confu}
In this chapter there has been an attempt to bring together  two
distinct  lines of research:
\begin{enumerate}
\item {\bf Lattice QCD} was applied to the multiquark systems
$[(Q\bar{Q})(Q\bar{Q})]$, $(Q\bar{q})$,  $[(Q\bar{q})(Q\bar{q})]$ and
$[(Q\bar{q})(\bar{Q}q)]$  in Secs.~\ref{QQQQLC}, \ref{EQq}, \ref{sectBB} and
\ref{sectBbB}.

\vspace{0.3cm}

\item {\bf Effective Potential Theories (EPTs)} ---  based on interquark
potentials with  four-quark form factors included --- were \mbox{developed}
in   Secs.~\ref{potmodel}--\ref{fmodelext} and
\ref{sectBBmod} to interpret phenomenologically  this lattice data.
\end{enumerate}
However, most of  the comparison between these two lines has been
  devoted to constructing EPTs that give some phenomenological
interpretation of the Lattice QCD {\it energies}.
For the $[(Q\bar{Q})(Q\bar{Q})]$ system the latter  concentrated on the
energies of the six geometries
in Fig.~\ref{Sixg} and this amounted in all to about 100 pieces of data.
In spite of four static quarks being a very unphysical system,
the Lattice QCD $\leftrightarrow $ EPT comparison showed that a 4-quark
potential seemed to be needed and that simply using a sum of 2-quark potentials
failed by generating far too much binding for large interquark
distances. On the other hand, for small interquark distances the use of only
2-quark potentials was sufficient --- the so-called weak coupling limit.
However, in Sect.~\ref{fmodelext} it was found that, if the model was
extended to explicitly include gluonic excited states, then much of the
attraction came from these excited states and that the need for a
4-quark potential was greatly reduced. {\it This observation for the
interaction between four
static quarks suggests that the usual
approach of simply using 2-quark potentials in multiquark problems could,
to some extent, be justified provided gluonic excited states are explicitly
included.}   

In the more physical case of
the  $[(Q\bar{q})(Q\bar{q})]$ system
 the EPTs are complicated by the presence of the light quark kinetic
energy and mass. Even so, the Lattice QCD $\leftrightarrow $ EPT
comparison in Fig.~\ref{Q2q2f1} still shows the same effect that
 simply using a sum of 2-quark potentials fails, if the basis states only
contain the gluon field in its ground state.

 This apparent need for a
4-quark potential can be viewed as a form factor --- a familiar and
successful technique in, for example, parametrizations of the
NN-potential~\cite{Bonn}. However, there the form factor is needed
to regularize the potential at {\it small} values of $r$, where the
potential can have $1/r^n$ singularities. It can, therefore, be simply
incorporated as a short-ranged vertex correction modelled from meson fields.
In contrast, the form factor
needed in the present model is introduced to eliminate problems ---
essentially the van der Waals effect --- at {\it large} values of $r$.
This form factor is, therefore, modelling a long-ranged effect which
could be excited gluon states, which become more effective the larger
the system.

\begin{figure}[t]
\centering
\includegraphics*[height=0.65\textwidth]{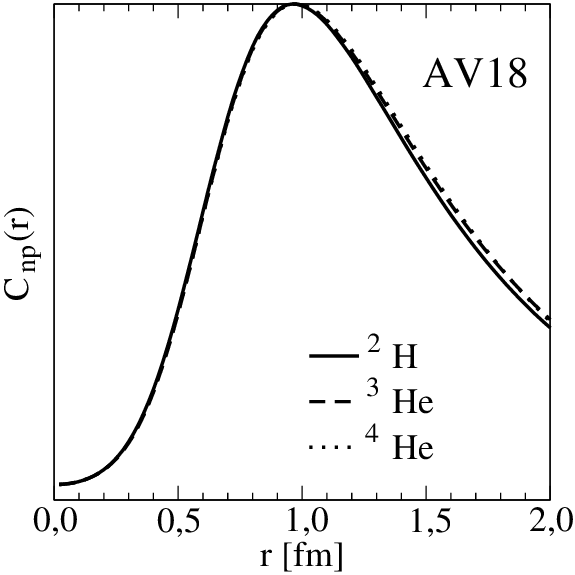}
\caption{Comparison of the neutron-proton radial correlations in the
deuteron, $^3He$ and $^4He$ \protect\cite{glock}. }
\label{npcor}
\end{figure}

Clearly for a better  ``Lattice QCD $\leftrightarrow $ EPT" comparison
more data is needed from a given quark system. This suggests that
comparisons should be made using other quantities in addition to the few
available lattice energies. The most obvious candidates are radial
distributions of the light quark, since --- being a function of
distance --- these
introduce many new pieces of data in a way that is more systematic than
the earlier choice of simply the six ``convenient" geometries of Fig.~\ref{Sixg}.
In addition, there can be several types of radial distribution:

\begin{itemize}
\item The charge (vector) density where the probe in Eq.~\ref{AC_3Q} is
$\Theta=\gamma_4$.
\item The matter (scalar) density where the probe  is $\Theta=1$  .
\item The pseudovector density with the probe $(\gamma
_{\mu}\gamma_5)$.
This is  needed for the $B^*B\pi$ coupling \cite{Divitiis} and was
exploited in Subsec.~\ref{sectBBlat}.
\item In addition to operators that probe the  radial distributions
of the light quark(s), it is also possible to map out the structure of the
 properties of the underlying gluon field. This is achieved  by using
different orientations of the elementary plaquette defined in
Eq.~\ref{elepla}.  Using purely spatial plaquettes the radial
\mbox{distributions} of the various components
of the colour {\it magnetic} field can be extracted
{\it e.g.} $U_{\Box }^{xy}\rightarrow B_z^2$, whereas those plaquettes
with a euclidean time superfix give the spatial distributions of the
various components of the colour {\it electric}  field {\it e.g.}
$U_{\Box }^{xt}\rightarrow E_x^2$.
In Ref.~\cite{4Qfluxmap1,4Qfluxmap2,4Qfluxmap3} these distributions were
calculated for two- and
four-static SU(2) quark systems, where the latter were restricted to the
corners of squares with sides upto 8 lattice spacings
({\it i.e.}\mbox{$\approx 1$ fm).} The lattice was
$20^3\times 32$  with $\beta=2.4$ ($a\approx 0.12$ fm).
There is a wealth of information in these flux profiles and they present
a formidable challenge to models that attempt to describe
them \cite{IP,BBZ}. However, the few  comparisons that have so far been 
made are very encouraging \cite{4Qfluxmap1}.

\end{itemize}
In principle, all of these distributions can be measured for the light quarks
in the 4-quark systems
$[(Q\bar{q})(Q\bar{q})]$ and  $[(Q\bar{q})(\bar{Q}q)]$  discussed in
Secs.~\ref{sectBB} and \ref{sectBbB}. However, this is probably too ambitious at
the present time for  meaningful  Lattice QCD $\leftrightarrow $ EPT
comparisons. First, the radial distributions in the 2-quark
$(Q\bar{q})$ system should be better understood --- the topic of
Sec.~\ref{sect.C3}.
It is possible that the radial distributions of
the $\bar{q}$ in the $[(Q\bar{q})(Q\bar{q})]$ and  $[(Q\bar{q})(\bar{Q}q)]$
systems are related to that in the simpler $(Q\bar{q})$ case, since
for the corresponding situation in light nuclei the neutron-proton radial
correlations in both $^3He$ and $^4He$ are very similar to that in the
deuteron --- see Fig.~\ref{npcor}.
This shows that, although the 3- and 4-nucleon calculations needed to
\mbox{extract} such correlations are very complicated, they result in some
simplicities. Possibly comparisons of correlations within few quark
systems could lead to similar simplifications and so enhance our
understanding of such systems. The work in Ref.~\cite{JonnaQQq} for the
$QQq$ system is a step in this direction.\footnote{Some recent lattice
calculations on doubly-charmed baryons have  concentrated on their 
masses\protect\cite{Flynn}. }

In Sec.~\ref{sect.C3} the charge and matter distributions of the
$\bar{q}$ in the $Q\bar{q}$ system ( $\approx B$-meson) were measured
and attempts made to fit these distributions with simple functions in
Subsec.~\ref{YEGfits} and also by using the Dirac equation in
Subsec.~\ref{Diracfit}. The latter can be viewed in two ways:
\begin{enumerate}
\item As simply an alternative form of parametrization of the two
distributions with no physical interpretation of the parameters needed.
\item  As the construction of an EPT, in which the parameters do have a
physical interpretation --- albeit phenomenological --- on which
extensions to multiquark systems can be based.
\end{enumerate}
This leads us to the main problem of how to set up an EPT in order to
understand multiquark systems. In the past the key word has been ``mimicing",
in which models based on potentials have been mimicing the successful
models of multi-nucleon systems. Unfortunately, since such models aim
directly at the {\it experimental} data, they may be attempting to
describe some mechanisms that are outside the scope of the model.
This I have called the
Nuclear-Physics-Inspired-Approach (NPIA) and it would correspond to Option 1 in
the comparison\\
\begin{center} NPIA $\stackrel{1}{\longleftrightarrow}$
Experimental Data   {\bf {\it versus}}
Lattice data $\stackrel{2}{\longleftrightarrow}$ QCDIA
\end{center}
However, I believe it is more reasonable to try to create
QCD-Inspired-Approaches (QCDIA) as in Option 2. Here the models attempt
to mimic directly details of the lattice data that have been obtained
under ``controlled conditions" and so, hopefully, do not contain
undesirable processes not included in the model.
In this way the data have
a better chance of deciding the {\it form} of the model. In contrast,
with the NPIA the form of the model tends to be decided beforehand with
the experimental data only leading to a tuning of the parameters.
An example of this was demonstrated by Fig.~\ref{Bolderfig}.
There it was seen that using only $(Q\bar{Q})$ correlations did not lead
to a flattening of the interquark potential $V(Q\bar{Q})$ --- even
though such a flattening
should arise with the onset of $(Q\bar{q})$ and $(\bar{Q}q)$ mesons being
created. Therefore, for an understanding of $V(Q\bar{Q})$ this defect was
considered to be a {\it negative }feature, since this implementation of 
LQCD clearly did not
agree with experiment. However, for model building this could be
envisaged as a {\it positive} feature, since LQCD is now generating {\it exact}
data from the QCD Lagrangian that can be interpreted by a model that is simply
 a linearly rising potential between a single $Q$ and a single $\bar{Q}$. This
intermediate stage model could, hopefully, be more easily  
 extended to the real life situation, when the more
complicated correlations in Fig.~\ref{FigBBb} are treated by LQCD.                
This intermediate stage  could be considered as yet another example
of the ``unphysical worlds" discussed in Sec.~\ref{bridges} in the context
of models with different numbers of colours, spatial dimensions, quark
masses {\it etc.}. This recalls my earlier work in 
$^{18}O$\cite{O184p2h,collrev}. There the
``real" $^{18}O$ needed to be described by ``4-particle 2-hole" states
in addition to the usual ``2-particle 0-hole" states. This is  analogous 
to the matrix of correlations in Fig.~\ref{FigBBb}. It would
have been very useful if there had been an ``experimental phase" of $^{18}O$
that
only needed \mbox{2-particle 0-hole} states for its description. In that case
the model for ``real" $^{18}O$ would have had less freedom.

So far the QCDIA has only been attempted for the $[(Q\bar{q})(Q\bar{q})]$
configuration in Sec.~\ref{sectBB}. There the $(Q\bar{q})$ lattice data
in Fig.~\ref{e2fit} was first fitted with a Schr\"{o}dinger equation to
give an effective light quark mass
$m_{q, \ {\rm effective}}\approx 400$ MeV. Using a variational method, this
Schr\"{o}dinger approach was then easily extended to the $Q^2\bar{q}^2$
system using the same interquark potential and $m_{q, \ {\rm
effective}}$ determined earlier from the  $Q\bar{q}$ data.
In principle, this could be extended to any system that can be described
in terms of interacting quark clusters.

 The problem with the above QCDIA is that it usually results in an
$m_{q, \ {\rm effective}}\ll 1$ GeV suggesting the need for a more
relativistic approach. This can be attempted at different levels:
\begin{enumerate}
\item At the one extreme we can use directly the Dirac equation to
describe, say, the  $Q\bar{q}$ system as in Subsec.~\ref{Diracfit}.
This had some success but we are then confronted with the problem of
extending the comparison to multi-quark systems --- a second step (2) that is
not \mbox{directly} possible for systems containing more than one light quark {\it
i.e.}
\begin{center} Lattice data for  $Q\bar{q}$ $\stackrel{1}{\longrightarrow}$
1-quark Dirac Eq. $\stackrel{2 \ ??}{\longrightarrow}$ Multi-quark case
\end{center}
\item Use some Semirelativistic Schr\"{o}dinger-like Equation (SRSE)
as described in Subsec.~\ref{Semrelfit}.
This is most easily formulated in momentum space with the kinetic
energy $E_{NR}=m_{q,e}+p^2/2m_{q,e}$
being replaced by  
\mbox{$E_R=\sqrt{p^2+m_{q,e}^2}$,} where
$m_{q,e}$ is an effective quark mass chosen to fit some piece(s) of
experimental or lattice data.
However, care must be taken to treat the potential terms to the same
semirelativistic degree by inserting appropriate factors
of  $m_{q,e}/E_p$ as in the 
Blankenbecler--Sugar equation \cite{Brown+J}. The second step (2) to multiquark systems
is then probably possible {\it i.e.}
\begin{center} Lattice data for  $Q\bar{q}$ $\stackrel{1}{\longrightarrow}$
SRSE  $\stackrel{2}{\longrightarrow}$ Multi-quark case
\end{center}
\item Thirdly, a compromise model emerges. First the  $Q\bar{q}$ lattice
data is fitted by a Dirac equation. This equation then generates other
observables that are interpreted in terms of a SRSE, which can be
extended to multi-quark systems. This is similar to the philosophy of
Bhaduri and Brack~\cite{BhBr}, who show how the Schr\"{o}dinger equation
with a quark effective mass of $m_{q, \ {\rm effective}}\approx 500$ MeV
is able to explain some of the results --- energies
and magnetic moments --- of a Dirac equation for a {\it zero} mass quark
\mbox{{\it i.e.} they} make the comparison
\begin{center} Lattice data for  $Q\bar{q}$ $\stackrel{1}{\longrightarrow}$
Dirac Eq. $\stackrel{2}{\longrightarrow}$ SRSE  $\stackrel{3}{\longrightarrow}$ Multi-quark
\end{center}
\end{enumerate}
So what are the ``Bridges from Lattice QCD to Nuclear Physics" as
advertized in the title of this chapter? So far, the only clear examples
involve mainly static (or heavy) quarks ($Q$) as in the extraction
of the string energy and the lattice spacing from $V_{QQ}$ with a
non-relativistic Schr\"{o}dinger equation (Sec.~\ref{simplebr}) and
 the energies of the various $Q^4$ geometries in
\mbox{Secs.~\ref{SectQ4}--\ref{fmodelext}.}
The introduction of light quarks as in the
$(Q\bar{q})$,  $[(Q\bar{q})(Q\bar{q})]$ and
$[(Q\bar{q})(\bar{Q}q)]$ systems  in Secs.~\ref{Heavylight} -- \ref{sectBbB}
leads to major complications and the only partial success is the
Schr\"{o}dinger description of the $(Q\bar{q}) +[(Q\bar{q})(Q\bar{q})]$
{\it energies} in Sec.~\ref{sectBB}. The real test of whether any 
``Bridge" exists is only now becoming possible with the advent of the
lattice data for radial distributions, since such distributions
are also at the centre of much of Nuclear Physics. Of course, the final
outcome could be that there is no useful bridge for treating multiquark
systems in the way we treat multinucleon systems. This would mean that
Lattice QCD would only ever be able to directly address problems involving
a few quarks --- perhaps upto the six quarks needed for the nucleon-nucleon
interaction --- but not show a general way for how to deal with multiquark systems.
This would result in the two  worlds of QCD and Nuclear Physics having
little direct connection with each other.

However, before such a pessimistic view  is adopted, we should remember that 
Nuclear Physics earlier had another two-world structure that lasted for many
years. Until
the 1960's there were essentially two models for the nucleus --- the
collective liquid-drop-like model that rarely mentioned the
nucleon-nucleon potential and, in contrast, the shell model based on this potential.
But with the advent of Brueckner theory it was shown how effective
interactions in many-nucleon systems could be constructed from the basic
nucleon-nucleon potential. This was followed up by the generation of
collective effects as interacting particle-hole states \cite{Brownbook}.
In this way a bridge was made between the basic  nucleon-nucleon
potential and collectivity --- but it took many years.

\section*{Acknowledgements}
\addcontentsline{toc}{section}{Acknowledgements}

The author wishes to thank Mika Jahma, Jonna Koponen, Timo L\"{a}hde
and Slawek Wycech for their invaluable help in preparing and commenting on this
manuscript.
The author, in his r\^{o}le as Editor, also wishes to thank
the other contributors to this Volume. Without their cooperation and
enthusiasm the volume would never have appeared.
\section*{Appendix A: Extensions of the $f$-Model from $2\times 2$ to
$6\times 6$}
\addcontentsline{toc}{section}{Appendix~A: Extensions of the $f$-Model from
$2\times 2$ to $6\times 6$}
\setcounter{equation}{0}
\renewcommand{\theequation}{A.\arabic{equation}}
\markright{Appendix A}
\subsection*{A.1: The $3\times 3$ extension of the unmodified 
two-body \mbox{approach} of
Subsec.~\protect\ref{Unmodified}}
\addcontentsline{toc}{subsection}{A.1: The $3\times 3$ extension of the unmodified two-body approach of
Subsec.~\protect\ref{Unmodified}}
Because the colour group  considered so far is SU(2),
there is no distinction between the group properties of quarks ($Q$) and
antiquarks ($\bar{Q}$).  Four such quarks can then be partitioned as
pairs
in {\it three} different ways
\begin{equation}
\label{ABC}
A=(Q_1Q_3)(Q_2Q_4), \ \ B=(Q_1Q_4)(Q_2Q_3) \ \ {\rm and} \ \
C=(Q_1Q_2)(Q_3Q_4),
\end{equation}
where each $(Q_iQ_j)$ is a colour singlet. However, these three basis
states are not orthogonal to each other. Also, remembering the fact that
the quarks are indeed fermions gives, in the weak coupling limit, the
condition in the Appendix of Ref.~\cite{GMS}
\begin{equation}
\label{A+B+C}
|A\rangle+|B\rangle+|C\rangle=0.
\end{equation}
Since $\langle A|A \rangle=\langle B|B \rangle=\langle C|C\rangle=1$,
we get --- in this limit --- the equalities
\\
$\langle A|B\rangle=\langle B|C\rangle=\langle A|C\rangle =-1/2$.

In SU(3) the partitioning problem is different since
in that case the three  partitions are
\begin{equation}
\label{ABC3}
A=(Q_1\bar{Q}_3)(Q_2\bar{Q}_4), \ \ B=(Q_1\bar{Q}_4)(Q_2\bar{Q}_3) \ \
{\rm and} \ \
C=[(Q_1Q_2)^d(\bar{Q}_3\bar{Q}_4)^d],
\end{equation}
where state $C$ is expressed in terms of either  colour
antitriplet ($d=\bar{3}$) or sextet ($d=6$) states and so cannot
appear asymptotically as two clusters.

If all three basis states in SU(2) are included,  then the matrix 
 below is singular for the obvious reason that
$|A\rangle+|B\rangle+|C\rangle=0$  {\it i.e.} 
\begin{equation}
\label{NVs}
\rm{det} \ {\bf N}=\rm{det}\left(\begin{array}{ccc}
1&-1/2&-1/2\\
-1/2&1&-1/2 \\
-1/2&-1/2&1\end{array}\right)=\rm{det}\left(\begin{array}{ccc}
1&1/2&1/2\\
1/2&1&-1/2 \\
1/2&-1/2&1\end{array}\right)=0.
\end{equation}
Earlier  this was interpreted to mean that it was
unnecessary
to include all three states and so the symmetry was broken by keeping
the two states with the lowest energy, let us say , $A$ and $B$.
A similar effect also occurred in the lattice simulations. There it was
found that the energy of the lowest state was always the same in both a
$2\times 2$ and $3\times 3$ description, providing $A$ or $B$ had the
lowest energy of the three possible partitions. In addition the energy of the second state was, in most cases,
more or less the same -- the largest
difference occurring with the tetrahedral geometry.

\subsection*{A.2: The $3\times 3$ extension of the $f$-model of
Subsec.~\protect\ref{mqi}}
\label{3*3}
\addcontentsline{toc}{subsection}{A.2: The $3\times 3$ extension of the
$f$-model of Subsec.~\protect\ref{mqi}}
 The $f$-model of Subsec.~\protect\ref{mqi}, by incorporating
multiquark interactions, had the good feature that,
when fitting the data ($E_0,E_1$) for a given square, only a single $\bar{f}$
was necessary to get a reasonable fit to both energies --- see Fig.~\ref{ffits}.
 Of course, $\bar{f}$ was dependent on the size
of the square, but a reasonable  parametrization was
\begin{equation}
\label{fIa}
f(Ia)=\exp(-b_s k_f S)\ \ {\rm (Version \ \ Ia)},
\end{equation}
where $S$ is the area of the square and $k_f\approx 0.5$. Earlier
in Eq.~\ref{f1} a slightly different notation was used.
The original hope was that, with $k_f$ determined from the squares and
nearby rectangles, the model
would automatically also fit other geometries with $S$ being the
``appropriate" area contained by the four quarks. When the four quarks lie
in a plane, the definition of $S$ is clear. However, in non-planar
cases the situation is more complicated. One possibility is to simply
take $S$ to be  the average of the sum of the four triangular areas defined by
the positions of the four quarks {\it i.e.} the faces of the tetrahedon. For
example, in the notation of Eq.~\ref{ABC}, the appropriate area $S(AB)$
for $f$ is
\begin{equation}
\label{Def:S}
S(AB)= 0.5[S(431)+S(432)+S(123)+S(124)],
\end{equation}
where $S(ijk)$ is the area of the triangle with corners at $i,j$ and
$k$. For planar geometries this simply reduces to the expected area, but
for non-planar cases this is only an approximation to $S(AB)$ -- a more
correct area being one that is not necessarily a combination of planar
areas  but of curved surfaces with
minimum areas. These possibilities are discussed in Ref.~\cite{fur+m}. It
would be feasible to incorporate this refinement here, since only a few
$(\approx 50)$ such areas are needed for the geometries in Fig.~\ref{Sixg}.
However, for a general situation, in
which the positions of the quarks are integrated over, it would become
impractical to use the exact value of $S(AB)$, since the expression for the
minimum area itself involves a
double integration. In contrast, the area used in Eq.~\ref{Def:S} is an
algebraic expression
and is, therefore, more  readily evaluated for any
geometry. The above model will be referred to as Version Ia.

This model Ia has only the one free parameter $k_f$ in $f({\rm Ia})$ of
Eq.~\ref{fIa}.
 Another possibility
with additional parameters $f_0,\ k_P$ is
\begin{equation}
\label{fIb}
f({\rm Ib})=f_0\exp(-b_s k_f S+\sqrt{b_s} k_P P) \ \ {\rm (Version \ \ Ib)},
\end{equation}
where $P$ is the perimeter bounding $S$. This form has been used in
Refs.~\cite{fur+m}.
However, as shown in Ref.~\cite{PP1},  this reduces in the continuum
limit to the same as Version Ia --- the differences at $\beta =2.4$
being mainly due to lattice artefacts.

Unfortunately, in the $2\times 2$ version both of these models have the feature
that, for regular tetrahedra, they are unable to reproduce a degenerate
ground state with a
{\it non-zero} energy, since  the two eigenvalues are
\begin{equation}
\label{E01}
E_0=-\frac{f/2}{1+f/2}[V_{CC}-V_{AA}] \ \ {\rm and} \ \
E_1=\frac{f/2}{1-f/2}[V_{CC}-V_{AA}],
\end{equation}
where, in the notation of Fig.~\ref{Sixg}, $V_{AA}=v_{13}+v_{24}$
and $V_{CC}=v_{14}+v_{23}$
and so for regular tetrahedra $V_{CC}=V_{AA}$, giving $E_0=E_1=0$.

\markright{Appendix A}
Prior to the work on tetrahedra the geometries considered had, at
most, two of the three possible partitions being degenerate in
energy ({\it e.g.}\ for squares) --- see Subsec.~\ref{mqi}.
In these cases, it is found that the lattice
energies $E_0$ and $E_1$ are essentially the same for the
three-basis-state calculation
$(A+B+C)$ and those two-basis-state calculations ($A+B,\ A+C$ and
effectively $B+C$) which involve the basis state with the {\it lowest}
unperturbed energy. This is one of the reasons why the $2\times 2$ version
of the $f$-model in Eq.~\ref{NVf} was quite successful for a
qualitative understanding of these cases. However, for tetrahedra
and the neighbouring geometries calculated in Subsec.~\ref{ComplQQ}, it now
seems plausible to extend the $f$-model to the corresponding $3\times 3$
version of
\begin{equation}
\label{Hamf2}
\left[{\bf V}(f)-\lambda_i(f) {\bf N}(f)\right]\Psi_i=0,
\end{equation}
in which
\begin{equation}
\label{NVf2}
{\bf N}(f)=\left(\begin{array}{ccc}
1&f/2&f'/2\\
f/2&1&-f''/2 \\
f'/2&-f''/2&1\end{array}\right)
\ \ {\rm and}
\end{equation}
\begin{equation}
\label{NVf3}
{\bf V}(f)=\left(\begin{array}{ccc}
v_{13}+v_{24} & fV_{AB}& f'V_{AC}\\
fV_{BA}&v_{14}+v_{23}&-f''V_{BC} \\
f'V_{CA}&-f''V_{CB}&v_{12}+v_{34}\end{array}\right),
\end{equation}
 where the negative sign in the BC matrix elements is of the same
origin as the one in Eqs.~\ref{NVT} and \ref{NVs}.

This apparently leads to the need for two more factors $f',f''$ defined
by
\begin{equation}
\label{Nf'}
\langle A|C\rangle =-f'/2 \ \ {\rm and} \ \ \langle B|C\rangle =-f''/2.
\end{equation}
However, with the parametrizations of $f$ as in Eqs.~\ref{fIa} or \ref{fIb}
and the
definition of $S$ as in Eq. \ref{Def:S}, it is seen that
$f'=f''=f$, since $S$ is simply proportional to the area of the faces of
the tetrahedron defined by the four quark positions and is {\em
independent} of the
state combination used. Therefore, the $3\times 3$ model has
{\it for all 4-quark geometries} a form where the
$N$ and $V$ matrices are
\begin{equation}
\label{N33}
{\bf N}(f)=\left(\begin{array}{ccc}
1&f/2&f/2\\
f/2&1&-f/2\\
f/2&-f/2&1\end{array}\right)
\ \ {\rm and} \ \
\end{equation}
\begin{equation}
{\bf V}(f)=\left(\begin{array}{ccc}
v_{AA} & fV_{AB}& fV_{AC}\\
fV_{BA}&v_{BB} & -fV_{BC}\\
fV_{CA}&-fV_{CB}&v_{CC} \end{array}\right).
\end{equation}
This extension from $2\times 2$ to $3\times 3$ has the good
 feature that all three basis states are now
 treated on an equal footing. This is convenient when considering
some general four-quark geometry, since it is then not necessary to
choose some favoured $2\times 2$ basis, which could well change as the
geometry develops from one form to another. In
the weak coupling limit ({\it i.e.}\ $f\rightarrow 1$) the $3\times 3$
matrix in Eq.~\ref{N33} becomes the singular matrix in Eq.~\ref{NVs}.
 However, in this
limit, each of the $2\times 2$ matrices corresponding to the three possible
partitions A+B, A+C and B+C gives the same results. Away from weak
coupling the $3\times 3$ matrix is no longer singular, but now the three
possible $2\times 2$ partitions do not necessarily give the same results.
In addition to this general problem as $f\rightarrow 1$, there are
also the following more specific unpleasant features:
\begin{enumerate}
\item For regular tetrahedra all eigenvalues are {\it zero} as
in the $2\times 2$ models. The reason for this is clear.
 There is only one energy
scale in the model, since all the $v_{ij}$ are the same. Therefore,
there
can not be any excitations.
\item For a linear geometry, since the ``appropriate" area as defined by
Eq.~\ref{Def:S} vanishes, we get $f=1$ {\it i.e.} we are back to the weak
coupling limit and a singular matrix.
\item For squares the model gives $E_1=-E_0$, whereas the predictions of
the $2\times 2$ version in Eqs.~\ref{E01} seem to be nearer the lattice data.
\item
The differences between using the various combinations of the
three partitions are often considerably larger than in the corresponding lattice
calculation.
\end{enumerate}
The most glaring problem is the fact that the $3\times 3$ model for the
tetrahedron gives three degenerate states with zero energy, because
there is only one scale in the model.
It is, therefore,
necessary to introduce a second
energy scale. However, any improvements in the model
have very limited choices, since there are only two different matrix
elements involved --- the diagonal ones all equal to $-E$ and the
off-diagonal ones all equal to $\pm0.5fE$. Therefore, the most
general modifications are to change the diagonal matrix elements to
$d_1-E$ and the off-diagonal ones to \mbox{$\pm 0.5f(d_2-E)$.} This results
in the eigenvalues
\begin{equation}
\label{eigvs2}
E_0=E_1=\frac{d_1+0.5fd_2}{1+0.5f} \ \ {\rm and} \ \
E_2=\frac{d_1-fd_2}{1-f}.
\end{equation}
At first sight it may appear that there is sufficient information to
now extract the new parameters $d_{1,2}$, since $f$ can be estimated
using the parameters (assumed to be universal) from other geometries
--- thus leaving two equations for $E_{0,1}$ and $E_2$ and the two
unknowns $d_{1,2}$. However, as said before, this is too much to
demand from the $f$-model, since in the lattice calculation the
third basis state in the complete A+B+C basis generally plays a
minor r\^{o}le in determining the values of $E_{0,1}$ and, therefore,
the third eigenvalue is presumably dominated by an excitation of the
gluon field. Furthermore, it is of interest to see that a similar
feature now arises with $E_2$ in Eq.~\ref{eigvs2}, since this
third state is removed in the weak coupling limit {\it i.e.}\
$E_2\rightarrow \infty$ as $f\rightarrow 1$. However, in its present
form, the $f$-model is only  expressed in terms of
the {\it lowest} energy gluon configurations, since the gluon field is not
explicitly in its formulation, but only appears {\bf implicitly} in
the form of the two-quark potentials and the $f$-factors. 
But already at this stage we see from the behaviour of $E_2$ that the
effect of {\it excited} gluon states seem to be  playing a r\^{o}le ---
the topic of the next subsection when the model is further extended to a
$6\times 6$ version.  
In view of
this, no quantitative attempt should be made to identify the second
excited state emerging from the lattice calculation as $E_2$ in this
$3\times 3$ version of the $f$-model.

In Ref.\ \cite{Lang} it was shown that the {\it two}-state model of
Eqs.~\ref{NVf} with the overlap factor $f=1$ agreed with
perturbation theory up to fourth order in the quark--gluon coupling
[{\it i.e.}\ to $O(\alpha^2)$] and gave $E_{0,1}$=0 for tetrahedra.
Therefore, the non-zero lattice results for small tetrahedra must be
of $O(\alpha^3)$ at least. Another aspect of this special situation
for tetrahedra is also seen --- when extracting or interpreting the
value of $E_1$ --- by the need for the third basis state both in the
lattice calculation  and in the
$f$-model, since in comparison with Eq.~\ref{eigvs2} the
two-basis-state version gives
\begin{equation}
\label{eigvs3}
E_0=\frac{d_1+0.5fd_2}{1+0.5f} \ \ {\rm and} \ \
E_1=\frac{d_1-0.5fd_2}{1-0.5f}
\end{equation}
{\it i.e.}\ both the two- and three- basis-state models have the same
ground state, but the former does not show the $E_0=E_1$ degeneracy.

The expressions in Eq.~\ref{eigvs2} are not particularly useful
unless there is a model for the parameters $d_{1,2}$. However, since
it is not the purpose at this stage to make a comprehensive study of
models covering all the 4-quark geometries considered in earlier
works \cite{GMS,GMP,GMP93,GMPS}, only a few general remarks will be
made here for the tetrahedron geometry. Models for the $d_{1,2}$
need extensions of the potential in Eq.~\ref{vcol}, so that for
the tetrahedron  {\it two} energy scales arise. Here several
ways of achieving this goal are suggested:

\vspace{0.3cm}

\noindent{\bf The effect of an isoscalar two-quark potential}.\\
As discussed in Ref.\ \cite{GMP93}, an isoscalar potential $w_{ij}$
can be introduced into $V_{ij}$ --- still ensuring $V_{ij}=v_{ij}$ for a
colour singlet two-quark system --- by extending the form in
Eq.~\ref{VQQ} to
\begin{equation}
\label{vcol1}
V_{ij}=-\frac{1}{3} {\bf \tau}_i\cdot{\bf \tau}_j
\left(v_{ij}-w_{ij}\right) +w_{ij} .
\end{equation}
In this case, $d_1=d_2=4w$, since all of the $w_{ij}$ are now equal
to $w$ and results in $E_0=E_1=E_2=4w$. Therefore, here  $w$ takes on
values that range from --0.0035 to
--0.0070 {\it i.e.}\ they have values much smaller than the corresponding
$v_{ij}=v$ in Eq.~\ref{vcol}. A similar feature was found in
Ref. \cite{GMP93}, when the form in Eq.~\ref{vcol1} was introduced
to improve the model fit for squares and rectangles. However, as
shown in Ref.\ \cite{Lang}, in perturbation theory all terms of
$O(\alpha^2)$ are included in the two state model of Eq.~\ref{NVf}
with $f=1$. Therefore, in the weak coupling limit
$w_{ij}$ must be of $O(\alpha^3)$ at least.

\vspace{0.3cm}

\noindent{\bf The effect of a three- or four-body potential.}\\
The $f$ factor is itself a four-body operator. However, it is
conceivable that additional multiquark effects arise. Some
perturbative possibilities are discussed in Ref.\ \cite{Lang}. There
it is shown that all three-quark terms arising from three gluon
vertices always vanish, but that the four-gluon vertex can
contribute to 2-,~3- and 4-quark terms at $O(\alpha^3)$. However, in
the tetrahedral case ($r=d$), cancellations result in this
particular 4-quark term also vanishing.

\vspace{0.3cm}

\noindent{\bf The effect of non-interacting three gluon exchange processes.}\\
These are also discussed qualitatively in Ref.\ \cite{Lang} and
contribute at $O(\alpha^3)$ to 2-,~3- and 4-quark potentials.

\vspace{0.3cm}
\markright{Appendix A}
\noindent{\bf The effect of two quark potentials where the gluon field is
excited.}\\
The first excited state [$V^*(r)$] of the two-quark potential
$V(r)$ is approximately given by $V^*(r)\approx V(r)+\pi /r$
--- see for example Refs.~\cite{CM1,CM2}. Therefore, if a fourth
state, based on such an excited state, is introduced into the
model, it will give attraction in the ground
state, since it is higher in energy than the three degenerate basis
states so far considered. Furthermore, as the size of the tetrahedron 
increases this
fourth state will approach the other three states, so that the
attraction felt in the ground state will increase --- a trend seen in
the tetrahedron results for $E_{0,1}$ in Fig.~\ref{f:energs}. This
possibility will be discussed more in the $6\times 6$ extension below.

\vspace{0.5cm}


The above isoscalar potential option now offers a reason for
\mbox{$E_0=E_1\not =0$.} But, unfortunately, $E_2$ is still equal to
$E_{0,1}$ since $d_1=d_2$. However, there is no reason to expect any
three or four body forces to also be purely isoscalars. In this
case, their contributions to $d_1$ and $d_2$ could be different and
through the presence of the $(1-f)$ factor in Eq.~\ref{eigvs2} any
estimates of $E_2$ could be very model dependent.
\subsection*{A.3: The $6\times 6$ extension of the $f$-model of
Subsec.~\protect\ref{mqi}}
\label{6*6}
\addcontentsline{toc}{subsection}{A.3: The $6\times 6$ extension of the
$f$-model of  Subsec.~\protect\ref{mqi}}
The above models both have trouble in describing regular tetrahedra.
In Refs.~\cite{GPplb426,GPprc57} an attempt is made to overcome this problem.
An interesting feature of the regular tetrahedron data is
that the lowest state becomes {\it more} bound as the tetrahedron
increases in size with the magnitude of $E_0$ increasing from
  --0.0202(8) to --0.028(3) as the $d^3$ cube
containing the tetrahedron increases from $d=2$ to $d=4$. This is
opposite
to what happens with squares, where the magnitude of $E_0$ decreases
from
$-0.0572(4)$ to $-0.047(3)$ as $d$ increases from 2 to 5. This indicates
that there could be
coupling to some higher state(s) that becomes more effective as the size
increases and suggests that these higher states contain gluon
excitation with respect to the $A,B,C$ configurations. Therefore, the
$3\times 3$ model in the previous subsection is further extended to a 
$6\times 6$ model 
 by adding three more states $A^*,B^*,C^*$,
where in analogy with Eq. \ref{ABC}
\begin{equation}
\label{A^*}
A^*=(Q_1 Q_3)_{E_u} (Q_2 Q_4)_{E_u} \ \  etc..
\end{equation}
Here $(Q_1 Q_3 )_{E_u}$ denotes a state where the gluon field is excited
to the lowest state with the symmetry of the $E_u$ representation of the
lattice symmetry group $D_{4h}$. Because it is an odd parity excitation,
$A^*,B^*,C^*$ must contain two such states in order to have the
same parity as $A,B,C$.
The excitation energy of an $E_u$ state over its ground state ($A_{1g}$)
counterpart is $\approx \pi/r$ for two quarks a distance $r$ apart.
As $r$ increases this excitation energy decreases making
the effect of the $A^*,B^*,C^*$ states more important, leading to the
effect mentioned above. Here we have
assumed that these states arise from a combination of excited states
with $E_u$. However, it is possible that they involve other excitations,
{\it e.g.}
\begin{equation}
A^*=(Q_1 Q_3)_{A'_{1g}} (Q_2 Q_4) \ \  etc.,
\end{equation}
where the $A'_{1g}$ state is a gluonic excitation with the {\it same} quantum
numbers as the ground state ($A_{1g}$). For this case  the following
formalism would be essentially the same. Another possibility, which is
not considered here, is that
the relevant excitations are flux configurations where all four quarks,
instead of two, are involved in  forming a colour singlet.
In the strong coupling approximation such states would reduce to
two-body singlets due to Casimir scaling of the string tensions, namely,
the string tension for a higher representation would be more than double
the value of the fundamental string tension, thus preventing junctions
of two strings in the fundamental and one in the higher representation.
This would happen both in SU(2) and SU(3), the only exception being the
unexcited $C$ state in SU(3), which would involve an antitriplet string
---  see Eq.~\ref{ABC3}.

For the regular tetrahedral case, in addition to 
$f=f'=f''$, there are now several  new matrix elements
that need to be discussed for
${\bf N(f)}$ and ${\bf V(f)}$ (Eqs.~\ref{Hamf2}--\ref{NVf3}):

\vspace{0.3cm}

a) With  the inclusion of the $A^*,B^*,C^*$ states and
the antisymmetry condition $|A^*\rangle+|B^*\rangle+|C^*\rangle=0$
analogous to Eq.~\ref{A+B+C},
there are now two more gluon overlap functions $f^{a,c}$ defined as
\[\langle A^*|B^*\rangle =\langle A^*|C^*\rangle=\langle B^*|C^*\rangle =-f^c/2  \ \ {\rm and} \ \ \]
\begin{equation}
\langle A^*|B\rangle =\langle A^*|C\rangle =.. \ \  etc. \ \ .. =-f^a/2.
\end{equation}
Here it is assumed that $f^{a,c}$ are both dependent on $S$ as defined
in
Eq.~\ref{Def:S}. Since $f^c$  involves only the excited states, it is
reasonable to expect it has a form similar to $f$ in Eq.~\ref{fIa} {\it i.e.}
\begin{equation}
\label{fc}
f^c=\exp(-b_sk_cS).
\end{equation}

\vspace{0.3cm}

b) By orthogonality $\langle A|A^*\rangle=\langle B|B^*\rangle=
\langle C|C^*\rangle=0$

\vspace{0.3cm}

c) In the weak coupling limit, from the
$|A^*\rangle+|B^*\rangle+|C^*\rangle=0$  condition,
we expect $\langle A|B^*\rangle=\langle B|C^*\rangle=.....=0$ at small
distances.
To take this into account  $f^a$ is  parametrized as
\begin{equation}
\label{fa}
f^a=(f^a_1+b_s f^a_2 S)\exp(-b_sk_aS).
\end{equation}
If all three parameters $f^a_1,f^a_2,k_a$ are
varied, it is found that $f^a_1$ is always
consistent with zero --- as
expected from the above condition that \mbox{$\langle
A|B^*\rangle=....=0$.} Therefore, usually
 $f^a_1$ is fixed at zero.

\vspace{0.3cm}
d) For the potential matrix ${\bf V(f)}$ the diagonal matrix elements,
after the lowest energy $V_{DD}$ amongst the
basis states is removed, are
\[\langle A^*|V-V_{DD}|A^*\rangle =v^*_{13}+v^*_{24}-V_{DD}, \ \  etc.,\]
where  $V_{DD}={\rm min}[V_{AA}=v_{13}+v_{24}, V_{BB}=v_{14}+v_{23},
V_{CC}=v_{12}+v_{34}]$
 and $v^*_{ij}\approx v_{ij}+\pi/r$ is the potential of the excited 
$E_u$ state, which is a quantity also
measured on the lattice along with the four-quark energies.

In the special case of regular tetrahedra
$V_{DD}=V_{AA}=V_{BB}=V_{CC}$ and  ${\bf V}$ reduces to the form
\begin{equation}
\label{KH2}
{\bf V}= \left[ \begin{array}{ccc|ccc}
V_{AA}&-fV_{AA}/2&-fV_{AA}/2&0&-f^aV_a/2&-f^aV_a/2 \\
-fV_{AA}/2&V_{AA}&-fV_{AA}/2&-f^aV_a/2 &0      &-f^aV_a/2     \\
-fV_{AA}/2&-fV_{AA}/2&V_{AA}&-f^aV_a/2 & -f^aV_a/2     &0          \\
 \hline
0 &  -f^aV_a/2  & -f^aV_a/2   &V_b&-f^cV_c/2&-f^cV_c/2          \\
-f^aV_a/2 &0    & -f^aV_a/2   &-f^cV_c/2 &V_b     &-f^cV_c/2
\\
-f^aV_a/2 & -f^aV_a/2   &0    &-f^cV_c/2 &   -f^cV_c/2   &V_b       \\
\end{array} \right],
\end{equation}
where $V_a \ , \ V_b \ , \ V_c$ can be expressed in terms of
$V_{AA}$ and $v^*(ij)$ plus some fine tuning
parameters.
As with all geometries
\begin{equation}
\label{KH1}
{\bf N}= \left[ \begin{array}{ccc|ccc}
1&-f/2&-f/2&0&-f^a/2&-f^a/2 \\
-f/2 &1   &-f/2&-f^a/2 &0     &-f^a/2    \\
 -f/2 & -f/2   &1   &-f^a/2 & -f^a/2     &0        \\     \hline
 0 &-f^a/2    & -f^a/2   &1&-f^c/2&-f^c/2         \\
 -f^a/2 &0    &  -f^a/2  &-f^c/2 &1     &-f^c/2            \\
 -f^a/2 & -f^a/2   &0    &-f^c/2 & -f^c/2     &1         \\
\end{array} \right].
\end{equation}
The full $6\times 6$ matrix $[{\bf V}-(E+V_{AA}){\bf N}]$ now factorizes
into
three $2\times 2$ matrices,
two of which are identical -- giving the observed degeneracy.
These two matrices have  determinants of the form
\begin{equation}
\left| \begin{array}{cc}
-E(1+f/2)&-f^a(E-V_a)/2    \\
 -f^a(E-V_a)/2  & \ \ -E(1+f^c/2)+V_b+f^cV_c/2 \\
\end{array} \right|=0,
\end{equation}
whereas the third $2\times 2$ matrix has the determinant
\begin{equation}
 \left| \begin{array}{cc}
-E(1-f)&f^a(E-V_a)    \\
 f^a(E-V_a)  & \ \ -E(1-f^c)+V_b-f^cV_c \\
\end{array} \right|=0.
\end{equation}
In this case the problem reduces to solving two quadratic
equations \mbox{for $E$.} However, away from the regular tetrahedron the
complete
6$\times$6 matrix needs to be treated directly.

By fitting simultaneously the energies $E_0$ and $E_1$ from the lattice
results for the geometries in Fig.~\ref{Sixg} an interquark potential
model can be constructed that is able to
explain, on the average, these  energies.
The full model utilized 6 basis states $A,B,C,A^*,B^*,C^*$ and
in its most general  \mbox{form}  has eight parameters.
However, in practice, only 3 of these ($k_f,k_a,f^a_2$) need
be considered as completely free when fitting the data.

The parameters that are of most interest are those connected
with the ranges of the various interactions, namely, $k_f$ and $k_a$.
Here  ``range" is defined  as $r_{f,a,c}=\sqrt{1/b_sk_{f,a,c}}$.
In the $2\times 2$ version, where $k_{a}$ is effectively infinite, we get
$k_f({\rm Ia})$=0.57(1) {\it i.e.} $r_f({\rm Ia})=5.0$ in lattice units
of 0.12 fm.
However,  when the excited states $A^*,B^*,C^*$ are
introduced, the
interaction between the basic states $A,B,C$ decreases by raising
$k_f$ to  1.51 giving $r_f =3.1$.
But at the same time this loss of binding by the
direct interaction between $A,B,C$ is compensated by their coupling to
the
$A^*,B^*,C^*$ states. This coupling in Eq.~\ref{fa} is found to have about the
{\it same}  range $r_a=5.1$ as $r_f({\rm Ia})$ in Eq.~\ref{fIa}, 
whereas {\it the direct
interaction between the $A^*,B^*,C^*$ states seems, in all fits, to be
satisfied with simply a two-quark description without any four-quark
correction ({\it i.e.} $k_c$=0) in Eq.~\ref{fc}}. 
The observation that $r_f({\rm Ia})\approx r_a$
suggests that the energy density has a range dictated by the longest
range available --- namely $r_a$. Therefore, when the
$A^*,B^*,C^*$ states are not explicitly present, as in Model Ia, the
only available range $r_f({\rm Ia})$ has to simulate the r\^{o}le of $r_a$.
In the binding energies the contributions from the
$A^*,B^*,C^*$ states rapidly dominate over those from the
$A,B,C$ states. For example, with squares of side $R$, the $A,B,C$
states contribute only 85, 40, 10\% to the binding energy for $R$=2,4,6
respectively. Of course, at the largest distances ($\approx 0.7$ fm)
the quenched approximation is expected to break down and the r\^{o}le
of quark-pair creation to become  important.
\markright{Appendix B}
\section*{Appendix B: Extension of the $f$-Model to  
$[(Q\bar{q})(Q\bar{q})]$ \mbox{Systems}}
\addcontentsline{toc}{section}{Appendix B: Extension of the $f$-Model to
 $[(Q\bar{q})(Q\bar{q})]$ Systems}
\setcounter{equation}{0}
\renewcommand{\theequation}{B.\arabic{equation}}
When only two of the four quarks are static the corresponding
matrices for
$Q({\bf r_1})Q({\bf r_2})\bar{q}({\bf r_3})\bar{q}({\bf r_4})$ can be
expressed in a similar form but where the matrix elements are now
{\it integrals} over the positions of the two light antiquarks. In the
notation of Fig.~\ref{configs} we
consider basis state $A$ to be the one realised as two separate
heavy-light mesons --- $[Q_1\bar{q}_3]$ and $[Q_2\bar{q}_4]$ --- when
the
distance ${\bf R}={\bf r_1}-{\bf r_2}$ between
the two heavy quarks becomes large. In this state the convenient
coordinates are then
${\bf s_1}={\bf r_3}-{\bf r_1}$ and ${\bf s_2}={\bf r_4}-{\bf r_2}$,
whereas
for the other partition $B$ the convenient
coordinates are
${\bf t_1}={\bf r_3}-{\bf r_2}={\bf s_1}+{\bf R}$ and ${\bf t_2}
={\bf r_4}-{\bf r_1}={\bf s_2}-{\bf R}.$

To describe this system in terms of an Effective Potential Theory 
the three ingredients
quoted in Subsec.~\ref{sect.EPT} are needed --- see Ref.~\cite{GKP2}
for more details of the specific example now to be described:

\vspace{0.3cm}

\noindent{\bf A wave equation.}\\
For simplicity, the system is considered  to be
non-relativistic resulting in a Schr\"{o}dinger-like equation
\begin{equation}
\label{KVN}
|{\bf K}'(R)+{\bf V}'(R)-E(4,R){\bf N}'(R)|\psi=0
\end{equation}
{\it i.e.} a Resonating Group equation as discussed by Oka and Yazaki
in Chapter~6 of Ref.~\cite{WWeise} and also Ref.~\cite{SMBSY}.
This is a generalisation of Eq.~\ref{Hamf} to non-static quarks and  can be
solved using a variational wave function taken to have the form~\cite{zou:86}
\begin{equation}
\label{wf4}
\psi({\bf r_i},f)=f^{1/2}({\bf r_1},{\bf r_2},{\bf r_3},{\bf r_4})
\sum^{N_4}_{i=1}\exp(-\tilde{\bf{X}}{\bf M}_i\bf{X}),
\end{equation}
where $\tilde{\bf{X}}=(\bf{s_1}, \  \bf{s_2}, \   \bf{R})$ and each
matrix
$\bf{M}_i$ has the form
\begin{equation}
\label{M}
  {\bf M}_i = \frac{1}{2}\left( \begin{array}{ccc}
a_i&b_i&c_i\\
b_i&d_i&e_i\\
c_i&e_i&g_i\\
\end{array} \right).
\end{equation}
Since the present problem considers the masses of the
light quarks to be equal, it is sufficient to use a simplified form of
$\bf{M}_i$ with $b_i=0, \ d_i=a_i$ and $e_i=c_i$. This is not necessary,
but it is expected to be the dominant term in such a symmetric case.
Already for $N_4=2$, this wave function is indeed
adequate for giving sufficiently accurate  four-quark binding energies.
Even this choice involves five free parameters $(a_1, \ c_1, \ a_2, \
c_2, \ g_2)$ in the variation -- with
$g_1$ being fixed at unity to set the overall normalisation.
In what follows the positions of the
light quarks are integrated over leaving matrix elements that are
functions of {\bf R}. In order to achieve this in any practical way it
is necessary to have a form for
$f({\bf r_1},{\bf r_2},{\bf r_3},{\bf r_4})$ that has a simple spatial
dependence. Here the very symmetrical form in Eq.~\ref{f2},
defined by a {\it single} parameter $k_f$, is used 
{\it i.e.}
\begin{equation} 
\label{f4Q} 
f=\exp\big[-k_f b_s \sum\limits_{i<j}r^2_{ij}\big] 
\end{equation} 
(N.B. Here $k_f = k/6$, where $k$ was defined in Eq.~\ref{f2}.)
It should be emphasised that this form of $f$ is purely for numerical
simplicity leading to analytical expressions for all matrix elements.
As in the static case $k_f$ is a free parameter, which should be adjusted
to fit the four-quark lattice energies.

The wave function in Eq.~\ref{wf4} is used for both states $A$ and $B$.
This is an approximation that appears to work well for the
$Q^2\bar{q}^2$ system, since $A$ and $B$ are similar in structure for
the $R$ values of interest here.
 
\vspace{0.3cm}

\noindent{\bf An interquark potential.}\\
This enters in three different contexts:\\
1) As $v(13), \  v(24), \ v(14), \  v(23)$ in  the $V_{Q\bar{q}}$
potential. This is taken to be of the
standard form in Eq.~\ref{VQQ}, namely
\begin{equation}
\label{v2fit}
aV(2,r)=-\frac{0.309(38)}{ r/a}+0.1649(36)r/a+0.629(25),
\end{equation}
which gives a string energy  of (445 MeV)$^{2}$ for $a=0.18$ fm.
This was obtained by fitting $V_{Q\bar{Q}}$ generated on a
$16^3\times 24$ lattice.
Here the emphasis was to get a good fit over the important range of
$r\sim (2-4)a$ and is in contrast to the potential in Ref.~\cite{Edwards},
which was designed to extract the string tension at large values of $r$. \\
2) As $v(34)$ for the  $V_{\bar{q}\bar{q}}$ potential. Here it is
assumed  to also be of the form in Eq.~\ref{v2fit}.\\
3) As $v(12)$ for  the  $V_{QQ}$ potential. This was calculated from the same
gauge configurations as the four-quark energies. In this case there was
no need to fit $V_{QQ}$ with a function of $R$, since it is only ever needed
at discrete values of $R$ -- the ones for which the four-quark energies
are calculated.
 
\vspace{0.3cm}

\noindent{\bf An effective quark mass $\mathbf{m_q}$.}\\
In this case $m_q$ can be determined beforehand by carrying out an EPT
analysis of the correponding {\it two-body} energies
in Fig.~\ref{ch6-spdf}. Using the potential in Eq.~\ref{v2fit},
a value of $m_q\approx 400$ MeV is able to give a good fit to the
spin-averaged energies with $L=0, \ 1, \ 2$ and 3. However, the results
are not strongly dependent on $m_q$ --- see Fig.~\ref{e2fit}.
A disturbing feature of this result is that $m_q\ll 1$ GeV indicating
the need for a relativistic approach --- as discussed in Sec.~\ref{Confu}.

\vspace{0.5cm}

In Eq.~\ref{KVN} the normalisation matrix ${\bf N}'(R, k_f)$ --- a
generalisation of ${\bf N}(f)$ in
Eq.~\ref{NVf} to non-static quarks in SU(3) --- can now be written as
\begin{equation}
\label{Nfs}
{\bf N'}(R, \ k_f)=\left( \begin{array}{ll}
 N(R, \ 0)   & \frac{1}{3}N(R, \ k_f) \\
\frac{1}{3}N(R, \ k_f) & N(R, \ 0)\\
\end{array} \right),
\end{equation}
where,  after integrating over $\bf{s_1}$ and $ \bf{s_2}$, 
$N(R, \ k_f)$ can be expressed as a sum of  terms of the form
\begin{equation}
\label{Nfi}
\frac{\pi^3}{(aX)^{3/2}}\exp\left[-(Z-\frac{Y^2}{X})R^2\right],
\end{equation}
where $a=0.5(a_i+a_j)+3k_f$, $c=0.5(c_i\pm c_j)+ 2k_f$,
$d=0.5(c_i\pm c_j)- 2k_f$,
$g=0.5(g_i+g_j)+4k_f$, $X=a-k_f^2/a$, $Y=c+k_fd/a$ and $Z=g-d^2/a$.

Since two of the quarks are not static there is now also a kinetic
energy matrix in Eq.~\ref{KVN}, namely,
\begin{equation}
\label{Kfs}
{\bf K'}(R, \ k_f)=\left( \begin{array}{ll}
 K_3(R,0)+K_4(R,0)   & \frac{1}{3}[K_3(R,k_f)+K_4(R,k_f)] \\
\frac{1}{3}[K_3(R,k_f)+K_4(R,k_f)] &  \ K_3(R,0)+K_4(R,0)\\
\end{array} \right),
\end{equation}
where, for example,
\begin{equation}
\label{Kfsp}
 K_3(R,k_f)=\int d^3s_1d^3s_2 \psi^{\star}(k_f)\left[-\frac{d^2}
{2m_q dr_3^2}\right]\psi(k_f).
\end{equation}
\markright{Bibliography} 
Again these integrals can be expressed in forms similar to that in
Eq.~\ref{Nfi}.

Finally, the  potential matrix ---  a generalisation of ${\bf V}(f)$ in
Eq.~\ref{NVf} to non-static quarks --- has the form
\begin{equation}
\label{Vfs}
{\bf V'}(R, \ k_f)=\left( \begin{array}{ll}
 \langle v(13),0\rangle+\langle v(24),0\rangle   &  \ \langle
V_{AB},k_f\rangle \\
\langle V_{AB},k_f\rangle & \langle v(14), 0\rangle+\langle v(23),
0\rangle \\
\end{array} \right).
\end{equation}
Here
\[\langle V_{AB},k_f\rangle =\langle V_{Q\bar{q}}\rangle-
\langle V_{\bar{q}\bar{q}}\rangle-\langle V_{QQ}\rangle,  \]
where
\begin{align}
\label{Vfsp}
\langle V_{Q\bar{q}}\rangle=& \frac{1}{3}\left[\langle v(13),k_f\rangle+
\langle v(24),k_f\rangle+\langle v(14),k_f\rangle+\langle
v(23),k_f\rangle\right],\nonumber \\
\langle V_{\bar{q}\bar{q}}\rangle=&\frac{1}{3}\langle v(34),k_f\rangle
\ \ \ {\rm and} \ \ \
\langle V_{QQ}\rangle=\frac{1}{3}N(R,k_f) V(2,R).
\end{align}
Here $N(R,k_f)$ is defined in Eq.~\ref{Nfs}, $V(2,R)$ is the
 potential between the two heavy quarks and,
for example,
\begin{equation}
\label{v13}
\langle v(13),k_f\rangle= \int d^3s_1d^3s_2
\psi^{\star}(k_f)V(s_1) \psi(k_f).
\end{equation}
For potentials of the form in Eq.~\ref{VQQ}, these integrals can be
expressed in terms of Error functions.
The energy $E(4,R,k_f)$ of the two heavy-light meson system is then
obtained by diagonalising Eq.~\ref{KVN}. Since this equation
is a $2\times 2$ determinant, a prediction can also be made for an
excited state $E^*(4,R,k_f)$ and the corresponding binding energy $B^*(4,R).$


\begin{thebibliography}{99}
\bibitem{pertref} {\it e.g.} R. K. Ellis, W.J. Stirling and B. R. Webber,
``QCD and collider physics", (Cambridge University Press, 1996);

E. Leader, ``An introduction to gauge theories and modern particle
physics: Vol. 2  \ CP-violation, QCD and hard processes", (Cambridge
University Press, 1996);

T. Muta, ``Foundations of QCD: An introduction to Perturbative Methods in
Gauge Theories (2nd Edition)", World Scientific Lecture Notes in
Physics -- Vol. 57 (World Scientific, 1998);

Y. L. Dokshitzer,  Phil. Trans. Roy. Soc. Lond. {\bf  A359}, 309 (2001)
 {\tt hep-ph/0106348};

P. Hoyer, Nucl. Phys. {\bf  A711}, 3 (2002),  {\tt hep-ph/0208181};

S. Capitani, Phys. Rept. {\bf  382}, 113 (2003), {\tt hep-lat/0211036}

\bibitem{lurev} M. L\"{u}scher, Annales Henri Poincare  {\bf  4}, S197 (2003),
 {\tt hep-ph/0211220};

M. Creutz, Nucl. Phys. Proc. Suppl.  {\bf 94}, 219 (2001), Latt00 and 
{\tt hep-lat/0010047};
``The Early Days of Lattice Gauge Theory", {\tt hep-lat/0306024};

C. Davies, ``Lattice QCD", Lectures given at 55th Scottish Universities
Summer School in Physics: Heavy Flavor Physics, St. Andrews, Scotland,
(Institute of Physics 2002, eds. C.T.H. Davies and S.M. Playfer, 2001) p.105,
{\tt hep-ph/0205181};
 
G. M\"{u}nster and M. Walzl, ``Lattice Gauge Theory --- a Short Primer", 
 Published in Zuoz 2000, Phenomenology of
gauge interactions p. 127, {\tt  hep-lat/0012005};

T. DeGrand, ``Lattice QCD at the end of 2003", {\tt  hep-ph/0312241} to
be published in Review for Int. J. Mod. Phys. A. (Worldscience)

\bibitem{ch6Aoki:2001hc} S.~Aoki {\it et al.}
 Nucl. Phys. Proc. Suppl. {\bf  106}, 230 (2002)
\bibitem{Lepage1} M. G. Alford, T. R. Klassen, G.P. Lepage,
 Phys. Rev. {\bf D58}, 034503 (1998), {\tt hep-lat/9712005};

G.P. Lepage, Nucl. Phys. Proc. Suppl. {\bf 47}, 3 (1996), also   Latt95
 and {\tt hep-lat/9510049}
\bibitem{clover1} B. Sheikholeslami and R. Wohlert, Nucl. Phys. {\bf
B259}, 572 (1985)
\bibitem{clover2} M. L\"{u}scher, S. Sint, R. Sommer, P. Weisz and U. Wolff,
Nucl. Phys. {\bf B491}, 323 (1997), {\tt hep-lat/9609035}
\bibitem{Kronfeld} A. S. Kronfeld, ``Uses of Effective Field Theory in
Lattice QCD" in: At the Frontiers of Particle Physics: Handbook of QCD,
(ed. M. Shifman) {\bf 4} \mbox{Chp. 39,} (World Scientific, Singapore, 2002),
{\tt hep-lat/0205021}
\bibitem{Kronfeld2} A. S. Kronfeld, eConf C020620:FRBT05 (2002),
{\tt hep-ph/0209231}
\bibitem{AlanI} A. C. Irving, Nucl. Phys. Proc. Suppl. {\bf 119}, 341 (2003),
Latt02 and {\tt hep-lat/0208065}
\bibitem{Farc} qq+q Collaboration, F. Farchioni, C. Gebert, I. Montvay
and L. Scorzato, ``On the price of light quarks",  {\tt hep-lat/0209142}
\bibitem{Creutz}  M. Creutz, ``Quarks, Gluons, and Lattices" (Cambridge
University, Cambridge, UK, 1983)
\bibitem{Montvay} I. Montvay and G. M\"{u}nster, ``Quantum Fields on a
Lattice" (Cambridge Monographs on Mathematical Physics, CUP 1994)
\bibitem{Rothe} H. J. Rothe, ``Lattice Gauge Theories", World Scientific
Lecture Notes in Physics - Vol. 59
(World Scientific Publishing Co., Singapore, 1997)
\bibitem{DetGo2} C. DeTar and S. Gottlieb, Physics Today, February 2004,
45
\bibitem{Neub2} H. Neuberger, ``Lattice Field Theory: past, present
and future", {\tt hep-ph/0402148} 
\bibitem{Symanzik} K. Symanzik, ``Recent Developments in Gauge Theories",
edited by G.'t Hooft {\it et al.} (Plenum, New York, 1980): Nucl. Phys. {\bf
B226}, 187, 205 (1983).
\bibitem{GasserL} J. Gasser and H. Leutwyler, Ann. Phys. {\bf 158},
142 (1984): Nucl. Phys. {\bf B250}, 465 (1985)
\bibitem{TT2002} A. W. Thomas, Nucl. Phys. Proc. Suppl. {\bf 119}, 50
(2003),  {\tt hep-lat/0208023}.
\bibitem{Soto} J. Soto, Quark Confinement and the Hadron Spectrum,
Proceedings of the 5th International Conference, Gargnano, Italy,
September 2002 (World Scientific 2003) p.227
\bibitem{vario} N. Brambilla, A. Pineda, J. Soto and A. Vario, Nucl.
Phys. {\bf B566}, 275 (2000);

 A.Vario, Quark Confinement and the Hadron
Spectrum, Proceedings of the 5th International Conference, Gargnano, Italy,
September 2002 (World Scientific 2003) p.73
\bibitem{Kolck} U. van Kolck, Nucl. Phys. {\bf A699}, 33 (2002);

U. van Kolck, L. J. Abu-Raddad and D. M. Cardamone,  {\tt nucl-th/0205058};

D. Phillips, Czech. J. Phys. {\bf 52}, B49 (2002),  {\tt nucl-th/0203040}
\bibitem{oller} J. A. Oller, AIP Conf. Proc. {\bf 660}, 116 (2003),
{\tt  nucl-th/0207086}
\bibitem{Weinberg} S. Weinberg, Physica {\bf A96}, 327 (1979);
Nucl. Phys. {\bf B363}, 3 (1991)
\bibitem{Argonne} S.C. Pieper and R. B. Wiringa, Ann. Rev. Nucl. Part.
Sci. {\bf 51}, 53 (2001)
\bibitem{Glen} N. Glendenning, ``Compact Stars" (Springer, 1996)
\bibitem{APR} A. Akmal, V. Pandharipande and D. Ravenhall, Phys. Rev.
{\bf C 58}, 1804 (1998)
\bibitem{Bonn} R. Machleidt and  I. Slaus,
 J. Phys. {\bf G27}, R69 (2001), {\tt  nucl-th/0101056};

 R. Machleidt, Nucl. Phys. {\bf A689}, 11 (2001)
{\tt nucl-th/0009055 }: Phys. Rev. {\bf C63}, 024001 (2001),
{\tt  nucl-th/0006014 }
\bibitem{Richardson} J. L. Richardson, Phys. Lett. {\bf 82B}, 272 (1979)
\bibitem{Cornell} E. Eichten, K.  Gottfried, T. Kinoshita, K. D. Lane and
T. M. Tan, Phys. Rev.  {\bf D21}, 203 (1980);

E. Eichten and F. Feinberg, Phys. Rev.  {\bf D23},
2724 (1981)
\bibitem{BhBr} R. K. Bhaduri and M. Brack, Phys. Rev. {\bf D25}, 1443
(1982)
\bibitem{BakerBZ} M. Baker, J. S. Ball and F. Zachariasen, Phys. Rev.
{\bf D51},  1968 (1995)
\bibitem{PGG} P. R. Page, T. Goldman and J. N. Ginocchio, Phys. Rev. Lett.
{\bf 86}, 204 (2001)
\bibitem{DiP} M. Di Pierro and E. Eichten, Phys. Rev. {\bf D64}, 114004 (2001)
\bibitem{crater} H. Crater, B. Liu and P. Van Alstine, ``Two-Body Dirac
Equations", {\tt hep-ph/0306291}
\bibitem{Brown+J} G. E. Brown and A. D. Jackson, ``The Nucleon-Nucleon
Interaction", (North--Holland Publishing Co., 1976)
\bibitem{WWeise} Editor W. Weise ``Quarks and Nuclei",
International Review of Nuclear Physics -- Vol. {\bf 1} 1984 (World
Scientific Publishing Co. Pte Ltd 1984)
\bibitem{SMBSY} T.~Sakai, J.~Mori, A.~J. Buchmann, K.~Shimizu and
K.~Yazaki, Nucl. Phys. {\bf A625}, 192 (1997),
{\tt nucl-th/9709054}
\bibitem{Oka2} M. Oka, ``Baryon-Baryon Interaction in the Quark Cluster
Model", {\tt hep-ph/0306173} 
\bibitem{coester} F. Coester, ``From Light Nuclei to Nuclear Matter ---
The R\^{o}le of Relativity", {\tt nucl-th/0111025};

F. Coester and W. N. Polyzou, ``Relativistic Quantum
Mechanics of Many Body Systems", {\tt nucl-th/0102050}
\bibitem{glozdan} L. Ya. Glozman and D. O. Riska, Phys. Rep. {\bf 268},
263 (1996), {\tt hep-ph/9505422}: Nucl. Phys. {\bf A603}, 326 (1996),
erratum {\it ibid} {\bf A620}, 510 (1997), {\tt hep-ph/9509269}
\bibitem{inst} S. Chernyshev, M. A. Nowak, and I. Zahed, Phys. \ Rev. \
{\bf D53}, 5176 (1996)
\bibitem{boost} H. Heiselberg and M. Hjorth-Jensen, Phys. Repts. {\bf
328}, 237 (2000)
\bibitem{shudan} E. Shuryak and D. O. Riska, {\it private communication}
\bibitem{Perkins} D. H. Perkins, ``An Introduction to High Energy Physics"
(Addison-Wesley Pub. Co., 1972)
\bibitem{Olsson}  S. Veseli and M.G. Olsson, Phys. Lett. {\bf B383}, 109 (1996),
{\tt hep-ph/9606257}
\markright{Bibliography}
\bibitem{Rebbi} C. Rebbi, Phys. Repts. {\bf 12C}, 1 (1974)
\bibitem{Wilson} K. Wilson, Phys. \ Rev. \  {\bf D10}, 2445  (1974)
\bibitem{sommerch6} R. Sommer, Nucl. Phys. {\bf B411}, 839 (1994),
{\tt  hep-lat/9310022}.
\bibitem{PP1} P.~Pennanen, Phys. \ Rev. \  {\bf D55}, 3958 (1997),
{\tt hep-lat/9608147}
\bibitem{Allton2} UKQCD Collaboration: C. R. Allton {\it et al.}, Phys. Rev.
{\bf D49}, 474 (1994), {\tt  hep-lat/9309002}

\bibitem{Edwards} R. G. Edwards, U. M. Heller and T. R. Klassen,
Nucl. Phys. {\bf B517}, 377 (1998), {\tt hep-lat/9711003}
\bibitem{GMS} A.M. Green, C. Michael and M.E. Sainio, Z. Phys. {\bf C67},
291 (1995), {\tt hep-lat/9404004}
\bibitem{JL}  A.M.~Green, J. Lukkarinen, P.~Pennanen and C.~Michael,
Phys. Rev. {\bf D53}, 261 (1996), {\tt hep-lat/9508002}
\bibitem{IP} N. Isgur and J.E. Paton,  Phys. \ Rev. \  {\bf D31},
2910 (1985)
\bibitem{BBZ} M. Baker, J.S. Ball and F. Zachariasen, Phys. \ Rev. \
{\bf D51}, 1968 (1995); Int. J. Mod. Phys.  {\bf A11}, 343 (1996)
\bibitem{MP98} UKQCD Collaboration: C.~Michael and J.~Peisa, Phys. Rev.
{\bf D58}, 034506 (1998), {\tt hep-lat/9802015}
\bibitem{GIJK} A.M.~Green, J.~Ignatius, M. ~Jahma and J.~Koponen, work in
progress
\bibitem{G+K+P+M} A.M.~Green, J.~Koponen, P.~Pennanen and C.~Michael,
Phys. \ Rev. \ {\bf D65}, 014512 (2002),
{\tt hep-lat/0105027}
\bibitem{G+K+P+M2} A.M.~Green, J.~Koponen, P.~Pennanen and C.~Michael,
Eur. Phys. J. {\bf C28}, 79 (2003), {\tt hep-lat/0206015}
\bibitem{GKP2} A. M. Green, J. Koponen and P.~Pennanen, Phys. Rev. {\bf D61},
014014 (2000), {\tt  hep-ph/9902249}
\bibitem{JonnaQQq}J. Koponen, work in progress
\bibitem{CMSU2} C.~Michael, Phys. Lett. {\bf B232}, 247 (1989),
Nucl. Phys. (Proc. Suppl) {\bf B 17}, 59 (1990)
\bibitem{SUNC} B. Lucini and M. Teper, JHEP {\bf 0106}, 050 (2001);

B. Lucini, M. Teper and U. Wenger, ``Features of SU(N) Gauge Theories",
{\tt  hep-lat/0309170}
\bibitem{Teper} M. Teper,
Phys. \ Rev. \ {\bf D59}, 014512 (1999);

B. Lucini and M. Teper, Phys. Rev. {\bf D66}, 097502 (2002)
\bibitem{anisolatt} F. Karsch, Nucl. Phys.  {\bf B205}, 285 (1982)
\bibitem{2+2}  G. Burgio {\it et al.}, Phys. Rev. {\bf D67}, 114502 (2003),
 {\tt  hep-lat/0303005} also Latt03 and {\tt  hep-lat/0309058}, 
{\tt  hep-lat/0310036}
\bibitem{anclover} CP-PACS, M. Okamoto {\it et al.}, Phys. \
Rev. \  {\bf D65}, 094508 (2002), {\tt  hep-lat/0112020}
\bibitem{anisotropic2} S. Hashimoto and M. Okamoto,
Phys. Rev.  {\bf D67}, 114503 (2003), {\tt hep-lat/0302012};

S. Sakai and A. Nakamura, `` Improved gauge action on an anisotropic
lattice", {\tt hep-lat/0311020}
\bibitem{Aoki} S. Aoki {\it et al.} (CP-PACS), Phys. Rev. Lett. {\bf 84}, 238
(2000),  {\tt  hep-lat/9904012}
\bibitem{Young} D. B. Leinweber,  A. W. Thomas, K. Tsushima and
S. V. Wright, Phys. \
Rev. \  {\bf D61}, 074502 (2000), {\tt  hep-lat/9906027};

R. D. Young, D. B. Leinweber, A. W. Thomas and
S. V. Wright, Phys. Rev.  {\bf D66}, 094507 (2002),
  {\tt  hep-lat/0205017}
\bibitem{Beane+S} S. R. Beane and M. J. Savage, Phys. Lett.  {\bf B535},
177 (2002), {\tt  hep-lat/0202013}
\bibitem{Beane+S2} S. R. Beane and M. J. Savage,
Phys. Rev. {\bf D67}, 054502 (2003), {\tt  hep-lat/0210046}
\bibitem{Lacock} P. Lacock and C. Michael, Phys. \ Rev. \  {\bf D52},
5213 (1995)
\bibitem{Foster} UKQCD Collaboration: M. Foster and C. Michael,
Phys. \ Rev. \  {\bf D59}, 074503 (1999),  {\tt hep-lat/9810021};

M.~Foster,  University of Liverpool PhD thesis 1998
\bibitem{Beane+SII} S. R. Beane and M. J. Savage,
Nucl. Phys. {\bf A717}, 91 (2003), {\tt  nucl-th/0208021}
\bibitem{epel} E. Epelbaum, Ulf-G. Meissner and W. Gl\"{o}ckle,
Nucl. Phys. {\bf A714}, 535 (2003), 
{\tt  nucl-th/0207089} and {\tt  nucl-th/0208040}
\bibitem{BBpmg}  UKQCD Collaboration: C.~Michael and P.~Pennanen,
Phys. Rev. {\bf D60}, 054012 (1999),
{\tt hep-lat/9901007}
\bibitem{lufinite} M. L\"{u}scher, Commun. Math. Phys.  {\bf 105}, 153 (1986) and
Nucl. Phys.  {\bf B354}, 531 (1991)
\bibitem{BBPS04} S.R. Beane, P.F. Bedaque, A. Parreno and  M.J. Savage,
``Two nucleons on a Lattice", {\tt   hep-lat/0312004}
\bibitem{Bedaq} P. Bedaque, ``Aharonov-Bohm effect and nucleon-nucleon
phase shifts on the lattice",  {\tt nucl-th/0402051}
\bibitem{pipiYam} CP-PACS Collaboration: T. Yamazaki {\it et al.},
``I=2 $\pi \pi$ Scattering Phase Shift with two Flavors of $O(a)$
Improved Dynamical Quarks", {\tt   hep-lat/0402025}
\bibitem{Hdib} I. Wetzorke and F. Karsch, Nucl. Phys. Proc. Suppl. 
{\bf 119}, 278 (2003), Latt02 and {\tt hep-lat/0208029}
\bibitem{Sasaki} S. Sasaki, `` Lattice study of exotic S=+1 baryon",
{\tt hep-lat/0310014}; 

F. X. Lee {\it et al.},`` A search for
Pentaquarks on the Lattice", poster at Lattice 2003 in Tsukuba
\bibitem{Nakano} T. Nakano {\it et al.}, Phys. Rev. Lett. {\bf 91}, 012002
(2003)
\bibitem{Balirev} G. Bali, Phys. Rept. {\bf 343}, 1 (2001),
{\tt hep-ph/0001312}
\bibitem{GMP}
A.M. Green, C. Michael and J.E. Paton,  Phys. Lett. {\bf B280}, 11 (1992)
\bibitem{GMP93} A.M.~Green, C.~Michael and J.E.~Paton, Nucl. Phys.
 {\bf A554}, 701 (1993), {\tt hep-lat/9209019}
\bibitem{GMPS}
A.M. Green, C. Michael, J.E. Paton and M.E. Sainio, Int. J. Mod. Phys. {\bf
E2}, 479 (1993), {\tt hep-lat/9301006}
\bibitem{flip} F. Lenz {\it et al.}, Ann. Phys. (N.Y.) {\bf 170}, 65
(1986);

K. Masutani, Nucl. Phys. {\bf A468}, 593 (1987)
\bibitem{CM1}
S. Perantonis, A. Huntley and C. Michael, Nucl. Phys. {\bf B326}, 544 (1989)
\bibitem{CM2}
S. Perantonis and C. Michael, Nucl. Phys. {\bf B347}, 854 (1990)
\bibitem{Booth} S.P. Booth {\it et al.}, Phys. Lett. {\bf B275}, 424 (1992)
\bibitem{DellaM} M. Della Morte {\it et al.}, `` Static quarks with
improved statistical precision", Latt03 and  {\tt hep-lat/0309080},
{\tt hep-lat/0307021}
\markright{Bibliography}
\bibitem{Hash} S. Hashimoto, Phys. Rev. {\bf D50}, 4639 (1994)
\bibitem{hyperb} A. Hasenfratz and F. Knechtli, Phys. Rev. {\bf D64},
034504 (2001); 

A. Hasenfratz, R. Hoffmann and F. Knechtli, Nucl. Phys. B
(Proc. Suppl.) {\bf 106}, 418 (2002)
\bibitem{hyperb2} K. Choi and W. Lee,``Penguin diagrams for the HYP
staggered fermions", Latt03 and {\tt hep-lat/0309070}; 

T. Bhattacharya {\it et al.},
``Calculating weak matrix elements using  HYP  staggered fermions",
Latt03 and {\tt hep-lat/0309105}; 

S. Bilson-Thompson and W. Lee,`` Description and
comparison of Fat7 and HYP fat links", {\tt hep-lat/0310056}
\bibitem{Ohta}
S. Ohta, M. Fukugita and A. Ukawa, Phys. Lett. {\bf B173}, 15 (1986).
\bibitem{Paton}
A.M. Green and J.E. Paton, Nucl. Phys. {\bf A492}, 595 (1989)
\bibitem{asypsca} C.~Michael, Phys. Lett. {\bf B283}, 103 (1992)
\bibitem{Gavela} M. B. Gavela {\it et al.}, Phys. Lett.  {\bf B82}, 431 (1979).
\bibitem{Morimatsu2} O. Morimatsu, A. M. Green and J.E. Paton, Phys.
Lett. {\bf B258}, 257 (1991)
\bibitem{Lang} J. T. A. Lang, J. E. Paton and A. M. Green, Phys. Lett.
 {\bf B366}, 18 (1996), {\tt hep-ph/9508315}
\bibitem{Morimatsu} O. Morimatsu, Nucl. Phys. {\bf A505}, 655 (1989)
\bibitem{Masud}
B. Masud, J.E. Paton, A.M. Green and G.Q. Liu, Nucl. Phys. {\bf A528}, 477 (1991)
\bibitem{MatSiv} H. Matsuoka and D. Sivers, Phys. \ Rev. \ {\bf D33},
1441 (1987)
\bibitem{Morimatsu3}
C. Alexandrou, T. Karapiperis and O. Morimatsu, Nucl. Phys. {\bf A518}, 723
(1990)
\bibitem{GLW} A.M. Green, G.Q. Liu and S. Wycech, Nucl. Phys. {\bf A509},
687 (1990)
\bibitem{GPplb426} A.M. Green and P. Pennanen, Phys. Lett. {\bf B426}, 243
(1998),  {\tt hep-lat/9709124}

\bibitem{GPprc57} A.M. Green and P. Pennanen, Phys. \ Rev. \ {\bf C57},
3384 (1998), {\tt hep-lat/9804003}

\bibitem{cbern} C. Bernard {\it et al.}, Phys. Rev. Lett. {\bf 81}, 4812
(1998), {\tt hep-ph/9806412}
\bibitem{PDG} D.E.~Groom {\it et al.}, Review of Particle Physics,
Eur. J. Phys. {\bf C15}, 1 (2000)
\bibitem{Nodestate} C. Weiser, Proceedings of the 28th International
Conference on High Energy Physics, Warsaw 1996, p.531 (Edited by I. Adjuk and
A. Wroblewski, World Scientific 1996) --- see also
http://www-ekp.physik.uni-karlsruhe.de/~weiser/proc\_warsaw/index.html
\bibitem{McN+M}
UKQCD Collaboration: C.~McNeile and C.~Michael,
Phys. Rev. {\bf D63}, 114503 (2001), {\tt hep-lat/0010019}
\bibitem{GKMMT03} A.M.~Green, J.~Koponen, C.~McNeile, C.~Michael and
G. Thompson, Phys. Rev. {\bf  D69}, 094505 (2004), {\tt hep-lat/0312007}
\bibitem{HJ1} H. J. Schnitzer, Phys.\ Rev.\  {\bf D18}, 3482 (1978)
\bibitem{HJ2} H. J. Schnitzer, Phys. Lett.  {\bf B226}, 171 (1989)
\bibitem{alex} C. Alexandrou, Ph. de Forcrand and A. Tsapalis,
Phys. Rev. {\bf D66}, 094503 (2002), {\tt hep-lat/0206026}
\bibitem{alex2} C. Alexandrou, Ph. de Forcrand and A. Tsapalis,
Nucl. Phys. Proc. Suppl. {\bf 119}, 422 (2003), Latt02 and {\tt hep-lat/0209067}.
\bibitem{Allton} UKQCD Collaboration: C.R.Allton {\it et al.},
Phys.\ Rev.\  {\bf D60}, 034507 (1999), {\tt hep-lat/9808016}
\bibitem{qmass}
UKQCD Collaboration: C.~McNeile and C.~Michael, Phys. \ Lett. \
{\bf B491}, 123 (2000), {\tt hep-lat/0006020}
\bibitem{Mur} V. D. Mur, V. S. Popov, Yu. A. Simonov and V. P. Yurov,
J. Exp. Theor. Phys. {\bf 78}, 1 (1994), {\tt  hep-ph/9401203}
\bibitem{luterm} M. L\"{u}scher, K. Symanzik and P. Weisz, Nucl. Phys.
{\bf B173}, 465 (1980)

M. L\"{u}scher, Nucl. Phys.  {\bf B180}, 317 (1981)
\bibitem{Paris} M. W. Paris, Phys. Rev. {\bf C68}, 025201 (2003),
 {\tt  nucl-th/0305020}
\bibitem{Kuzs} D. S. Kuzmenko and Yu. A. Simonov, 
Phys. Atom. Nucl. {\bf 66}, 950 (2003), Yad. Fiz.  {\bf 66}, 983 (2003), 
{\tt hep-ph/0202277} and {\tt hep-ph/0302071}
\bibitem{Ichie} H. Ichie, V. Bornyakov, T. Streuer and G. Schierholz,
Nucl. Phys. {\bf A721}, 899 (2003),  {\tt hep-lat/0212036} also
{\tt hep-lat/0304008};

H. Suganuma, T.T. Takahashi and H. Ichie, ``Detailed Lattice-QCD Study for
the Three-Quark Potential and Y-type  Flux-Tube Formation",
 {\tt hep-lat/0312031};

T. Takahashi,  H. Suganuma and H. Ichie, ``Y-Type Flux tube formation in
Baryons",    {\tt hep-lat/0401001}
\bibitem{Zeng} J. Zeng, J. W. Van Orden and W. Roberts,
Phys. Rev. {\bf D52},   5229 (1995), {\tt hep-ph/9412269}
\bibitem{Timo} T. A. L\"{a}hde, C. J. Nyf\"{a}lt and D. O. Riska, Nucl.
Phys. {\bf A674}, 141 (2000), {\tt hep-ph/9908485}
\bibitem{Bethe-S} E.E. Salpeter and H. Bethe, Phys. Rev. {\bf 84}, 1232
(1951)
\bibitem{Salpeter} E.E. Salpeter, Phys. Rev. {\bf 87}, 328 (1952)
\bibitem{Gbreit} G. Breit, Phys. Rev. {\bf 34}, 553 (1929)
 \bibitem{Hallref} W. Lucha and F. F. Sch\"{o}berl, Int. J. Mod. Phys. 
{\bf A14}, 2309 (1999), {\tt hep-ph/9812368};

R. L. Hall, W. Lucha and F. F. Sch\"{o}berl,
Int. J. Mod. Phys. {\bf A18}, 2657 (2003), {\tt hep-th/0210149}

\bibitem{SRule} K.C.~Bowler, L.~Del Debbio, J.M.~Flynn, G.N.~Lacagnina,
V.I.~Lesk, C.M.~Maynard and D.G.~Richards, Nucl. Phys. {\bf B619}, 507 (2001),
 {\tt hep-lat/0007020};

A.A.~Khan {\it et al.} CP-PACS, Phys. Rev. {\bf D65}, 054505 (2002),
Erratum-ibid. {\bf D67}, 059901 (2003),  {\tt hep-lat/0105015};

S.~Aoki {\it et al.} CP-PACS, Nucl. Phys. Proc. Suppl.  {\bf 106}, 780 (2002),
Latt01 and {\tt hep-lat/0110128}
\bibitem{vol:76}
M.~B. Voloshin and L.~B. Okun, Pisma Zh. Eksp. Teor. Fiz. {\bf 23}, 369
(1976)

\bibitem{jaf:76}
R.~L. Jaffe, Phys. Rev. {\bf D15}, 267 (1976);
C.~W. Wong and K.~F. Liu, Phys. Rev. {\bf D21}, 2039 (1980).

\bibitem{gut:79}
F.~Gutbrod, G.~Kramer and C.~Rumpf, Zeit. Phys. {\bf C1}, 391 (1979)
\bibitem{tor:91}
N.~A. T\"ornqvist, Phys. Rev. Lett. {\bf 67}, 556 (1991);
Z. Phys. {\bf C61}, 525 (1994), {\tt hep-ph/9310247}
\bibitem{BeaneQQ} D. Arndt, S. Beane and M. Savage,
Nucl. Phys.  {\bf A726}, 339 (2003), {\tt hep-lat/0304004}.
\markright{Bibliography}
\bibitem{boy:96}
M.~M. Boyce, ``String inspired QCD and E(6) models", Ph.D. thesis,
Carleton  University, 1996, {\tt  hep-ph/9609433}

\bibitem{lip:86}
H.~J. Lipkin, Phys. Lett. {\bf B172}, 242 (1986)

\bibitem{zou:86}
S.~Zouzou, B.~Silvestre-Brac, C.~Gignoux and J.-M. Richard, Z. Phys.
{\bf C30},  457 (1986);

J.-M. Richard, ``Hadrons with two heavy quarks", Proc. Conf. on
future of high sensitivity charm experiments, Batavia (1994),
  {\tt hep-ph/9407224}
\bibitem{BBold} D. Richards, D. Sinclair and D. Sivers, Phys. Rev. {\bf D42},
3191 (1990)
\bibitem{BBcanada} C. Stewart and R. Koniuk, Phys. Rev. {\bf D57}, 5581
(1998),
{\tt hep-lat/9803003}

\bibitem{mih:97}
A.~Mihaly, H.~R. Fiebig, H.~Markum and K.~Rabitsch, Phys. Rev. {\bf
D55}, 3077  (1997);

H.~R. Fiebig, H.~Markum, A.~Mihaly, K.~Rabitsch and R.~M. Woloshyn,
Nucl. Phys.
  Proc. Suppl. {\bf 63}, 188 (1998), {\tt hep-lat/9709152}

\bibitem{BBpmg2}P.~Pennanen, A. M. Green and  C.~Michael,
Nucl. Phys. Proc. Suppl. {\bf 73}, 351 (1999);

C. Michael, Proceedings of Confinement III, Newport News, VA (1998),
{\tt hep-ph/9809211}

\bibitem{Divitiis} G.M.~de Divitiis, L.~Del Debbio, M.~Di Pierro,
J.M.~Flynn, C.~Michael and J.~Peisa, JHEP 9810, 010 (1998), {\tt
hep-lat/9807032}

\bibitem{GKP3} A. M. Green, J. Koponen and P.~Pennanen,
Nucl. Phys. Proc. Suppl. {\bf 83}, 292 (2000), {\tt  hep-ph/9908016}

\bibitem{UKQCDmp} UKQCD Collaboration: C.~Michael and P.~Pennanen,
work in progress.

\bibitem{Barnes1} T.~Barnes, N.~Black, D.J.~Dean and E.S.~Swanson,
Phys. Rev. {\bf C60}, 045202 (1999), {\tt nucl-th/9902068};

S.~Pepin, F.~Stancu , M.~Genovese and  J.M.~Richard,
Phys. Lett. {\bf B393}, 119 (1997)
\bibitem{Bolder00} B. Bolder {\it et al.}, Phys. Rev. {\bf D63}, 074504 (2001)  
\bibitem{phi:98b}
O.~Philipsen and H.~Wittig, Phys. Rev. Lett. {\bf 81}, 4056 (1998),
{\tt hep-lat/9807020};

{\bf ALPHA} Collaboration, F.~Knechtli and R.~Sommer, Phys. Lett. {\bf
B440},  345 (1998), {\tt  hep-lat/9807022}

\bibitem{ste:99}
P.~W. Stephenson, Nucl. Phys. {\bf B550}, 427 (1999),
{\tt hep-lat/9902002};

O.~Philipsen and H.~Wittig, Phys. Lett. {\bf B451}, 146 (1999),
  {\tt hep-lat/9902003};

P.~de~Forcrand and O.~Philipsen, Phys. Lett.  {\bf B475}, 280 (2000), 
{\tt hep-lat/9912050};
 
S. Kratochvila and P.~de~Forcrand, Nucl. Phys. {\bf B671}, 103 (2003),
{\tt  hep-lat/0306011}


\bibitem{tro:98}
H.~Trottier, Phys. Rev. {\bf D60}, 034506 (1999), {\tt hep-lat/0209048}

\bibitem{ppcmsb} UKQCD Collaboration: P.~Pennanen and C.~Michael,
``String breaking in zero-temperature lattice QCD", {\tt hep-lat/0001015}

\bibitem{ppcmsb0} UKQCD Collaboration: P.~Pennanen,  C.~Michael and
A. M.~Green, Nucl. Phys. Suppl. {\bf 83}, 200 (2000),
{\tt hep-lat/9908032}

\bibitem{LeY} A. L. Yaouanc, L. Oliver, O. Pene and J. C. Raynal,
Phys. Rev. {\bf D9}, 1415 (1974)
\bibitem{IPK} N. Isgur, R. Kokoski and J. Paton, Phys. Rev. Lett. {\bf 54},
869 (1985)
\bibitem{4Qfluxmap1} A. M. Green, C.~Michael and P. S. Spencer,
Phys. Rev. {\bf D55}, 1216 (1997), {\tt hep-lat/9610011}
\bibitem{4Qfluxmap2} P.~Pennanen, A. M. Green and  C.~Michael,
Phys. Rev. {\bf D56}, 3903 (1997), {\tt hep-lat/9705033}
\bibitem{4Qfluxmap3} P.~Pennanen, A. M. Green and  C.~Michael,
Phys. Rev.  {\bf D59}, 014504(1999), {\tt hep-lat/9804004}
\bibitem{glock} W. Gl\"{o}ckle {\it et al.}, Acta Phys. Pol.  {\bf B32},
3053 (2001), {\tt nucl-th/0109070}
\bibitem{Flynn} UKQCD collaboration, J. M. Flynn, F. Mescia 
and A. S. B. Tariq, JHEP {\bf 0307}, 066 (2003), {\tt hep-lat/0307025}
\bibitem{O184p2h} G.E.~Brown and A.~M. Green, Nucl. Phys. {\bf 85}, 87 (1966)  
\bibitem{collrev} A.~M. Green, Repts. Prog. Phys., {\bf 28}, 113 (1965)
\bibitem{Brownbook} G.E.~Brown, ``Unified Theory of Nuclear Models and
Forces", (North-Holland Publishing, Amsterdam, 1967)

\bibitem{fur+m} S.~Furui, A.~M. Green and B.~Masud, Nucl. Phys.
{\bf A582},  682 (1995),  {\tt hep-lat/0006003}

\end{thebibliography}
\end{document}